\documentclass[onecolumn,amsmath,amssymb,nofootinbib,12pt]{article}
\usepackage{jheppub}
\usepackage{ifpdf}

\usepackage{graphicx,subcaption}
\graphicspath{ {figures/} }
\usepackage{amsfonts}
\usepackage{amsmath}
\usepackage{amssymb}
\usepackage{mathtools}
\usepackage{tensor}
\usepackage{epsfig}
\usepackage{pdfpages}
\usepackage{bbm}
\usepackage{graphicx,epstopdf}
\usepackage[numbers]{natbib}
\usepackage[makeroom]{cancel}
\usepackage{hyperref}
\usepackage{tikzsymbols}
\hypersetup{pdftex,colorlinks=true,allcolors=blue}
\usepackage{array}
\usepackage[export]{adjustbox}

\usepackage[normalem]{ulem}

\newcommand{\Rbb}{\mathbb{R}}

\newcommand{\Zbb}{\mathbb{Z}}
\newcommand{\Ocal}{\mathcal{O}}

\newcommand{\Brane}{{\rm brane}}
\newcommand{\brane}{\mt{brane}}
\newcommand{\bulk}{\mt{bulk}}
\newcommand{\ren}{\mt{ren}}
\newcommand{\area}{A}
\newcommand{\gen}{\mt{gen}}
\newcommand{\eff}{\mt{eff}}

\newcommand{\hyperF}{\tensor[_2]{F}{_1}}
\newcommand{\CFTR}{\mathbf{R}}
\newcommand{\braneR}{R}
\newcommand{\AdS}{\mathrm{AdS}}

\newcommand{\RTsurf}{\Sigma_\xR} 

\newcommand{\AdSdmetric}{{g}^{\AdS_d}} 
\newcommand{\RTmetric}{h}
\newcommand{\QESmetricNoz}{\mathfrak{h}}

\newcommand{\curvK}{\mathcal{K}}
\newcommand{\inducedK}{\tilde{\curvK}}
\newcommand{\inducedR}{\tilde{R}}
\newcommand{\inducedg}{\tilde{g}}

\newcommand{\extr}{{\rm ext}}
\newcommand{\islands}{\mathrm{islands}}

\newcommand{\IR}{\mt{IR}}
\newcommand{\cT}{c_\mt{T}}
\newcommand{\overscript}[2]{\overset{\scriptscriptstyle{(#2)}}{#1}{}}

\newcommand{\xR}{\mathbf{R}}
\newcommand{\xV}{\mathbf{V}}
\newcommand{\EE}{\mt{EE}}
\newcommand{\tilh}{{\tilde h}}

\newcommand{\RN}[1]{%
	\textup{\uppercase\expandafter{\romannumeral#1}}%
}
\newcommand{\labell}[1]{\label{#1}} 

\definecolor{limegreen}{rgb}{0.2, 0.8, 0.2}
\definecolor{deepcarrotorange}{rgb}{0.91, 0.41, 0.17}
\definecolor{darkviolet}{rgb}{0.58, 0.0, 0.83}
\definecolor{cyan}{rgb}{0.0, 0.72, 0.92}
\definecolor{plum}{rgb}{0.67, 0.0, 0.55}

\newcommand{\rcm}[1]{\textcolor{red}{\bf [[Rob: #1]]}}

\newcommand{\vc}[1]{\textcolor{magenta}{\bf [[Vincent: #1]]}}
\newcommand{\dn}[1]{\textcolor{limegreen}{\bf [[Dominik: #1]]}}

\newcommand{\hd}[1]{\noindent{\bf #1}\ }

\newcommand{\eg}{{\it e.g.,}\ }
\newcommand{\ie}{{\it i.e.,}\ }
\newcommand{\reef}[1]{(\ref{#1})}
\newcommand{\ssc}{\scriptscriptstyle}
\newcommand{\mt}[1]{\textrm{\tiny #1}}

\newcommand{\s}{z_\mt{B}}

\newcommand{\lb}{\ell_\mt{B}}
\newcommand{\lamb}{\lambda_{b}}
\newcommand{\lgb}{\lambda_\mt{GB}}
\newcommand{\thb}{\theta_\mt{CFT}}
\newcommand{\zeb}{\zeta_\mt{CFT}}
\newcommand{\SCFT}{\Sigma_\mt{CFT}}
\newcommand{\puv}{P_\mt{UV}}
\newcommand{\pb}{P_\mt{B}}
\newcommand{\pbo}{P_\mt{B,0}}
\newcommand{\veps}{\varepsilon}
\newcommand{\sgen}{S_\mt{gen}}

\newcommand{\eps}{\epsilon}

\newcommand{\mK}{{\cal K}} 
\newcommand{\mR}{{\cal R}} 
\newcommand{\ric}{{\tilde R}}

\newcommand{\ads}[1]{{\mt{AdS}_{\tiny #1}}}
\newcommand{\tg}{{\tilde g}}
\newcommand{\tR}{{\tilde R}}
\newcommand{\tdr}{{\tilde r}}

\newcommand{\beq}{\begin{equation}}
\newcommand{\eeq}{\end{equation}}
\newcommand{\beqa}{\begin{eqnarray}}
\newcommand{\eeqa}{\end{eqnarray}}

\newcommand{\bea}{\begin{eqnarray}}
\newcommand{\eea}{\end{eqnarray}}

\renewcommand{\(}{\left(}
\renewcommand{\)}{\right)}
\renewcommand{\[}{\left[}
\renewcommand{\]}{\right]}

\newcommand{\mO}{\mathcal{O}}

\newcommand{\Gbr}{G_\mt{brane}}
\newcommand{\Gbk}{G_\mt{bulk}}
\newcommand{\Geff}{G_\mt{eff}}
\newcommand{\leff}{\ell_\mt{eff}}

\newcommand{\RNum}[1]{\uppercase\expandafter{\romannumeral #1\relax}}

\usepackage[thinlines]{easytable}
\usepackage{textcomp}

\title{\boldmath Quantum Extremal Islands Made Easy, Part I:\\
Entanglement on the Brane}

\author[a,b]{Hong Zhe Chen,}
\author[a]{Robert C. Myers,}
\author[a]{Dominik Neuenfeld,}
\author[c]{Ignacio A. Reyes}
\author[a,b]{and Joshua Sandor}

\affiliation[a]{Perimeter Institute for Theoretical Physics, Waterloo, ON N2L 2Y5, Canada}
\affiliation[b]{Dept.~of Physics $\&$ Astronomy, University of Waterloo, Waterloo, ON N2L 3G1, Canada}
\affiliation[c]{Max-Planck-Institut f\"ur Gravitationsphysik, Am M\"uhlenberg 1, 14476 Potsdam, Germany}

\emailAdd{hchen2@pitp.ca}
\emailAdd{rmyers@pitp.ca}
\emailAdd{dneuenfeld@pitp.ca}
\emailAdd{ignacio.reyes@aei.mpg.de}
\emailAdd{jsandor@perimeterinstitute.ca}

\abstract{Recent progress in our understanding of the black hole information paradox has lead to a new prescription for calculating entanglement entropies, which involves special subsystems in regions where gravity is dynamical, called \textit{quantum extremal islands}. We present a simple holographic framework where the emergence of quantum extremal islands can be understood in terms of the standard Ryu-Takayanagi prescription, used for calculating entanglement entropies in the boundary theory. Our setup describes a $d$-dimensional boundary CFT coupled to a ($d$--1)-dimensional defect, which are dual to global AdS${}_{d+1}$ containing a codimension-one brane. Through the Randall-Sundrum mechanism, graviton modes become localized at the brane, and in a certain parameter regime, an effective description of the brane is given by Einstein gravity on an AdS${}_d$ background coupled to two copies of the boundary CFT. Within this effective description, the standard RT formula implies the existence of quantum extremal islands in the gravitating region, whenever the RT surface crosses the brane. This indicates that islands are a universal feature of effective theories of gravity and need not be tied to the presence of black holes.}

\begin{document}
\maketitle
\flushbottom





\section{Introduction}\label{sec:introduction}
%

Almost half a century ago, it was discovered that black holes behave as quantum objects, with an associated temperature, entropy and other thermodynamic properties \cite{Hawking:1974sw,Hawking:1974rv,Hawking:1976de,Bekenstein:1972tm,Bekenstein:1973ur}. One realization of these ideas is the Bekenstein-Hawking (BH) formula, which states that the black hole entropy is a quarter of its horizon area measured in Planck units, \ie $S_\mt{BH}=A/4 G_\mt{N}$. These concepts gained a wider scope in the context of the AdS/CFT correspondence, where the Ryu-Takayanagi (RT) prescription \cite{Ryu:2006ef,Ryu:2006bv,Hubeny:2007xt,Rangamani:2016dms} applies the same geometric expression to extremal bulk surfaces in evaluating the entanglement entropy for generic subregions on the boundary theory. Indeed, this was later derived as a special case of the generalized gravitational entropy in \cite{Lewkowycz:2013nqa}.

However, as pointed out by Hawking early on \cite{Hawking:1976ra}, a standard semiclassical analysis seemingly leads to an inconsistency in describing the time evolution of black holes. If a pure state of matter collapses to form a black hole, which is then allowed to completely evaporate via Hawking radiation, the final quantum state appears to be mixed, contradicting unitary evolution. This is the black hole information paradox.  On the other hand, arguments from the AdS/CFT correspondence suggest that unitarity should remain valid, \eg \cite{Polchinski:2016hrw,Harlow:2014yka}. There, one expects that after an initial rise of the entanglement entropy of the Hawking radiation, subtle correlations between the quanta emitted at early and late times lead to a purification of the final state and a decrease in the late-time entropy. This qualitative behaviour of the entropy is known as the Page curve \cite{Page:1993wv} -- see also \cite{Harlow:2014yka}.

As emphasized with the generalized second law \cite{Bekenstein:1974ax} (see also \cite{Wall:2009wm,Wall:2011hj}), the geometric BH entropy is naturally combined with the entanglement entropy of quantum fields outside the event horizon to produce a finite quantity known as the generalized entropy. In the context of holographic entropy, this leads to an extension of the RT prescription to include quantum corrections in the bulk \cite{Faulkner:2013ana,Engelhardt:2014gca}
\begin{align}
	\label{eq:sgen_intro}
	 S_\EE(\xR) = {\rm min}\left\{\extr\,
 S_\gen(\xV)\right\}={\rm min}\left\{\extr
  \( \frac{A(\xV)}{4 G_\mt{N}} + S_\mt{QFT}\)\right\}\,,
\end{align}
where $\xV$ is a bulk surface homologous to the boundary subregion $\xR$, while $S_\mt{QFT}$ is the entropy of the quantum fields on a partial Cauchy surface extending from $\xV$ to $\xR$ on the asymptotic boundary. The surface which extremizes the generalized entropy in the above expression is then referred to as a Quantum Extremal Surface (QES) \cite{Engelhardt:2014gca}. Further, the `min' indicates that in the situation where there is more than one extremal surface, one chooses that which yields the minimum value for $S_\gen(\xV)$. 

This approach produced some surprising new insights with holographic models of black hole evaporation \cite{Almheiri:2019psf, Penington:2019npb, Almheiri:2019hni}. In particular, at late stages in the evaporation, the quantum term can compete with the classical BH contribution in eq.~\reef{eq:sgen_intro} to produce new saddle points for the QES, which could describe the late-time phase of the Page curve. Perhaps the biggest surprise is that the Page curve can be reproduced from saddlepoint calculations in semi-classical gravity, \ie in a situation where the details of the black hole microstates or of the encoding of information in the Hawking radiation are still not revealed. Further, the evaluation of the entanglement entropy of the Hawking radiation is seen to be encapsulated by the so-called `island rule' \cite{Almheiri:2019hni},
\beq
 S_\EE(\xR) ={\rm min}\left\{\extr
  \(  S_\mt{QFT}(\braneR \cup \text{islands}) + \frac{A\(\partial(\islands)\)}{4 G_\mt{N}}\)\right\}\,.
 \labell{wonderA}
\eeq
That is, the entropy of the radiation collected in a nongravitating reservoir is evaluated as the contributions from the quantum fields in the reservoir but possibly also on a quantum extremal island (QEI) in the gravitating region, \ie a separate region near the black hole, as well as a geometric BH contribution from the boundary of the island. In the early phase of the Hawking evaporation, extremizing this expression yields the empty set for the island, \ie there is no island. However, at late times, a QEI appears to reduce the radiation's entropy and yields the expected late-time behaviour of the Page curve. These results have sparked further progress with a variety of new investigations, \eg \cite{Almheiri:2019yqk, Almheiri:2019psy, Almheiri:2019qdq, Penington:2019kki, Akers:2019nfi, Rozali:2019day, Chen:2019uhq, Bousso:2019ykv, Gautason:2020tmk, Hartman:2020swn, Marolf:2020xie, Hollowood:2020cou, Anegawa:2020ezn, Hashimoto:2020cas, Sully:2020pza, Balasubramanian:2020hfs, Alishahiha:2020qza, Geng:2020qvw, Krishnan:2020oun}. 

In this paper, we aim to explore the island formula \eqref{wonderA} in further generality. Recall that the latter was motivated by the `doubly holographic' model presented in \cite{Almheiri:2019hni}, who in turn began with the two-dimensional model of \cite{Almheiri:2019psf}. The latter consists of a bath, \ie a two-dimensional CFT on a half line, and a pair of quantum mechanical systems, which are assumed to be holographically dual to Jackiw-Teitelboim (JT) gravity on AdS$_2$ coupled to the same CFT as in the bath. Hence if the quantum mechanical systems begin in a thermofield double state, the dual description is given by a two-sided AdS$_2$ black hole. If the boundary of the bath is then coupled to one of the quantum  systems, \ie to the asymptotic boundary of one side of the black hole, the black hole begins to evaporate as Hawking radiation leaks into the bath. Now the insight of \cite{Almheiri:2019hni} was to examine the case where the two-dimensional CFT is itself \textit{holographic}, and so can be replaced with a locally AdS$_3$ bulk. The boundary of this bulk geometry has two components: the asymptotically AdS boundary, on which the bath lives, and the Planck brane, where the JT gravity is supported. This third perspective on the system has the advantage that the generalized entropy in eq.~\reef{eq:sgen_intro} or \reef{wonderA} is realized completely geometrically. That is, the entanglement entropy of the boundary CFT is computed by RT surfaces in the three-dimensional bulk, and the geometric BH contribution is given by the usual expression for JT gravity. Further, calculations in this doubly holographic model produce the expected Page curve, with RT surfaces ending on the Planck brane manifesting the island rule \eqref{wonderA}. 

In the present paper, we generalize this doubly holographic model to higher dimensions as follows (see also figure \ref{fig:threetales}): We consider a $d$-dimensional holographic CFT coupled to a codimension-one conformal defect. As usual, the gravitational dual corresponds to an asymptotically AdS$_{d+1}$ spacetime, containing a codimension-one brane anchored on the asymptotic boundary at the position of the defect. The gravitational backreaction of the brane warps the geometry creating localized graviton modes in its vicinity, as per the usual Randall-Sundrum (RS) scenario \cite{Randall:1999ee,Randall:1999vf,Karch:2000ct}. Hence at sufficiently long wavelengths, the system can then also be described by an effective theory of Einstein gravity coupled to (two copies of) the holographic CFT on the brane, all coupled to the CFT on the static boundary geometry.\footnote{Some tuning of the parameters characterizing the brane is required to achieve this effective description. Note that the fact that the RS gravity on the brane has a finite cutoff \cite{Randall:1999ee,Randall:1999vf} makes conspicuous that this is only an effective theory.} To better emulate the previous model with JT gravity \cite{Almheiri:2019hni}, we also consider introducing an intrinsic Einstein term to the brane action, analogous to the construction of Dvali, Gabadadze and Porrati (DGP) \cite{Dvali:2000hr}.\footnote{Without the DGP term, our construction resembles that in \cite{Rozali:2019day} in many respects. Our model resembles the setup in \cite{Almheiri:2019hni} even more closely if we make a $\mathbb Z_2$ orbifold quotient across the brane.}  In any event, this more or less standard holographic model can be viewed from three perspectives in analogy with \cite{Almheiri:2019hni}: the bulk gravity perspective, with a brane coupled to gravity in an asymptotically AdS$_{d+1}$ space; the boundary perspective, with the boundary CFT coupled to a conformal defect; and the brane perspective, with a region where the holographic CFT couples to Einstein gravity and another region where the same CFT propagates on a fixed background geometry.

\begin{figure}[h]
	\def\svgwidth{1\linewidth}
	\centering{
		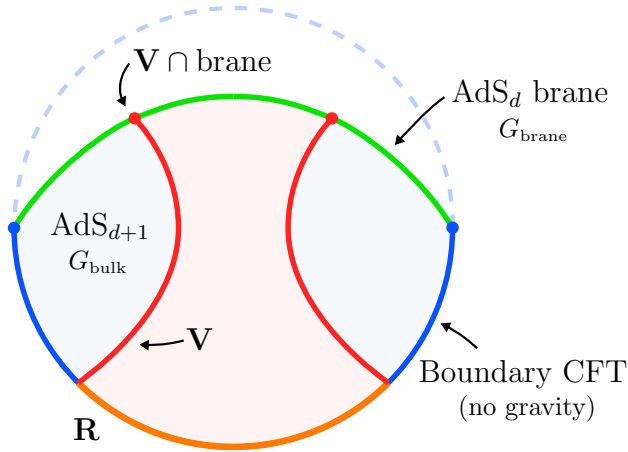
		\caption{A sketch of our holographic setup illustrating the various elements appearing in eq.~\eqref{eq:sad0}, which manifests the island rule in our analysis. }
                \label{fig:first} 	}
\end{figure}

From the bulk gravity perspective, the entanglement entropy is realized in a completely geometric way in terms of the areas of RT surfaces, with a contribution in the bulk and another contribution on the brane. That is, we have an extension of the usual RT prescription with
\beq
 S_\EE(\xR) = {\rm min}\left\{\extr\,
 S_\gen(\xV)\right\}={\rm min}\left\{\extr
  \(
  \frac{A(\xV)}{4 G_\bulk} + \frac{A(\xV \cap {\rm brane})}{4 G_\brane}\)\right\}\,,
 \label{eq:sad0}
\eeq
where again, where $\xV$ is a bulk surface homologous to the boundary subregion $\xR$ (see figure \ref{fig:first}). Note that the brane contribution seems natural here, we will argue for its presence by extending the derivation in \cite{Myers:2010tj}. In contrast to eq.~\reef{eq:sgen_intro}, we are not considering quantum field contributions in the AdS$_{d+1}$ bulk. However, from the brane perspective, the usual RT term, \ie the first term on the right-hand side of eq.~\reef{eq:sad0}, is interpreted as the leading planar contribution of the boundary CFT to $S_\EE(\xR)$, and the island rule \eqref{wonderA} is realized in situations where the RT surface cross over the brane. 

We emphasize the underlying \textit{simplicity} of our holographic model. In particular, the elements of construction are more or less standard, and the entropies are evaluated with the geometric formula for holographic entanglement entropy. Hence we generalize the island rule to any number of dimensions but also cast it in a framework where many of its features follow simply from the properties of the RT prescription -- and in fact, can be understood analytically. In particular, we will be able to address several issues which appeared puzzling in \cite{Almheiri:2019hni}. Other recent analyses in higher dimensions were undertaken numerically in \cite{Almheiri:2019psy}, in an effective theory in flat space \cite{Hashimoto:2020cas} and using a Randall-Sundrum-inspired toy model in \cite{Geng:2020qvw}.

The remainder of this paper is organized as follows: In section \ref{sec:branegravity}, we begin by studying a certain class of $d$-dimensional branes embedded in AdS$_{d+1}$. We show how the Randall-Sundrum gravity induced on the brane is equivalent to the bulk description of the brane embedded in the higher dimensional geometry. In section \ref{face}, we elucidate the different holographic perspectives of this system as described above, \ie we can describe the system as a $d$-dimensional boundary CFT coupled to a conformal defect, a $d$-dimensional CFT which contains a region with dynamical gravity, or a ($d$+1)-dimensional theory of gravity coupled to a codimension-one brane. Section \ref{HEE} investigates the relation between the appearance of quantum extremal islands using eq.~\reef{wonderA} and the bulk picture using eq.~\reef{eq:sad0} with RT surfaces crossing the brane. In the same section, we present some explicit calculations explicitly illustrating appearance of such QEI for $d=3$. Section \ref{sec:discussion} concludes with a discussion of our results. In appendix \ref{generalE}, we extend  the arguments in \cite{Myers:2010tj} to support the appearance of the brane contribution to the generalized entropy in eq.~\reef{eq:sad0}. Appendix \ref{bubble} examines a surprising class of spherical RT surfaces, which can be supported at finite size by the brane. 

We must note that most of our discussion is quite general and not necessarily linked to the physics of black holes. In fact, the explicit calculations in section \ref{sec:examples} evaluate the entanglement entropy of entangling regions (with components on either side of the conformal defect) in the vacuum state of the boundary system.\footnote{Further, let us note that the formation of QEIs on branes in the `Einstein gravity regime' require us to introduce somewhat unconventional couplings. That is, we must consider a negative Newton's constant on the brane and/or a Gauss-Bonnet interaction in the four-dimensional bulk gravity.} This illustrates that QEIs are not a feature exclusive to the black hole information problem, but may play a role in more general settings where gravity and entanglement are involved. Nevertheless, it is indeed possible within our model to also discuss black holes. In a forthcoming publication \cite{QEI}, we will apply the methods developed here to the case of eternal black holes coupled to a thermal bath in higher dimensions, similar to \cite{Almheiri:2019yqk}.


\section{Brane Gravity}\label{sec:branegravity}
%

As described in the introduction, we are studying a holographic system where the boundary theory is a $d$-dimensional CFT which lives on a spherical cylinder $R\times S^{d-1}$ (where the $R$ is the time direction). Further, this CFT is coupled to a (codimension-one) conformal defect positioned on the equator of the sphere. Hence, the defect spans the geometry $R\times S^{d-2}$ and supports a ($d-1$)-dimensional CFT. The bulk description of this system involves an asymptotically AdS$_{d+1}$ spacetime with a codimension-one brane spread through the middle of the space (and extending to the position of the defect at asymptotic infinity). In this setup, the brane has an AdS$_d$ geometry and further, we consider the case in which the brane has a substantial tension and backreacts on the bulk geometry. If the brane tension is appropriately tuned, the backreaction produces Randall-Sundrum gravity  supported on the brane \cite{Randall:1999vf,Randall:1999ee}, \ie in the backreacted geometry, new (normalizable) modes of the bulk graviton are localized near the brane inducing an effective theory of dynamical gravity on the brane. In the following, we review the bulk geometry produced by the backreaction of the brane, and also the gravitational action induced on the brane.

\subsection{Brane Geometry}\label{BranGeo}

In the bulk, we have Einstein gravity with a negative cosmological constant in $d+1$ dimensions, \ie
\beq
I_\mt{bulk} = \frac{1}{16 \pi G_\mt{bulk}}\int d^{d+1}x\sqrt{-g}
\[{R}(g) + \frac{d(d-1)}{L^2} \] \,,
\label{act2}
\eeq
where $g_{ab}$ denotes the bulk metric, and we are ignoring the corresponding surface terms here \cite{PhysRevLett.28.1082,Gibbons:1976ue,Emparan:1999pm}.
We also introduce a codimension-one (\ie $d$-dimensional) brane in the bulk gravity theory. The brane action is simply given by
\beq\label{braneaction}
I_\mt{brane} = -T_o\int d^dx\sqrt{-\tilde{g}}\,.
\eeq
where $T_o$ is the brane tension and $\tilde g_{ij}$ denotes the induced metric on the brane.

Away from the brane, the spacetime geometry locally takes the form of AdS$_{d+1}$ with the curvature scale set by $L$. As described above, the induced geometry on the brane will be an AdS$_d$ space, and so it is useful to consider the following metric where the AdS$_{d+1}$ geometry is foliated by AdS$_d$ slices
\beq\label{metric}
ds^2 
= d\rho^2 + \cosh^2\left({\rho}/{L}\right)\, g_{ij}^{\mt{AdS}_d}\,dx^{i}dx^{j}\,.
\eeq
Implicitly here, $L$ also sets the curvature of the AdS$_d$ metric, \eg in global coordinates,
\beq\label{metric2}
g_{ij}^{\mt{AdS}_d}\,dx^{i}dx^{j}=L^2\left[-\cosh^2\!\tdr\,dt^2+d\tdr^2+\sinh^2\!\tdr\,d\Omega_{d-2}^2
\right]\,.
\eeq
With the above choices, we approach the asymptotic boundary with $\rho\to\pm\infty$, or with fixed $\rho$ and $\tdr\to\infty$. In the latter case, we arrive at the equator of the boundary $S^{d-1}$, where the conformal defect is located. For the following, it will be convenient to replace $\rho$ with a Fefferman-Graham-like coordinate \cite{FG,Fefferman:2007rka},
\beq\label{zrho}
z = 2 L e^{-\rho/L}\,,
\eeq
with which the metric \reef{metric} becomes
\beq\label{metric3}
ds^2=\frac{L^2}{z^2}\left[dz^2 +  \left(1 + \frac{z^2}{4\,L^2}\right)^2 g_{ij}^{\AdS_d}\,dx^{i}dx^{j} \right]\,.
\eeq
In these coordinates we approach the asymptotic boundary with $z\to0$ and with $z\to\infty$. Below, we will focus on the region near $z\sim 0$.

\begin{figure}[h]
	\def\svgwidth{1\linewidth}
	\centering{
		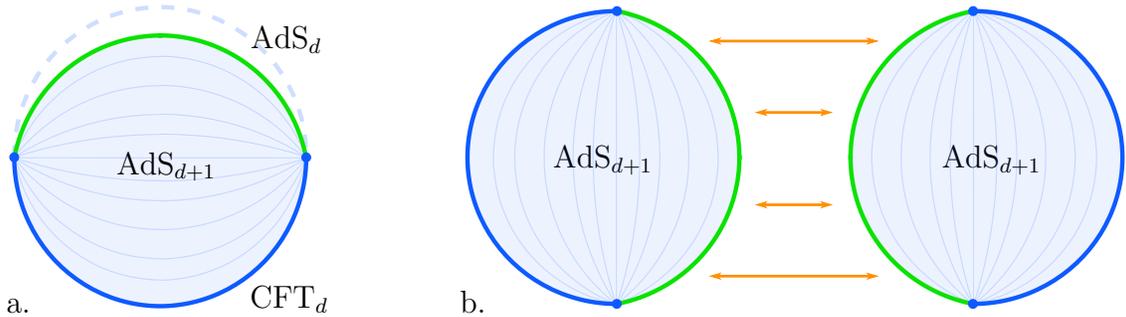
		\caption{Panel (a): Our Randall-Sundrum construction involves foliating  with AdS$_d$ slices. Then identical portions of two such AdS$_{d+1}$ geometries are glued together along an common AdS$_d$ slice. Panel (b): The jump in the extrinsic curvature across the interface between the two geometries is supported by a(n infinitely) thin brane. The brane is represented by a green line in the figures and the bulk AdS$_{d+1}$ spacetime is blue with a $d$-dimensional CFT at the asymptotic boundary.} \label{fig:brane2}
	}
\end{figure}

As described above, the brane spans an AdS$_d$ geometry in the middle of the backreacted spacetime. Following the usual Randall-Sundrum approach, we construct the desired solution
by cutting off the AdS$_{d+1}$ geometry at some $z=\s$, and then
complete the space by gluing this geometry to another copy of itself
-- see figure \ref{fig:brane2}.
Then the Israel junction conditions (\eg see \cite{israel1966singular,Misner:1974qy}) fix $\s$ by relating the discontinuity of the extrinsic curvature across this surface to the stress tensor introduced by the brane, \ie
\beq\label{Israel1}
 \Delta{K}_{ij}-\tilde{g}_{ij}\,\Delta{K}_{k}{}^{k} = 8 \pi \Gbk\, S_{ij} = - 8 \pi \Gbk T_o\,\tilde{g}_{ij}\,,
\eeq
where $\Delta{K}_{ij}={K}^+_{ij} - {K}^-_{ij}= 2{K}_{ij}$, given the symmetry of our construction.
The extrinsic curvature is calculated as \cite{Misner:1974qy}
\beq \label{extrinsic}
{K}_{ij} =\frac{1}{2}\frac{\partial g_{ij}}{\partial n}\bigg|_{z=\s} =-\frac{z}{2L} \frac{\partial g_{ij}}{\partial z} \bigg|_{z=\s} = \frac{1}{L}\frac{4 L^2 -\s^2}{4L^2 +\s^2}\,\tilde{g}_{ij}\,,
\eeq
where $\partial_n = -\frac{z}{L}\partial_z$ is an outward directed unit normal vector. Further, we are using the notation introduced above where $\tilde g_{ij}$ corresponds to the induced metric on the surface $z=\s$, \ie on the brane. Combining eqs.~\reef{Israel1} and \reef{extrinsic}, we arrive at
\beq\label{positionbrane}
\frac{4L^2 -\s^2}{4L^2 +\s^2} = \frac{4\pi \Gbk L\, T_o}{d-1}\,.
\eeq

Now if we consider $\s\ll L$, it will ensure that the defect is well approximated by the holographic gravity theory on the brane -- see the discussion in the next subsection. In this regime, we can solve eq.~\reef{positionbrane} in a small $\s$ expansion, and to leading order, we find that
\beq\label{position}
\s^2\simeq z_\mt{0}^2 = 2L^2\left(1-\frac{4 \pi \Gbk L T_o}{d-1}\right)\,.
\eeq
Hence to achieve this result, we must tune the expression in brackets on the right to be small, \ie
\beq\label{tune}
\veps\equiv 1-\frac{4 \pi \Gbk L T_o}{d-1}\ll 1\,.
\eeq
As the notation suggests, we can think of this quantity $\veps$ as an expansion parameter in solving for the brane position from eq.~\reef{positionbrane}.
A useful check of our calculations below will come from carrying the solution to the next order, \ie
$\s^2= z_\mt{0}^2+\delta[ \s^2]_\mt{2}+\cdots$ with
\beq\label{secondorder}
\delta[ \s^2]_\mt{2} =\frac{(d-1)L}{4 \pi \Gbk T_o}\,\veps^2
\ = \frac{(d-1)L}{4 \pi \Gbk T_o}\left(1-\frac{4 \pi \Gbk L T_o}{d-1}\right)^2\,.
\eeq

To conclude, we consider the intrinsic geometry of the brane. As we noted above, the curvature scale of $g_{ij}^{\mt{AdS}_d}$ is simply $L$, and hence given the full bulk metric \reef{metric3}, we can read off the curvature scale of the surface $z=\s$ as
\beq\label{curve1}
\lb=\frac{L^2}{\s}\left(1 + \frac{\s^2}{4\,L^2}\right)\,.
\eeq
Note that since we are considering $\s/L\ll 1$, it follows that
$\lb/L\gg 1$, \ie the brane is weakly curved. Using eq.~\reef{position}, we can solve for $\lb$ to leading order in the $\veps$ expansion to find
\beq\label{curve2}
\frac{L^2}{\lb^2}\simeq
2\,\veps\ =2\left(1-\frac{4 \pi \Gbk L T_o}{d-1}\right)\,.
\eeq
It will be useful to have the following expressions for the Ricci tensor and scalar evaluated for the brane geometry, and these are compactly written using eq.~\reef{curve1} as
\beq\label{Ricky2}
\tilde{R}_{ij}(\tilde g)=-\frac{d-1}{\lb^2}\, \tilde{g}_{ij}\,,\qquad \tilde{R}(\tilde g)=-\frac{d(d-1)}{\lb^2}\,.
\eeq

\subsection{Gravitational Action on the Brane}\label{indyaction}

As noted above, following the usual Randall-Sundrum scenario \cite{Randall:1999vf,Randall:1999ee,Karch:2000ct}, new (normalizable) modes of the bulk graviton are localized near the brane in the backreacted geometry, and this induces an effective theory of dynamical gravity on the brane. The gravitational action can be determined as follows:
First, one considers  a Fefferman-Graham (FG) expansion near the boundary of an asymptotic AdS geometry \cite{FG,Fefferman:2007rka}. Then integrating the bulk action (including the Gibbons-Hawking-York surface term \cite{PhysRevLett.28.1082,Gibbons:1976ue}) over the radial direction out to some regulator surface produces a series of divergent terms, which through the FG expansion can be associated with various geometric terms involving the intrinsic curvature of the boundary metric. Usually in AdS/CFT calculations, a series of boundary counterterms are added to the action to remove these divergences, as the regulator surface is taken to infinity \cite{Emparan:1999pm}. In the present braneworld construction, the regulator surface is replaced by the brane, which remains at a finite radius, and no additional counterterms are added. Rather the `divergent' terms become contributions to the gravitational action of the brane theory, and hence the latter from previous discussions of the boundary counterterms \cite{Emparan:1999pm}, \ie
\begin{multline}\label{diver1}
I_\mt{diver}=\frac{1}{16\pi \Gbk}\int d^dx \sqrt{-\tilde{g}}\left[\frac{2(d-1)}{L}+\frac{L}{(d-2)}\tilde{R}
\right.\\
+\left. \frac{L^3}{(d-4)(d-2)^2} \left(\tilde{R}^{ij}\tilde{R}_{ij}-\frac{d}{4(d-1)}\,\tilde{R}^2\right) +\cdots\right]\,.
\end{multline}

Several comments are in order at this point: First of all, we note that the above expression is written in terms of the induced metric $\tilde g_{ij}$ on the brane (as in \cite{Emparan:1999pm}) rather than the boundary metric $\overscript{g}{0}_{ij}$ that enters the FG expansion. Using the standard results, \eg \cite{Skenderis:2002wp,deHaro:2000vlm}, we can relate the two with
\beq\label{relate}
\tilde g_{ij}(x_k) = \frac{L^2}{\s^2}\,\overscript{g}{0}_{ij}(x_k) +  \overscript{g}{1}_{ij}(x_k)+\frac{\s^2}{L^2}\, \overscript{g}{2}_{ij}(x_k)+\cdots\,,
\eeq
where the higher order terms can be expressed in terms of the curvatures of $\overscript{g}{0}_{ij}$, \eg
\beq\label{oneg}
\overscript{g}{1}_{ij} = -\frac{L^2}{d-2}\left(R_{ij}\big[\overscript{g}{0}\big] -\frac{\overscript{g}{0}_{ij}}{2(d-1)}\,R\big[\overscript{g}{0}\big]\right)\,.
\eeq
In other words, the two metrics are related by a Weyl scaling and a field redefinition. Further, we see a factor of $(d-2)$ appearing in the denominator of the second term, \ie the Einstein-Hilbert term, in eq.~\reef{diver1}. Hence this expression only applies for $d\ge3$ and must be reevaluated for $d=2$, which we do in section \ref{sec:two-d}. Similar factors, as well as a factor of $d-4$, appear in the denominator of the third term, which again indicates that this expression must be reconsidered for $d=4$.

In any event, the gravitational action on the brane is given by combining the above expression with the brane action \reef{braneaction},
\beq\label{totaction}
I_\mt{induced} = 2\, I_\mt{diver} +  I_\mt{brane}\,,
\eeq
where the factor of two in the first term accounts for integrating over the bulk geometry on both sides of the brane.
The combined result can be written as
\beqa
I_\mt{induced}&=&\frac{1}{16 \pi G_\mt{eff}}\int d^{d}x\sqrt{-\tilde{g}}
\[\frac{(d-1)(d-2)}{\ell_\mt{eff}^2} + \tilde{R}(\tilde{g})\right]
\labell{act3}\\
&&\qquad\quad
+\frac{1}{16 \pi G_\mt{RS}}\int d^{d}x\sqrt{-\tilde{g}}\left[ \frac{L^2}{(d-4)(d-2)}\(\tilde{R}^{ij}\tilde{R}_{ij}-
\frac{d}{4(d-1)} \tilde{R}^2\)+\cdots\]\,,
\nonumber
\eeqa
where
\beq
\frac{1}{G_\mt{eff}}=\frac{1}{G_\mt{RS}}=\frac{2\,L}{(d-2)\,G_\mt{bulk}}\,,
\qquad\qquad
\frac{1}{\ell_\mt{eff}^2}=\frac{2}{L^2}\left(1-\frac{4 \pi \Gbk L T_o}{d-1}\right)\,.
\label{Newton2}
\eeq
In the present discussion $G_\mt{eff}$ and $G_\mt{RS}$ are equal, but by adding terms to the brane action this can change. We will explain this in section \ref{sec:DGP}. Comparing eqs.~\reef{curve2} and \reef{Newton2}, we see that $\ell_\mt{eff}$ (which sets the cosmological constant term in $I_\mt{induced}$) precisely matches the leading order expression for the brane curvature $\lb$. Hence if we only consider the first two terms in eq.~\reef{act3}, the resulting Einstein equations would reproduce the leading expression (in the $\veps$ expansion) for the curvatures in eq.~\reef{Ricky2}. Further, it is a straightforward exercise to show that if the contribution of the curvature squared terms is also included in the gravitation equations of motion, the curvature is shifted to precisely reproduce the $\veps^2$ term in eq.~\reef{Ricky2}. Hence rather than using the Israel junction condtions, we could determine the position of the brane in the backreacted geometry by first solving the gravitational equations of the brane action \reef{act3} and then finding the appropriate surface $z=\s$ with the corresponding curvature. More generally, the fact that these two approaches match was verified by \cite{deHaro:2000wj},\footnote{See also earlier discussions, \eg \cite{Shiromizu:1999wj,Verlinde:1999fy,Gubser:1999vj}.} which argued the bulk Einstein equations combined with the Israel junction conditions are equivalent to the brane gravity equations of motion.\footnote{We note that the brane graviton acquires a small mass through interactions with the CFT residing there \cite{Karch:2000ct,Karch:2001jb,Porrati:2001gx}. However, this mass plays no role in the following as it is negligible in the regime of interest, \ie $L/\ell_\mt{eff}\ll 1$ -- see further discussion in section  \ref{face}. This point was emphasized in \cite{Geng:2020qvw}.}

Of course, the gravitational approach only provides an effective approach in the limit that $\ell_\mt{eff}\gg L$ since otherwise the contributions of the higher curvature terms cannot be ignored.
For example, if the curvatures are proportional to $1/\ell_\mt{eff}^2$ at leading order, then the curvature squared term is suppressed by a factor of $L^2/\ell_\mt{eff}^2$ relative the first two terms. Similarly the higher order curvature terms denoted by the ellipsis in eq.~\reef{act3} are further suppressed by a further factor of $L^2/\ell_\mt{eff}^2$ for each additional curvature appearing these terms. From eq.~\reef{curve2}, we can write $\frac{L^2}{\ell_\mt{eff}^2}=2\veps$ and hence we see that the gravitational brane action and the resulting equations of motion can be organized in the same small $\veps$ expansion discussed in the previous section.\footnote{Note that we have distinguished the gravitational couplings in the Einstein terms and in the higher curvature interactions, \ie in the first and second lines of eq.~\reef{act3}, even though $G_\mt{eff}=G_\mt{RS}$ here. However, this distinction will become important in section \ref{sec:DGP}.}

Recall that we can give a holographic description of this system involving (two weakly interacting copies of) the boundary CFT living on the brane. However, this CFT has a finite UV cutoff because the brane resides at a finite radius in the bulk, \eg see \cite{deHaro:2000vlm,Emparan:2006ni,Myers:2013lva}. The action \reef{diver1} is then the induced gravitational action resulting from integrating out the CFT degrees of freedom. The UV cutoff is usually discussed in the context of the boundary metric $g^{\ssc (0)}_{ij}$, where the short distance cutoff would be given by $\delta\simeq\s$.  However, recall that the gravitational action \reef{act3} is expressed in terms of the induced metric $\tilde g_{ij}$ and so the conformal transformation in eq.~\reef{relate} yields $\tilde\delta\simeq L$ for this description of the brane theory. Therefore the $\veps$ expansion corresponds to an expansion in powers of the short distance cutoff, \ie $\veps\sim\tilde \delta^2/\ell_\mt{eff}^2$.

%

\subsection{The case of two dimensions}\label{sec:two-d}

Recall that the curvature terms in the induced action \reef{diver1} have coefficients with inverse powers of $(d-2)$ and so we must reconsider the calculation of this brane action for $d=2$, \ie when the bulk space is (locally) AdS$_3$ and the induced geometry on the brane is AdS$_2$. This section sketches the necessary calculations, which are largely the same as those performed in higher dimensions, but with a few important differences.

Let us add that in contrast to the induced action, the calculations in section \ref{BranGeo}, where the position of the brane is determined using the Israel junction conditions, need no modifications for $d=2$.
Therefore we can simply substitute $d=2$ into eqs.~\reef{position} and \reef{secondorder} for the brane position to find
\beq\label{2dposition}
\s^2 \simeq 2L^2\veps+ \frac{L}{4 \pi \Gbk T_o}\,\veps^2+\cdots\,, \qquad{\rm with}\quad
\veps=1-4 \pi \Gbk L T_o \,.
\eeq
Of course, we must be able to reproduce the same result using the new induced gravity action.

\subsubsection*{Integration of bulk action}

As discussed in section \ref{indyaction}, one can determine the structure of the terms in the induced action by a careful examination of the FG expansion near the asymptotic boundary \cite{Skenderis:1999nb,deHaro:2000wj,deHaro:2000vlm}. However, we can take the simpler route here, since in two dimensions the Riemann curvature has a single component and therefore the entire induced action can be expressed in terms of the Ricci scalar $\ric(\tilde g)$. Therefore, we evaluate the on-shell bulk action and match the boundary divergences to an expansion in $\ric(\tilde g)$. That is, we substitute the metric \reef{metric3} into the bulk action \reef{act2} plus the corresponding  Gibbons-Hawking-York surface term \cite{PhysRevLett.28.1082,Gibbons:1976ue} and integrate over the radial direction $z$. The result can be expressed as a boundary integral with a series of divergences as $\s\to0$,\footnote{This expression also includes ${\cal O}(\s^2)$ contributions, which are necessary to match eq.~\reef{2dposition} to ${\cal O}(\veps^2)$ in the following. Further, note that we are ignoring the contributions coming from asymptotic boundaries at  $z\to\infty$.}
\beq\label{induced act2}
I_\mt{diver}=\frac{L}{16 \pi \Gbk}\int d^2x\sqrt{-g^{\ads2}} \left[\frac{1}{\s^2} +\frac{1}{L^2}\,\log\Big(\frac{\s}{L}\Big) -\frac{\s^2}{16L^4} +\cdots\right]\,.
\eeq
Now we rewrite the above expression in terms of the induced metric and the corresponding Ricci scalar combining eqs.~\reef{metric3}, \reef{curve1} and \reef{Ricky2}, which yield
\beq\label{AdS3}
\sqrt{-\tilde{g}} =  \frac{L^2}{\s^2}\left(1 + \frac{\s^2}{4L^2}\right)^2 \sqrt{-g^{\ads2}} \,, \qquad\quad
\ric = -2\,\frac{\s^2}{L^4}\left(1+\frac{\s^2}{4L^2}\right)^{-2}\,.
\eeq
Using these expressions, 
the induced action becomes\footnote{Our derivation of eq.~\eqref{ind-action2} will miss terms involving derivatives of $\tilde{R}$ as these vanish for the constant curvature geometry of our brane. However,  such terms will only appear at higher orders, \ie in the `$\cdots$' (other than the total derivative $\tilde\Box\tilde{R}$).} 
\beq\label{ind-action2}
I_\mt{diver}=\frac{L}{16 \pi \Gbk}\int d^2x\sqrt{-\tilde{g}} \Big[\frac{2}{L^2} -  \frac12\, \tilde{R} \,\log\left(-\frac{L^2 }{2}\tR\right) + \frac{1}{2}\,\tilde{R}+\frac{L^2}{16}\,\tilde{R}^2 +\cdots\Big]\,.
\eeq

The most striking feature of this induced action is the term proportional to $\ric\log|\ric|$. The appearance of this logarithm is related to the conformal anomaly \cite{Henningson:1998gx,Henningson:1998ey,Burgess:1999vb}, and points towards the fact that the corresponding gravitational action in nonlocal,\footnote{Similar nonlocalities appear in the curvature-squared or four-derivative contributions with $d=4$, or more generally in the interactions with $d/2$ curvatures for higher (even) $d$. Hence they do not play a role in higher dimensions if we work in the regime where the induced action \reef{act3} is well approximated by Einstein gravity coupled to a cosmological constant.} as we discuss next.  Further, since the Einstein-Hilbert term is topological in two dimensions, it turns out that this unusual action is precisely what is needed to match the dynamics of the bulk gravity described above, \ie the position of the brane in eq.~\reef{2dposition}.

The logarithmic contribution  should correspond to that coming from the nonlocal Polyakov action \cite{Skenderis:1999nb}. Schematically, we would have
\beq
I_\mt{bulk}\simeq I_\mt{Poly}= -\frac{\alpha\,L}{16 \pi \Gbk}\int d^2x\sqrt{-\tilde{g}} \, \tilde{R}\,\frac1{\tilde \Box}\,\tilde{R} \,,
\label{induced-action3}
\eeq
where we have introduced an arbitrary constant $\alpha$ here but it will be fixed by comparing with the divergences in the integrated action. Of course, $\frac1{\tilde \Box}\,\tilde{R}$ indicates a convolution of the Ricci scalar with the scalar Green's function, but there are subtleties here in dealing with constant curvatures. The latter are ameliorated by making the action \reef{induced-action3} local by introducing a auxiliary field $\phi$ (\eg see \cite{Skenderis:1999nb,Alvarez:1982zi}),
\beq
 I_\mt{Poly}=\frac{\alpha\,L}{8\pi \Gbk}\int d^2x\sqrt{-\tilde{g}} \, \left[
 -\frac12\,\tilde g^{ij}\tilde\nabla_i\phi\tilde\nabla_j\phi
 +\phi\,\tilde R + \chi\,e^{-\phi}
\right] \,.
\label{PolyAct2}
\eeq
where $\chi$ is a fixed constant.\footnote{The last term is needed to take care of zero mode problem \cite{Alvarez:1982zi}. Examining the equation of motion \reef{eom4}, one can think of $\phi$ as a conformal factor relating the metric $\tg_{ij}$ to a canonical constant curvature metric $\hat g_{ij}$, \ie $\tg_{ij}=e^{\phi}\hat g_{ij}$ with $\hat R(\hat g)=\chi$ \cite{Alvarez:1982zi,Frolov:1996hd}.
Hence we choose $\chi$ to be negative to match the sign of $\tilde R$. Further, note that with the interaction $\chi e^{-\phi}$ in the action \reef{PolyAct2}, $\phi$ becomes an interacting field \cite{Skenderis:1999nb}.
\label{ZZZ}}

The equation of motion resulting from eq.~\reef{PolyAct2}  is
\beq
0=\tilde\Box \phi +\tilde R - \chi\,e^{-\phi}\,,
\label{eom4}
\eeq
which has a simple solution when $\tilde R$ is a constant, namely,
\beq
 \phi=\phi_0 = \log(\chi/\tilde R)\,.
\label{sol4}
\eeq
Evaluating the Polyakov action with $\phi = \phi_0$ yields
\beq
 I_\mt{Poly}\big|_{\phi=\phi_0}=-\frac{\alpha\,L}{8\pi \Gbk}\int d^2x\sqrt{-\tilde{g}} \, \left[
 \tilde R \,\log(\tilde R/\chi) - \tilde R\, \right] \,.
\label{PolyAct3}
\eeq
Comparing this expression with the log term in eq.~\reef{ind-action2}, we fix $\alpha=\frac14$ and $\chi=-\frac2{L^2}$.

Varying the action \reef{PolyAct2} with respect to the metric, we find the corresponding contribution to the `gravitational' equations of motion
\beqa
T^\mt{Poly}_{ij}=-\frac2{\sqrt{-g}}\,\frac{\delta I_\mt{Poly}}{\delta g^{ij}}
&=&\frac{L}{32\pi \Gbk}\Big[
\tilde\nabla_i\phi\tilde\nabla_j\phi
 +2\,\tilde\nabla_i\tilde\nabla_j\phi
\label{eom5}\\
&&\qquad\qquad\left.
-\tg_{ij}\left(\frac12\,(\tilde\nabla\phi)^2
 +2\,\tilde\Box \phi- \chi\,e^{-\phi} \right)
\right]\,,
\nonumber
\eeqa
where we have used $\tilde R_{ij} -\frac12\tg_{ij}\tilde R=0$ for $d=2$ to eliminate the terms linear in $\phi$ (without any derivatives). Now substituting $\phi_0$, we find that this expression reduces to
\beq
T^\mt{Poly}_{ij}\big|_{\phi=\phi_0}=
\frac{L}{32\pi \Gbk}\,\tg_{ij}\, \tilde R\,,
\label{onshell5}
\eeq
which we will substitute into evaluating the equations of motion below to fix the position of the brane. As an aside, we can take the trace of the above expression to find
that it reproduces the trace anomaly, \eg \cite{Duff:1977ay,Duff:1993wm}
\beq\label{trace}
\langle T^i{}_i \rangle = \frac{c}{24\pi}\,\tilde R \,,
\eeq
where we recall that $c=\frac{3L}{2\Gbk}$ for the boundary CFT. In our case, the trace anomaly will be twice as large, since there are two copies of the CFT living on the brane.

The induced action $I_\mt{induced} = 2\,I_\mt{diver} + I_\mt{brane}$ can be written as
\beq\label{induct}
I_\mt{induced} = \frac{1}{16 \pi G_\mt{eff}}\int d^2x\sqrt{-\tilde{g}} \Big[\frac{2}{\ell_\mt{eff}^2} -  \tilde{R} \,\log\left(-\frac{L^2 }{2}\tR\right)+\tR +\frac{L^2}{8}\,\tilde{R}^2 +\cdots\Big]\,,
\eeq
where $\ell_\mt{eff}$ is given by the expression in eq.~\reef{Newton2} with $d=2$, \ie
\beq\label{lobster}
\frac{L^2}{{\ell}_\mt{eff}^2}=2\( 1-4 \pi \Gbk L T_o\)\,,
\eeq
however, we have set $\Geff =  \Gbk/L$ here. The metric variation then yields the following equation of motion
\beq\label{gamble3}
0=\frac{2}{\ell_\mt{eff}^2}\,\tilde{g}_{ij}+\tilde{g}_{ij} \,
\tilde{R} + \frac{L^2}8\,\tR\(\tg_{ij}\,\tR-4 \tR_{ij}\)+\cdots\,,
\eeq
where we dropped the terms involving derivatives of curvatures arising from the variation of the $\tR^2$ term.
To leading order, we find $\tR\sim -2/\ell_\mt{eff}^2 = -4\veps/L^2$ in agreement with eqs.~\reef{curve2} and \reef{Ricky2}. Hence, the gravitational equations of motion again fix the (leading-order) position of the brane for $d=2$, and further it is a straightforward exercise to match to second order corrections in eq.~\reef{2dposition} using the curvature-squared contributions in eq.~\reef{gamble3}.

\subsubsection*{Adding JT gravity}

Much of the recent literature on quantum extremal islands examines models involving two-dimensional gravity, \eg \cite{Almheiri:2019psf, Almheiri:2019hni, Almheiri:2019yqk, Chen:2019uhq, Penington:2019kki, Almheiri:2019qdq, Chen:2019iro}, however, the gravitational theory in these models is Jackiw-Teitelboim (JT) gravity \cite{Jackiw:1984je,Teitelboim:1983ux}. One can incorporate JT gravity into the current model by dropping the usual tension term \reef{braneaction}, and instead using the following brane action\footnote{Alternatively, one could simply add $I_\mt{JT}$ to the usual tension term. With this approach, an extra source term appears in eq.~\reef{fulleom}, but it can be eliminated by shifting the dilaton in a manner similar to eq.~\reef{shift1}.} 
\beq\label{braneact2}
I_\mt{brane}= I_\mt{JT} + I_\mt{ct}\,,
\eeq
where the JT action takes the usual form,
\beq\label{JTee}
I_\mt{JT} =\frac{1}{16\pi G_\mt{brane}}\int d^2x\sqrt{-\tilde{g}}\left[\Phi_0\,\tilde{R}+ \Phi\left(\tilde{R}+\frac{2}{\ell^2_\mt{JT}}
\right)\right]\,.
\eeq
Here, as in previous actions, we have ignored the boundary terms associated with the JT action, \eg see \cite{Maldacena:2016upp}, and we have introduced the dilaton $\Phi$. Recall that $\Phi_0$ is simply a constant and so the first term is topological but contributes to the generalized entropy. In eq.~\reef{braneact2}, we have also included a counterterm 
\beq\label{count123}
I_\mt{ct}=-\frac{1}{4\pi \Gbk L}\int d^2x\sqrt{-\tilde{g}}\,,
\eeq
which is tuned to cancel the induced cosmological constant on the brane. This choice ensures that the JT gravity \reef{JTee} couples to the boundary CFT in the expected way, \eg as in  \cite{Almheiri:2019psf,Maldacena:2016upp} -- see further comments below.

The full induced action now takes the form
\begin{align}
&&I_\mt{induced}=\frac{1}{16 \pi G_\mt{eff}}\int d^2x\sqrt{-\tilde{g}} \Big[ -  \tilde{R} \,\log\left(-\frac{L^2 }{2}\tR\right) +\frac{L^2}{8}\,\tilde{R}^2 +\cdots\Big]
\nonumber\\
&&\qquad+\frac{1}{16\pi G_\mt{brane}}\int d^2x\sqrt{-\tilde{g}}\left[\tilde\Phi_0\,\tilde{R}+ \Phi\left(\tilde{R}+\frac{2}{\ell^2_\mt{JT}}
\right)\right]\,,
\label{fullindyact}
\end{align}
where we have combined the two topological contributions in the second line with\footnote{In \cite{Almheiri:2019psf}, $\tilde \Phi_0$ would also absorb a logarithmic constant $-2\log(L/z_\mt{B})$, which would be accompanied by a shift in the prefactor in the argument of the logarithmic term in eq.~\reef{fullindyact}, \ie $2/L^2\to2/ z^2_\mt{B}$.}
\beq\label{shift0}
\tilde \Phi_0 =\Phi_0 + G_\mt{brane}/G_\mt{eff}\,.
\eeq

Now, with the JT action \reef{JTee}, the dilaton equation of motion fixes $\tilde R=-2/\ell^2_\mt{JT}$, \ie the brane geometry is locally AdS$_2$ everywhere with $\ell_\mt{B}= \ell_\mt{JT}$. Then the position $\s$ of the brane is fixed by eq.~\reef{curve1} and implicitly we assume that $\ell_\mt{JT}\gg L$, which ensures that $\s\ll L$ as in our previous discussions. The gravitational equation of motion coming from the variation of the metric becomes
\beq\label{fulleom}
-\nabla_{i}\nabla_{j}\Phi+\tilde{g}_{ij}\(\nabla^2\Phi-\frac{\Phi}{\ell^2_\mt{JT}}\)= 8\pi G_\brane\, \widetilde T^\mt{CFT}_{ij}
= -\frac{\Gbr}{ \Geff}\, \frac{1 }{\hat{\ell}_\mt{eff}^2}\,\tilde{g}_{ij} \,,
\eeq
where $\hat{\ell}_\mt{eff}$ is the effective curvature scale produced by $\ell_\mt{JT}$.
That is, in the case without JT gravity, we can combine eqs.~\reef{positionbrane}, \reef{curve1} and \reef{Newton2} to find
\beq\label{curve33}
\frac{L^2}{{\ell}_\mt{eff}^2}= f\!\(\frac{L^2}{\ell_\mt{B}^2}\)\equiv 2\(1-\sqrt{1-\frac{L^2}{\ell_\mt{B}^2}}\,\)  \,.
\eeq
We can understand this expression as the gravitational equation of motion coming from the two-dimensional action \reef{induct}, where a Taylor expansion of the right-hand side for $L/\ell_\mt{B}\ll1$ corresponds to varying the curvature terms and subsequently substituting $\tilde R_{ij}=-\frac{1}{\ell^2_\mt{B}}\,\tilde g_{ij}$, as in eq.~\reef{Ricky2}.  Now in the JT equation of motion \reef{fulleom}, the effective curvature scale $\hat{\ell}_\mt{eff}$ satisfies ${L^2}/{\hat{\ell}_\mt{eff}^2}= f\!\({L^2}/{\ell_\mt{JT}^2}\)$. We have indicated in eq.~\reef{fulleom} that the left-hand side corresponds to the stress tensor of the boundary CFT which lives on the brane. In the present arrangement,\footnote{In  more interesting scenarios, \eg with evaporating black holes as in \cite{Almheiri:2019psf,Almheiri:2019hni,Chen:2019uhq}, it is more appropriate to work directly with the CFT's stress tensor, rather than replacing these degrees of freedom by an effective gravity action after integrating out the CFT.} this takes a particularly simple form, with $T^\mt{CFT}_{ij}\propto \tilde{g}_{ij}$. Of course, this source term in eq.~\reef{fulleom} can be easily absorbed by shifting the dilaton, 
\beq\label{shift1}
\tilde\Phi\equiv \Phi- \frac{\Gbr}{ \Geff}\, \frac{\ell^2_\mt{JT} }{\hat{\ell}_\mt{eff}^2}\,,
\eeq
so that $\tilde\Phi$ satisfies the usual source-free equation studied in \eg \cite{Maldacena:2016upp}.

At this point, we observe that  the trace of eq.~\reef{fulleom} yields on the right-hand side,
\beq\label{almost}
\langle \big[\widetilde T^\mt{CFT}\big]^i{}_i \rangle = -\frac{L}{ 4\pi\Gbk}\, \frac{1 }{\hat{\ell}_\mt{eff}^2}=-\frac{L}{ 4\pi\Gbk}\, \frac{1 }{\ell_\mt{JT}^2}\(1+\frac14\,\frac{L^2}{\ell_\mt{JT}^2}+\frac18\,\frac{L^4}{\ell_\mt{JT}^4}+\cdots\)\,,
\eeq
where in the final expression, we are Taylor expanding $f(L^2/\ell_\mt{JT}^2)$ assuming $L^2/\ell_\mt{JT}^2\ll 1$, as above. Noting that $\tilde R=-2/\ell^2_\mt{JT}$ and comparing to eq.~\reef{trace},\footnote{Recall that the central charge here is twice that appearing in eq.~\reef{trace} because the brane supports two (weakly interacting) copies of the boundary CFT.} we see that the expected trace anomaly has recieved a infinite series of higher order corrections. We can interprete the latter as arising from the finite UV cutoff on the brane, recalling that $\tilde\delta\simeq L$ as discussed at the end of section \ref{indyaction}.

%

\subsection{DGP Gravity on the Brane} \label{sec:DGP}
The previous discussion of $d=2$ motivates that it is interesting to add an intrinsic gravity term to the brane action. Here, we extend this discussion to higher dimensions, \ie extend the brane action to include an Einstein-Hilbert term. Of course, this scenario can be viewed as a version of Dvali-Gabadadze-Porrati (DGP) gravity \cite{Dvali:2000hr} in an AdS background. Hence, it combines features of both RS and DGP gravity theories. We discuss the modifications of the brane dynamics and the induced action below, but it also produces interesting modifications of the generalized entropy, as discussed in sections \ref{HEE} and appendices \ref{generalE} and \ref{bubble}.

We write the extended brane action, replacing eq.~\reef{braneaction}, as
\beq\label{newbran}
I_\mt{brane} = -(T_o-\Delta T)\int d^dx \sqrt{-\tilde{g}} + \frac{1}{16\pi \Gbr}\int d^dx \sqrt{-\tilde{g}} \tilde{R}\,.
\eeq
In general, for a fixed brane tension, the position of the brane will be modified with the additional Einstein-Hilbert term. Hence we have parametrized the full brane tension as $T_o-\Delta T$ and the contribution $\Delta T$ will be tuned to keep the position of the brane fixed. This choice will facilitate the comparison of the generalized entropy between different scenarios in the following.

As in section \ref{BranGeo}, the position of the brane can be determined using the Israel junction conditions \reef{Israel1}. Hence we begin by evaluating the brane's stress tensor,
\beq\label{stressbran}
S_{ij}\equiv -\frac{2}{\sqrt{-\tilde g}}\,\frac{\delta I_\mt{brane}}{\delta \tg^{ij}}=-\tilde{g}_{ij}(T_o-\Delta T) -\frac{1}{8\pi \Gbr}\left(\ric_{ij}-\frac12\tg_{ij}\,\ric\right)\,.
\eeq
As commented above, we choose $\Delta T$ to cancel the curvature contributions in this expression, \ie the stress tensor reduces to $S_{ij}=-T_o\,\tilde{g}_{ij}$. With this tuning, the Israel junction conditions in eq.~\reef{Israel1} are unchanged as the analysis which follows from there. Therefore the brane position and curvature remain identical to those determined in eqs.~\reef{positionbrane} and \reef{curve1}. This allows use to determine the desired tuning as
\beq\label{tune3}
\Delta T = \frac{(d-1)(d-2)}{16\pi \Gbr\,\ell_\mt{B}^2}\ 
\simeq \frac{(d-1)(d-2)}{8\pi \Gbr\,L^2}\,\veps\,.
\eeq
We have used eq.~\reef{curve2} to show that the shift in the brane tension is small in the $\veps$ expansion. 

We return to the induced gravitational action on the brane that takes the same form as in eq.~\reef{act3} but with the effective Newton's constant in eq.~\reef{Newton2} replaced by
\beq
\frac{1}{G_\mt{eff}}=\frac{2L}{(d-2)\,G_\mt{bulk}}+\frac{1}{\Gbr}\,.
\label{Newton33}
\eeq
By construction, $\ell_\mt{eff}$ and the position of the brane are unchanged. Note that the gravitational couplings in the Einstein terms and in the higher curvature interactions, \ie in the first and second lines of eq.~\reef{act3},  are now distinct. That is, $G_\mt{eff}$ no longer equals $G_\mt{RS}$. 

In the following, it will be useful to define the ratio
\beq\label{newdefs}
\lamb=\frac{G_\mt{RS}}{\Gbr}
\qquad{\rm with}\quad \frac{1}{G_\mt{RS}}=\frac{2L}{(d-2)\,G_\mt{bulk}}\,,
\eeq
where $G_\mt{RS}$ is the induced Newton's constant on an RS brane appearing in eq.~\reef{Newton2}, while the dimensionless ratio $\lamb$ controls the relative strength of the Newton's constants in the bulk and on the brane. With these definitions, the induced Newton's constant on the DGP brane, in eq.~\reef{Newton33}, can be rewritten as
\beq\label{Newton34}
\frac{1}{G_\mt{eff}}=\frac{1}{G_\mt{RS}}\(1+\lamb\)\,.
\eeq\\

Of course, one can also consider other modifications of the brane action beyond adding the Einstein-Hilbert term in eq.~\reef{newbran} -- see discussion in the next subsection and \cite{domino}. Further, we will discuss adding topological gravitational terms on the brane or in the bulk in sections \ref{HEE} and \ref{sec:discussion}. In particular, we will see in section \ref{sec:examples} that adding a Gauss-Bonnet term to the four-dimensional bulk gravity theory yields another tuneable parameter which, for a certain parameter range, makes it possible to find quantum extremal islands in the absence of black holes.

\section{Three perspectives: Bulk/Brane/Boundary}\label{face}
%

Our setup can be interpreted from three different `holographic' perspectives, which are analogous to the three descriptions of \cite{Almheiri:2019hni}, suitably generalised to arbitrary dimensions. A set of analogous descriptions for gravity on a brane in higher dimensions was discussed in the context of the Karch-Randall model \cite{Karch:2000ct}, and in fact, these are the models discussed here with the addition of the DGP term \reef{newbran}. In this section we review each of the dual descriptions, and explore their relation. 

\begin{figure}[h]
	\def\svgwidth{0.9\linewidth}
	\centering{
\begingroup%
  \makeatletter%
  \providecommand\color[2][]{%
    \errmessage{(Inkscape) Color is used for the text in Inkscape, but the package 'color.sty' is not loaded}%
    \renewcommand\color[2][]{}%
  }%
  \providecommand\transparent[1]{%
    \errmessage{(Inkscape) Transparency is used (non-zero) for the text in Inkscape, but the package 'transparent.sty' is not loaded}%
    \renewcommand\transparent[1]{}%
  }%
  \providecommand\rotatebox[2]{#2}%
  \newcommand*\fsize{\dimexpr\f@size pt\relax}%
  \newcommand*\lineheight[1]{\fontsize{\fsize}{#1\fsize}\selectfont}%
  \ifx\svgwidth\undefined%
    \setlength{\unitlength}{841.88976378bp}%
    \ifx\svgscale\undefined%
      \relax%
    \else%
      \setlength{\unitlength}{\unitlength * \real{\svgscale}}%
    \fi%
  \else%
    \setlength{\unitlength}{\svgwidth}%
  \fi%
  \global\let\svgwidth\undefined%
  \global\let\svgscale\undefined%
  \makeatother%
  \begin{picture}(1,0.51178451)%
    \lineheight{1}%
    \setlength\tabcolsep{0pt}%
    \put(0.71482651,0.00964993){\color[rgb]{0,0,0}\makebox(0,0)[lt]{\lineheight{1.25}\smash{\begin{tabular}[t]{l}f.\end{tabular}}}}%
    \put(0.71482651,0.21659723){\color[rgb]{0,0,0}\makebox(0,0)[lt]{\lineheight{1.25}\smash{\begin{tabular}[t]{l}c.\end{tabular}}}}%
    \put(0.36558402,0.00964993){\color[rgb]{0,0,0}\makebox(0,0)[lt]{\lineheight{1.25}\smash{\begin{tabular}[t]{l}e.\end{tabular}}}}%
    \put(0.36558402,0.21659723){\color[rgb]{0,0,0}\makebox(0,0)[lt]{\lineheight{1.25}\smash{\begin{tabular}[t]{l}b.\end{tabular}}}}%
    \put(0,0){\includegraphics[width=\unitlength,page=1]{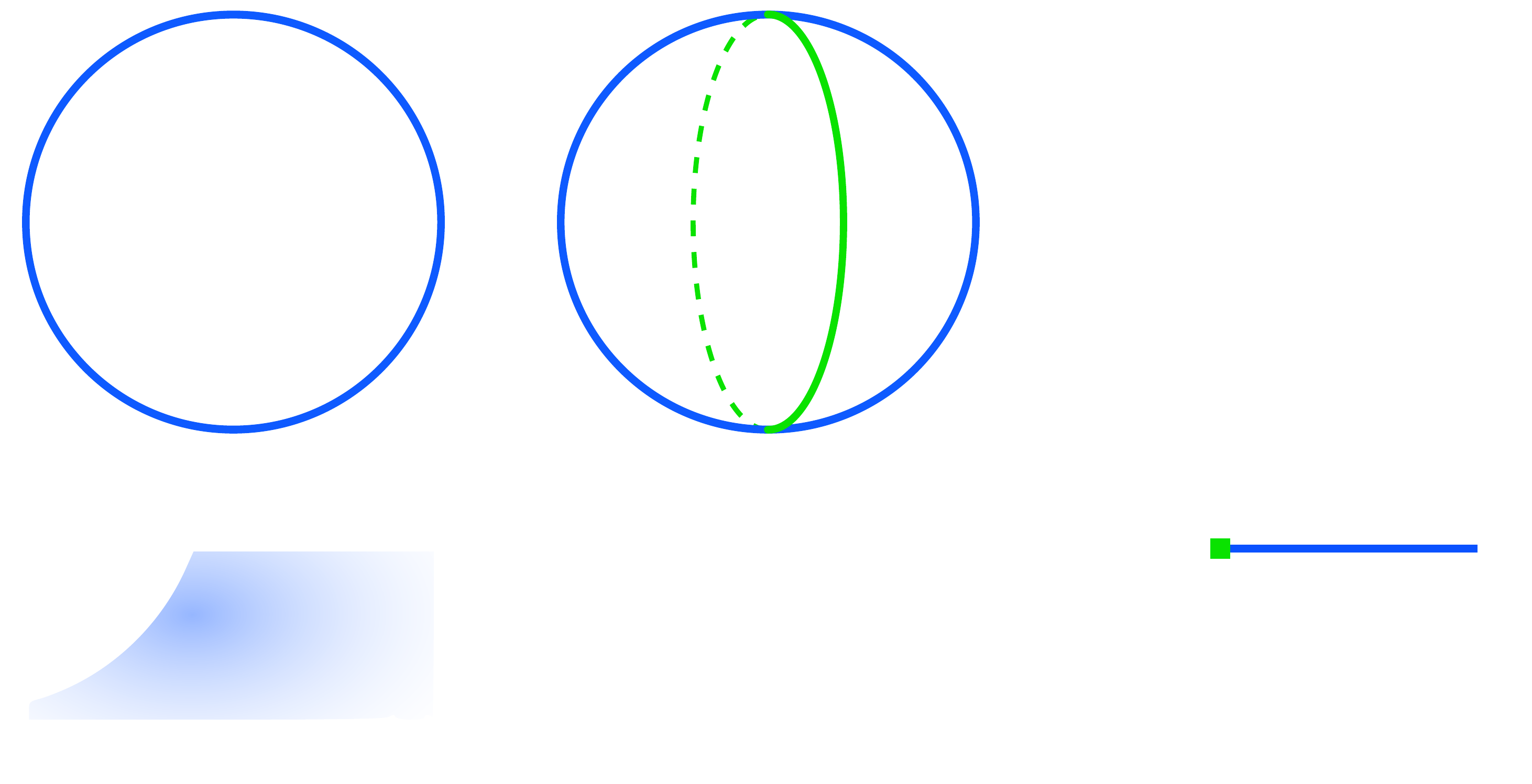}}%
    \put(0.01744126,0.21701045){\color[rgb]{0,0,0}\makebox(0,0)[lt]{\lineheight{1.25}\smash{\begin{tabular}[t]{l}a.\end{tabular}}}}%
    \put(0,0){\includegraphics[width=\unitlength,page=2]{ThreeTales_EditcopyJS4.pdf}}%
    \put(0.01744079,0.00964962){\color[rgb]{0,0,0}\makebox(0,0)[lt]{\lineheight{1.25}\smash{\begin{tabular}[t]{l}d.\end{tabular}}}}%
    \put(0,0){\includegraphics[width=\unitlength,page=3]{ThreeTales_EditcopyJS4.pdf}}%
  \end{picture}%
\endgroup%

		\caption{This figure shows the relation between a time-slice in our construction and the holographic setup of \cite{Almheiri:2019hni}. The top row illustrates three perspectives with which the system discussed here can be described, while the bottom row displays the analogous descriptions for the model in \cite{Almheiri:2019hni}. The comparison can be made more precise by performing a $\mathbb Z_2$ orbifold quotient across the bulk brane/conformal defect in the top row. \\
			\textbf{a.} Bulk gravity perspective, with an asymptotically AdS$_{d+1}$ space (shaded blue) which contains a co-dimension one Randall-Sundrum brane (shaded grey).\\ 
			\textbf{b.} Brane perspective, with dual CFT$_d$ on the asymptotic boundary geometry (blue) and also extending on the AdS$_d$ region (shaded green) where gravity is dynamical.\\
			\textbf{c.} Boundary perspective, with the holographic CFT$_d$ on $S^{d-1}$ (blue) coupled to a codimension-one conformal defect (green).\\
			\textbf{d.} AdS$_3$ formulation with two boundary components: the flat asymptotic boundary (straight black line) and a ``Planck brane'' (curved black line) with an AdS$_2$ geometry.\\ 
			\textbf{e.} The holographic CFT extends over a region with a fixed metric (blue) and an AdS$_2$ region with JT gravity (green).\\
			\textbf{f.} The microscopic description as a two-dimensional BCFT (blue) coupled to a quantum mechanical system at its boundary (green).\\
		}
		\label{fig:threetales}
	}
\end{figure}

First, consider the {\it bulk gravity perspective} corresponding to the geometric picture portrayed in section \ref{BranGeo}: we have an AdS$_{d+1}$ bulk region where gravity is dynamical, containing a DGP brane with tension running through the middle of the spacetime  -- see figure \ref{fig:threetales}a. The induced geometry on the brane is AdS$_d$. In the second picture, we integrate out the bulk action from the asymptotic boundary where gravity is frozen up to the brane, giving rise to Randall-Sundrum gravity \cite{Randall:1999vf,Randall:1999ee,Karch:2000ct} on the brane. From the resulting {\it brane perspective}, the CFT$_d$ is then supported in a region with dynamical gravity (\ie the brane) and another non-dynamical one (\ie the asymptotic boundary) -- figure \ref{fig:threetales}b. Finally, the third description makes full use of the AdS/CFT dictionary, by using holography {along} the brane. This {\it boundary perspective} describes the system as a CFT$_d$ coupled to a conformal defect that is located at the position where the brane intersects the asymptotic boundary -- see figure \ref{fig:threetales}c. 

A holographic system was presented in \cite{Almheiri:2019hni} to describe the evaporation of two-dimensional black holes in JT gravity. This system has three descriptions analogous to those above. Of course, it also includes certain elements that we did not introduce in our model, \ie end-of-the-world branes to give a holographic description of conformal boundaries separating various components  \cite{Takayanagi:2011zk,Fujita:2011fp} and performing a $\mathbb Z_2$ orbifold quotient across the Planck brane, \ie the brane supporting JT gravity. However, the essential ingredients are the same as above. The boundary perspective in \cite{Almheiri:2019hni} describes the system as a two-dimensional holographic conformal field theory with a boundary, at which it couples to a (one-dimensional) quantum mechanical system -- figure \ref{fig:threetales}f. With the brane perspective, the quantum mechanical system is replaced by its holographic dual, the Planck brane supporting JT gravity coupled to another copy of the two-dimensional holographic CFT -- see figure \ref{fig:threetales}e. Finally, the bulk gravity perspective replaces the holographic CFT with three-dimensional Einstein gravity in an asymptotically AdS$_3$ geometry. Because of the $\mathbb Z_2$ orbifolding, the latter effectively has two boundaries, the standard asymptotically AdS boundary and the dynamical Planck brane -- see figure \ref{fig:threetales}d.

This initial model \cite{Almheiri:2019hni} raised a number of intriguing puzzles. For example, as emphasized in \cite{Almheiri:2019yqk}, implicitly two different notions of the radiation degrees of freedom are being used: one being the semi-classical approximation and the other one in the purely quantum theory. Here, we will explain some details of the higher dimensional construction which allow us to provide a resolution of several of these questions in section \ref{sec:discussion}.

\begin{figure}[h]
	\def\svgwidth{1\linewidth}
	\centering{
\begingroup%
  \makeatletter%
  \providecommand\color[2][]{%
    \errmessage{(Inkscape) Color is used for the text in Inkscape, but the package 'color.sty' is not loaded}%
    \renewcommand\color[2][]{}%
  }%
  \providecommand\transparent[1]{%
    \errmessage{(Inkscape) Transparency is used (non-zero) for the text in Inkscape, but the package 'transparent.sty' is not loaded}%
    \renewcommand\transparent[1]{}%
  }%
  \providecommand\rotatebox[2]{#2}%
  \newcommand*\fsize{\dimexpr\f@size pt\relax}%
  \newcommand*\lineheight[1]{\fontsize{\fsize}{#1\fsize}\selectfont}%
  \ifx\svgwidth\undefined%
    \setlength{\unitlength}{825bp}%
    \ifx\svgscale\undefined%
      \relax%
    \else%
      \setlength{\unitlength}{\unitlength * \real{\svgscale}}%
    \fi%
  \else%
    \setlength{\unitlength}{\svgwidth}%
  \fi%
  \global\let\svgwidth\undefined%
  \global\let\svgscale\undefined%
  \makeatother%
  \begin{picture}(1,0.30787878)%
    \lineheight{1}%
    \setlength\tabcolsep{0pt}%
    \put(0.70308444,0.06171944){\color[rgb]{0,0,0}\makebox(0,0)[lt]{\lineheight{1.25}\smash{\begin{tabular}[t]{l}$\mu_\mt{B}$\end{tabular}}}}%
    \put(0,0){\includegraphics[width=\unitlength,page=1]{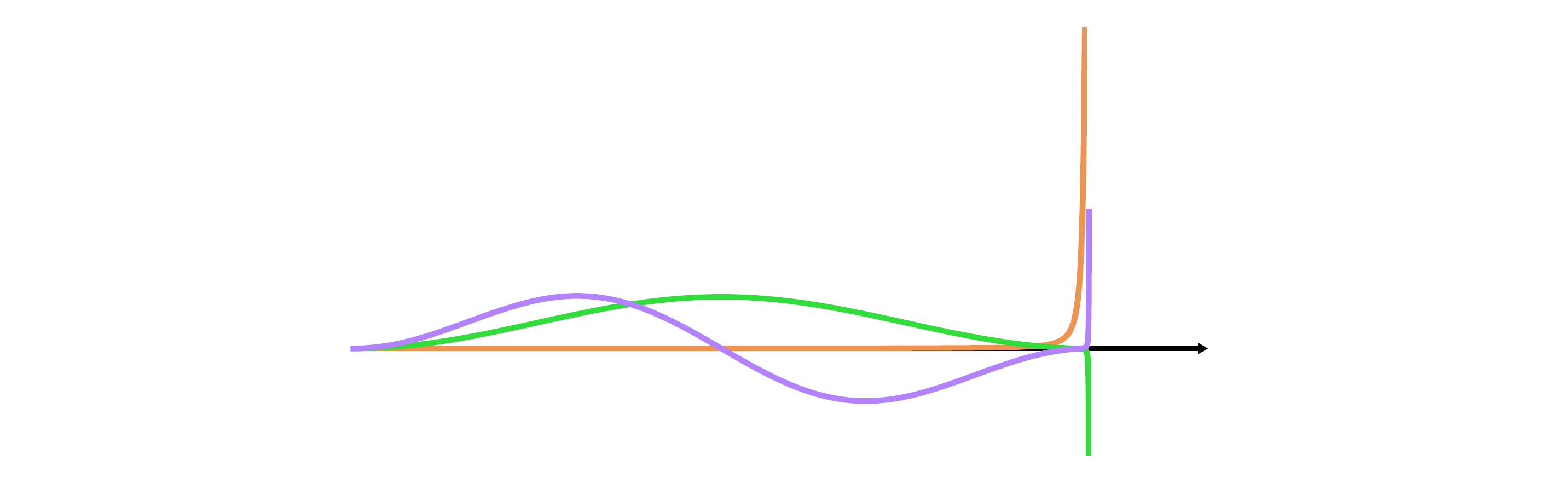}}%
    \put(0.23307041,0.28180926){\color[rgb]{0,0,0}\makebox(0,0)[lt]{\lineheight{1.25}\smash{\begin{tabular}[t]{l}$\psi(\mu)$\end{tabular}}}}%
    \put(0,0){\includegraphics[width=\unitlength,page=2]{bulkmodes_JSedit.pdf}}%
    \put(0.7411379,0.06171957){\color[rgb]{0,0,0}\makebox(0,0)[lt]{\lineheight{1.25}\smash{\begin{tabular}[t]{l}$\pi$\end{tabular}}}}%
    \put(0.22913163,0.06049422){\color[rgb]{0,0,0}\makebox(0,0)[lt]{\lineheight{1.25}\smash{\begin{tabular}[t]{l}0\end{tabular}}}}%
    \put(0.7764831,0.07889303){\color[rgb]{0,0,0}\makebox(0,0)[lt]{\lineheight{1.25}\smash{\begin{tabular}[t]{l}$\mu$\end{tabular}}}}%
  \end{picture}%
\endgroup%

		\caption{This figure illustrates the spatial profile of the first few normalized graviton modes in the presence of a large tension brane, and a $\mathbb Z_2$ orbifolding across the brane. We use the spatial coordinate $\mu$, related to $\rho$ in eq.~\reef{metric} by $\cot \mu = \sinh\rho/L$. The tension is adjusted such that the location of the brane is at $\mu = \mu_\mt{B}$ with $\mu_\mt{B}\lesssim\pi$. As discussed in the main text, the presence of the brane creates new bulk modes (orange), which are highly localized at the brane, and which play the role of a (nearly massless) graviton on the brane. The remaining bulk modes appear as KK modes in the brane theory.   }
                \label{fig:boundstate} 	}
\end{figure}

\paragraph{Bulk gravity perspective:} As discussed in section \ref{BranGeo}, the system has a bulk description in terms of gravity on an asymptotically AdS$_{d+1}$ spacetime containing a codimension-one brane, which splits the bulk into two halves -- see figure \ref{fig:threetales}a. The brane is characterized by the tension $T_o$ and also the DGP coupling $1/\Gbr$, introduced in eqs.~\eqref{braneaction} and \eqref{newbran}, respectively. We can use the Israel junction conditions \reef{Israel1} to determine the location of the brane as embedded in the higher dimensional space. The backreaction causes warping around the brane, and after a change of coordinates, tuning the brane tension can be understood as moving the brane further into a new asymptotic AdS region, as seen in eq.~\reef{positionbrane} or \reef{position}. For large brane tension, \ie with $\veps \ll 1$, the spectrum of graviton fluctuations in the bulk is almost unchanged with respect to the modes in empty AdS space. However, a new set of graviton states also appear localized at the brane \cite{Randall:1999vf,Randall:1999ee}, as illustrated figure \ref{fig:boundstate}. These are created by the nonlinear coupling of gravity to the brane. Unlike in the Randall-Sundrum model with a flat or de Sitter brane, the new graviton modes are not actually massless on the brane, but merely very light states whose wavefunction peaks around the brane \cite{Karch:2000ct,Karch:2001jb}. The remaining bulk graviton modes appear as a tower of  Kaluza-Klein states, from the point of view of the theory on the brane, with masses of ${\cal O}(1/\ell_\mt{eff})$ set by the curvature scale of the $d$-dimensional AdS geometry on the brane. These results have been studied in quite some detail \cite{Karch:2000ct,Karch:2001jb,Porrati:2001db,Miemiec:2000eq,Schwartz:2000ip,Porrati:2001gx} for Randall-Sundrum branes, but it is interesting to examine how the spectrum is modified by the DGP term \reef{newbran}. We will make some qualitative statements about this question below, but leave a detailed quantitative discussion and the interpretation of this mechanism from the point of view of the CFT for future work \cite{domino}. 

\paragraph{Brane perspective:} This second perspective, discussed in section \ref{indyaction}, effectively integrates out the spatial direction between the asymptotically AdS boundary and the brane to produce an effective action \reef{act3} for Randall-Sundrum/DGP gravity on the brane, with the new localized graviton state playing the role of the $d$-dimensional graviton. Hence, we are left with a $d$-dimensional theory of gravity coupled to (two copies of) the dual CFT on the brane -- see figure \ref{fig:threetales}b. As discussed in the description of the bulk perspective, amongst the new localized bulk modes, we have an almost massless graviton but also a tower of massive Kaluza-Klein states with masses of ${\cal O}(1/\ell_\mt{eff})$.  In section \ref{indyaction}, we demonstrated the consistency between the bulk gravity perspective and the brane perspective by observing how the equations of motion of the new effective action fix the brane position in the ambient spacetime. Of course, the bulk physics is also dual to the dual CFT on the asymptotic AdS$_{d+1}$ boundary, and so this description is completed by coupling the gravitational and CFT degrees of freedom on the brane to the  CFT on the fixed boundary geometry. We refer to that latter as the {\it bath} CFT. Next, we discuss how different parameters in the brane perspective are related to bulk parameters. 

There are four independent parameters which characterize the gravitational theory on the brane: the curvature scale $\ell_\mt{eff}$, the effective Newton's constant $G_\mt{eff}$, the central charge of the boundary CFT $\cT$, and the effective short-distance cutoff $\tilde\delta$. These emerge from the bulk theory through the four parameters characterizing the latter: the bulk curvature scale $L$, the bulk Newton's constant $\Gbk$, the brane Newton's constant $\Gbr$ and the brane tension $T_o$.\footnote{Recall that $\Delta T$ is determined by these parameters in eq.~\reef{tune3}, as well as eqs.~\reef{positionbrane} and \reef{curve1}.} From eq.~\reef{Newton2}, we see that $\ell_\mt{eff}$ is determined by a specific combination of $T_o$, $\Gbk$ and $L$. Similarly, $G_\mt{eff}$ is determined by $\Gbr$, $\Gbk$ and $L$ in eq.~\reef{Newton33}. The central charge of the boundary CFT is given by the standard expression $\cT\sim L^{d-1}/G_\mt{bulk}$, \eg see \cite{Buchel:2009sk}. 

Lastly, as discussed in section \ref{indyaction},  the theory on the brane comes with a short-distance cutoff $\tilde \delta$ \cite{deHaro:2000vlm,Emparan:2006ni,Myers:2013lva} at which the description of the brane theory in terms of (two copies of) the boundary CFT coupled to Einstein gravity breaks down. Following a standard bulk analysis, one would see that correlators of local operators (with appropriate gravitational dressings) now longer exhibit the expected CFT behaviour at short distances of order 
\beq
\tilde\delta_\mt{CFT}\sim L\,.
\label{ctoff2}
\eeq
We denote this cutoff with the subscript `CFT' to emphasize that the description of the matter degrees of freedom on the brane as a local $d$-dimensional CFT is failing at distances smaller than this short-distance cutoff. However, we stress that there is another scale $\tilde \delta_\mt{GR}$, which is the distance at which the approximation of Einstein gravity on the brane breaks down. The simple parameter counting above shows that this cannot be an independent scale. For the brane perspective, the true cutoff $\tilde \delta$ where the description in terms of the dual CFT coupled to Einstein gravity fails is 
\beq\label{ctoff}
\tilde \delta={\rm max}\left\{\tilde \delta_\mt{CFT}\,,\ \tilde \delta_\mt{GR}\right\}\,.
\eeq
We now discuss how $\tilde \delta_\mt{GR}$ is related to the other scales in the brane theory.

Recall that integrating out the bulk degrees of freedom produces a series of higher curvature terms in the effective action \reef{act3}, and hence demanding that $d$-dimensional Einstein gravity provides a good approximation of the brane theory introduces constraints. The suppression of these higher curvature corrections requires that the ratio $L/\ell_\mt{eff}$ be small. However, if we examine eq.~\reef{act3} carefully and note the distinction $G_\mt{eff}\ne G_\mt{RS}$, then suppressing the curvature-squared terms requires that
\beq
\frac{1}{1+\lamb}\,\frac{L^2}{\ell_\mt{eff}^2}
\ll 1\,,
\label{lmfao}
\eeq
using eq.~\reef{Newton34}. Note that for fixed bulk and boundary curvature scales, this implies a lower bound on the DGP term, such that $\lamb$ cannot be arbitrarily close to $-1$. For a pure RS brane with no additional DGP gravity, \ie $\lambda_b=0$, we conclude that the cutoff below which we find Einstein gravity coincides with the CFT cutoff $\tilde\delta_\mt{GR} \sim \tilde\delta_\mt{CFT}\sim L$. More generally then, the above expression suggests that the DGP term \reef{newbran} affects a shift producing a new short-distance cutoff for gravity,
\beq
\tilde\delta_\mt{GR} \sim \frac{L}{\sqrt{1+\lamb}}\ \sim
\frac{\tilde\delta_\mt{CFT}}{\sqrt{1+\lamb}}\,.
\label{haiku}
\eeq
Hence the true cutoff \reef{ctoff} depends on the sign of $\lamb$ -- we return to this point below. We should note that this result only applies for $d>4$. For $d=4$, the coefficient of the curvature-squared term is logarithmic in the cutoff, while for $d=2$ or $3$, this interaction is not associated with a UV divergence.

While the above are UV effects, there are also IR effects resulting from having a large number of matter degrees of freedom propagating on the brane, as explained in \cite{Dvali:2007hz,Dvali:2007wp,Reeb:2009rm}. The usual regime of validity for QFT in semiclassical gravity lies at energy scales below the Planck mass, or at distance scales larger than $G_\eff^{1/(d-2)}$. However, the boundary CFT has a large number of degrees of freedom, as indicated by the large $\cT$, and hence the semiclassical description of gravity in fact breaks down much earlier. A direct way to see this breakdown \cite{Dvali:2007wp} is to consider the computation of the (canonically normalized) graviton two-point function. In the high energy approximation, \ie ignoring the AdS geometry, we have here:\footnote{This propagator argument can also be applied for the higher curvature terms discussed above. For example, the curvature-squared terms gives a perturbative correction: $\langle h(p) \,h(-p)\rangle \sim p^{-2}\[1 + \frac{L^2}{1+\lamb}\, p^{2}+\cdots\]$. Hence this approach yields the same result for the cutoff in eq.~\reef{haiku}.}
\beq
\langle h(p)\, h(-p)\rangle \sim p^{-2}\[1 + \cT\, G_\eff\, p^{d-2}+\cdots\]\,.
\label{Dvali1}
\eeq
The leading correction arises from a diagram involving the external gravitons coupling to the CFT stress tensor two-point function. We see that such corrections are only suppressed relative to the `tree-level' result for momenta below a cutoff scale of order $(\cT G_\eff)^{-1/(d-2)}$. 
For our model, the gravitational theory of the brane can therefore only be treated semiclassically for distance scales larger than
\beq
\tilde\delta_\mt{GR} \sim (\cT G_\eff)^{1/(d-2)} \sim \frac{L}{(1+\lambda_b)^{1/(d-2)}}\sim \frac{\tilde\delta_\mt{CFT}}{(1+\lambda_b)^{1/(d-2)}}\,.
\label{Dvali2}
\eeq
Again, for a pure RS brane with $\lambda_b=0$, the cutoffs for Einstein gravity and the CFT agree, yielding $\tilde\delta \sim L$. However, the addition of a DGP gravity term modifies the cutoff, but in a manner distinct from eq.~\reef{haiku}, produced by the higher curvature terms. Note that the above result applies for $d\ge 3$.

The distinction between these two cutoffs indicates that these are really two different physical phenomena contributing to the breakdown of Einstein gravity in the brane perspective. Note that $\lamb>0$, in both eqs.~\reef{haiku} and \reef{Dvali2}, the effect is to produce a shorter cutoff scale, however, the second limit \reef{Dvali2} is the first to contribute (where we are assuming $d>4$). However, this result is smaller that $\tilde\delta_\mt{CFT}$ and hence from eq.~\reef{ctoff}, we find
\beq\label{ctoffplus}
\lamb>0\ \ :\qquad \tilde \delta \sim \tilde\delta_\mt{CFT}\sim L\,.
\eeq
On the other hand with $\lamb<0$, the cutoff $\tilde\delta_\mt{GR}$ is pushed to larger distance scales. In this case, eq.~\reef{haiku} is the first to modify the gravitational physics on the brane as we move to smaller distances. Further since this result is now larger than the CFT cutoff, in this regime, eq.~\reef{ctoff} yields
\beq\label{ctoffminus}
\lamb<0\ \ :\qquad \tilde \delta \sim \tilde\delta_\mt{GR}\sim \frac{L}{\sqrt{1+\lamb}}\,.
\eeq

Let us also note that the latter effect, \ie CFT corrections to the graviton propagator, are also responsible for the mass of the brane graviton \cite{Porrati:2001db}. It is interesting to note that if we take the high energy limit of the corrections to the graviton propagator, eq.~\eqref{Dvali1}, we can estimate a mass correction for low energy gravitons mode of roughly
\begin{align}
\label{eq:mass_correction}
\frac{\cT G_\eff}{\ell_\mt{eff}^{d}} \sim \frac{1}{(1 + \lambda_b)\, \ell_\mt{eff}^{\,2}} \left( \frac{L}{\ell_\mt{eff}}\right)^{d-2},
\end{align}
where we have substituted the $d$-dimensional AdS scale as a lower bound on the momentum. The scaling with the d-dimensional cosmological constant $- \frac 1 {\ell^2_\text{eff}}$ agrees with predictions in the Karch-Randall model \cite{Miemiec:2000eq, Schwartz:2000ip}. However, we caution the reader that the above argument by which we obtained the scaling is heuristic at best. Importantly, whether or not the graviton actually obtains a mass correction depends on the boundary conditions of the matter fields in AdS and can therefore not be determined by a local argument alone
\cite{Porrati:2001db}. However, taking eq.~\eqref{eq:mass_correction} at face value, we also see that a negative DGP coupling increases the mass scale, and vice versa for a positive coupling. This can be confirmed explicitly from bulk calculations \cite{domino}. 

\paragraph{Boundary perspective:} As the preceding discussion has made clear, the theory obtained by integrating out the bulk between the asymptotic boundary and the brane, has an effective description of the brane in terms of a local $d$-dimensional CFT coupled to Einstein gravity up to some cutoff \reef{ctoff}. However, the standard rules of AdS/CFT also allow for a fully microscopic description of the system in terms of the boundary theory. This is obtained by integrating out the bulk -- including the brane -- and the result  is given by the bath CFT on the fixed $d$-dimensional boundary geometry coupled to a ($d-1$)-dimensional conformal defect (positioned where the brane reaches the asymptotic boundary, \ie the equator of the boundary sphere) -- see figure \ref{fig:threetales}c. 

The bath CFT is characterized by the central charge $\cT\sim L^{d-1}/G_\mt{bulk}$, while the defect is characterized by its defect central charge $\tilde{c}_\mt{T}\sim \ell_\mt{eff}^{d-2}/G_\mt{eff}$. We note that in the absence of a DGP term, increasing the brane tension increases the defect central charge $\tilde c_\mt{T}$. Further, we note that the ratio of these two charges is given by
\beq\label{eq:eq:large_central_charges}
\frac{\tilde c_\mt{T}}{\cT} \sim \left(\frac{\ell_\mt{eff}}{L}\right)^{d-2} (1+\lambda_b) \,.
\eeq
Following the standard AdS/CFT dictionary, the ratio $\ell_\mt{eff}/L$ also  translates to a ratio of couplings in the defect and bath CFTs,\footnote{Remember that the AdS/CFT dictionary tells us that $G_N\sim \ell_\mt{AdS}^{d-1}/ N_{dof}$ and $\lambda_\text{Hooft}\sim (\ell_\mt{AdS}/\ell_\mt{s})^{d}$.}
\beq\label{ratou}
{\tilde \lambda}/{\lambda}\sim{\ell_\mt{eff}}/{L}  \,.
\eeq
Since we do not have a particular string construction in mind here, $\lambda$ should be thought of some positive power of the `t Hooft coupling of the bath CFT, while $\tilde\lambda$ will be some (different) positive power of the analogous coupling for the defect CFT.

Now the parameters in this boundary description must be constrained if we want to be in the regime where the brane perspective is valid. In particular, the latter requires that the brane curvature scale must be much larger than the effective cutoff, \ie
\beq
\ell_\mt{eff}/\tilde\delta \gg1\,.
\label{ratou3}
\eeq
Now as described above, the cutoff has a separate form depending on whether $\lamb$ is positive or negative. Eq.~\reef{ctoffplus} applies for $\lamb>0$, which then yields $\ell_\mt{eff}/L \gg1$. Hence we must have ${\tilde \lambda}/{\lambda}\gg1$ and also $\tilde c_\mt{T}/\cT\gg1$ since $1+\lamb>1$ in this case. Similarly for $\lamb<0$, combining eqs.~\reef{ctoffminus} and \reef{ratou3} yields $\ell_\mt{eff}/L \gg1/\sqrt{1+\lamb}$. In this case, $1+\lamb<1$ and it is straightforward to again show that the ratios must be constrained in the same manner. Hence for either sign of $\lamb$, we have 
\beq
{\tilde \lambda}/{\lambda}\gg1\qquad{\rm and}\qquad \tilde c_\mt{T}/\cT\gg1\,.
\label{ratou4}
\eeq
The large ratio of the central charges can also be heuristically understood requiring that energy and information are only leaking very slowly from the dynamical gravity region into the bath \cite{Rozali:2019day}. It has been argued that this ratio also sets the Page time \cite{Rozali:2019day}. With the boundary perspective, this can be understood as a requirement which ensures that the degrees of freedom on the defect and the CFT only slowly mix.

Lastly, the $d$-dimensional graviton can be understood as a field dual to the lightest operator appearing in the boundary OPE expansion of the CFT stress energy tensor \cite{Aharony:2003qf}. At weak coupling, one would naively assume that the lightest operator has dimension $\Delta = d$. However, due to strong coupling effects it becomes possible that a negative anomalous dimension of roughly $-1$ is obtained, so that the corresponding operator can act as the holographic dual to a $d$-dimensional graviton. The mass of the lightest state then signals that the anomalous dimension is not quite $-1$, such that the dimension of the boundary operator dual to the graviton is $\Delta \geq d-1$.


\section{Holographic EE on the Brane}\label{HEE}

In this section, we shall look for `quantum extremal islands' using the holographic setup described in the previous sections. Of course, quantum extremal islands have recently proven especially enlightening in the context of the black hole information paradox in two-dimensional JT gravity, where the emergence of these islands has signalled a transition to a phase where entropy of the Hawking radiation decreases over time, \eg \cite{Almheiri:2019psf, Almheiri:2019hni, Almheiri:2019yqk, Chen:2019uhq, Penington:2019kki, Almheiri:2019qdq, Chen:2019iro}. Some preliminary investigations of quantum extremal islands in higher dimensions also appeared in \cite{Penington:2019npb,Almheiri:2019psy}. In a companion paper \cite{QEI}, we will use the holographic model developed here to further extend these discussions to consider the black holes in arbitrary dimensions. However, in our present discussion black holes are not involved. Rather, we are simply considering the holographic entanglement entropies for certain regions in the vacuum of the boundary CFT coupled to the conformal defect. In situations to be discussed below, we find that the corresponding RT surfaces cross the brane in the bulk and this can be interpreted in terms of the appearance of a quantum extremal island in the effective theory of gravity on the brane.

In section \ref{sec:setupholo} we will describe the regions we are considering and the possible RT surfaces. Section \ref{sec:enzyme} discusses the extremization procedure of the RT surface in the presence of a brane and derives the conditions an RT surface needs to obey in our setting. Further, section \ref{sec:genera} shows that the leading contribution of the RT surface close to the brane can be understood as the Dong-Wald entropy on the brane, as seen from the brane perspective. In section \ref{sec:examples} we show an explicit calculation in $d=3$ and investigate the choices of parameters necessary to obtain quantum extremal islands. In particular, there, we will consider adding a DGP coupling to the bulk theory.

\subsection{Holographic setup}
\label{sec:setupholo}
\begin{figure}
	\def\svgwidth{0.8\linewidth}
	\centering{
\begingroup%
  \makeatletter%
  \providecommand\color[2][]{%
    \errmessage{(Inkscape) Color is used for the text in Inkscape, but the package 'color.sty' is not loaded}%
    \renewcommand\color[2][]{}%
  }%
  \providecommand\transparent[1]{%
    \errmessage{(Inkscape) Transparency is used (non-zero) for the text in Inkscape, but the package 'transparent.sty' is not loaded}%
    \renewcommand\transparent[1]{}%
  }%
  \providecommand\rotatebox[2]{#2}%
  \newcommand*\fsize{\dimexpr\f@size pt\relax}%
  \newcommand*\lineheight[1]{\fontsize{\fsize}{#1\fsize}\selectfont}%
  \ifx\svgwidth\undefined%
    \setlength{\unitlength}{566.92913386bp}%
    \ifx\svgscale\undefined%
      \relax%
    \else%
      \setlength{\unitlength}{\unitlength * \real{\svgscale}}%
    \fi%
  \else%
    \setlength{\unitlength}{\svgwidth}%
  \fi%
  \global\let\svgwidth\undefined%
  \global\let\svgscale\undefined%
  \makeatother%
  \begin{picture}(1,0.5)%
    \lineheight{1}%
    \setlength\tabcolsep{0pt}%
    \put(0,0){\includegraphics[width=\unitlength,page=1]{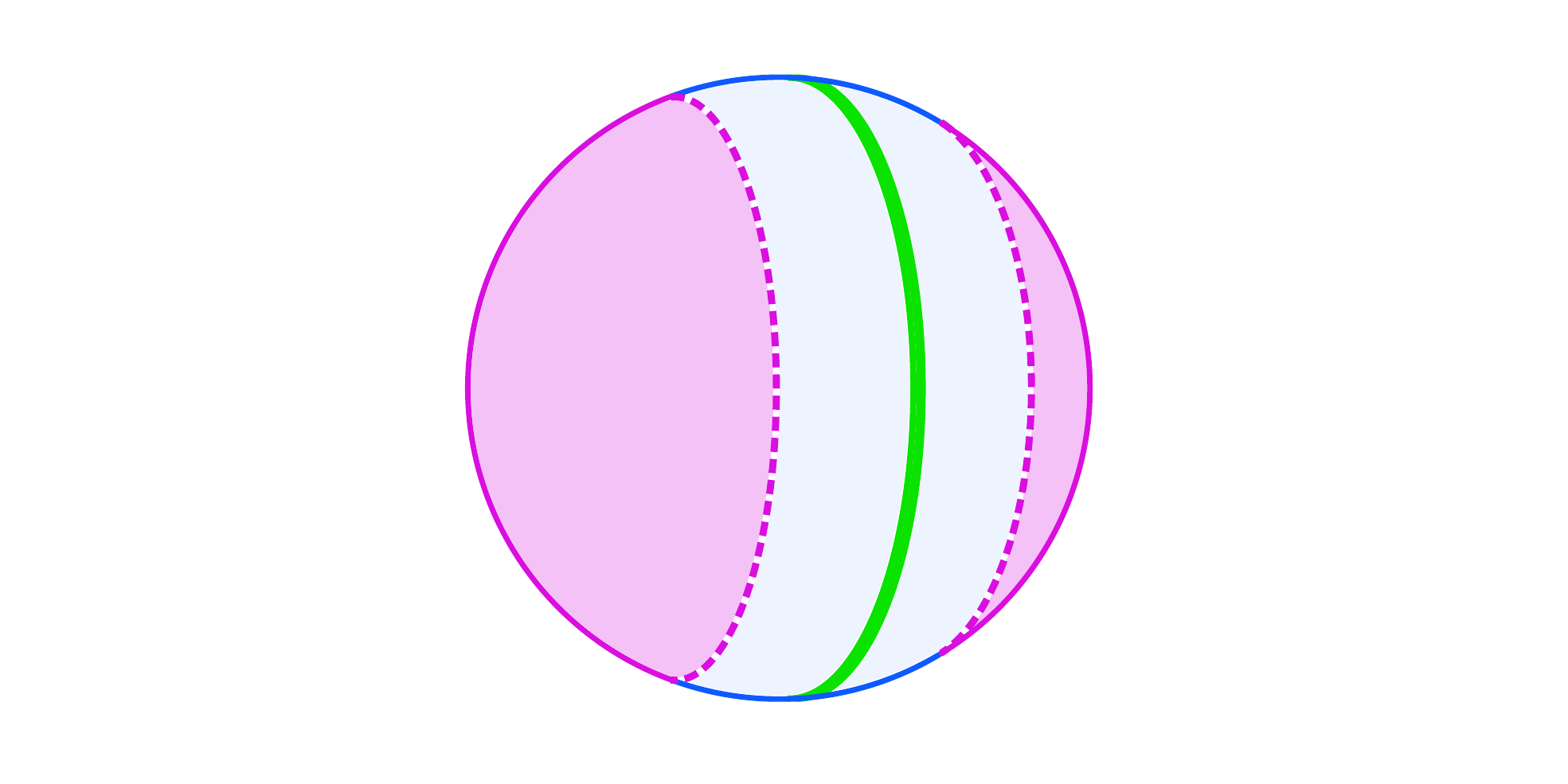}}%
    \put(0.66793671,0.03961742){\color[rgb]{0,0,0}\makebox(0,0)[lt]{\lineheight{1.25}\smash{\begin{tabular}[t]{l}Conformal Defect\end{tabular}}}}%
    \put(0,0){\includegraphics[width=\unitlength,page=2]{sec4.pdf}}%
    \put(0.43022644,0.47181392){\color[rgb]{0,0,0}\makebox(0,0)[lt]{\lineheight{1.25}\smash{\begin{tabular}[t]{l}$\theta_\mt{CFT}$\end{tabular}}}}%
    \put(0.5504454,0.46387848){\color[rgb]{0,0,0}\makebox(0,0)[lt]{\lineheight{1.25}\smash{\begin{tabular}[t]{l}$\theta_\mt{CFT}$\end{tabular}}}}%
    \put(0.38321579,0.23437525){\color[rgb]{0,0,0}\makebox(0,0)[lt]{\lineheight{1.25}\smash{\begin{tabular}[t]{l}\textbf{R}\end{tabular}}}}%
    \put(0,0){\includegraphics[width=\unitlength,page=3]{sec4.pdf}}%
    \put(0.21997648,0.1112373){\color[rgb]{0,0,0}\makebox(0,0)[lt]{\lineheight{1.25}\smash{\begin{tabular}[t]{l}$\Sigma_\mt{CFT}$\end{tabular}}}}%
    \put(0,0){\includegraphics[width=\unitlength,page=4]{sec4.pdf}}%
  \end{picture}%
\endgroup%

		\caption{A timeslice of our CFT setup. A conformal defect running along the equator separates the two halves of \textbf{R} and its corresponding engangling surface $\Sigma_\mt{CFT}$.} \label{EEprob}
	}
\end{figure}
\begin{figure}[h]
	\def\svgwidth{0.5\linewidth}
	\centering{
\begingroup%
  \makeatletter%
  \providecommand\color[2][]{%
    \errmessage{(Inkscape) Color is used for the text in Inkscape, but the package 'color.sty' is not loaded}%
    \renewcommand\color[2][]{}%
  }%
  \providecommand\transparent[1]{%
    \errmessage{(Inkscape) Transparency is used (non-zero) for the text in Inkscape, but the package 'transparent.sty' is not loaded}%
    \renewcommand\transparent[1]{}%
  }%
  \providecommand\rotatebox[2]{#2}%
  \newcommand*\fsize{\dimexpr\f@size pt\relax}%
  \newcommand*\lineheight[1]{\fontsize{\fsize}{#1\fsize}\selectfont}%
  \ifx\svgwidth\undefined%
    \setlength{\unitlength}{650bp}%
    \ifx\svgscale\undefined%
      \relax%
    \else%
      \setlength{\unitlength}{\unitlength * \real{\svgscale}}%
    \fi%
  \else%
    \setlength{\unitlength}{\svgwidth}%
  \fi%
  \global\let\svgwidth\undefined%
  \global\let\svgscale\undefined%
  \makeatother%
  \begin{picture}(1,0.77692308)%
    \lineheight{1}%
    \setlength\tabcolsep{0pt}%
    \put(0,0){\includegraphics[width=\unitlength,page=1]{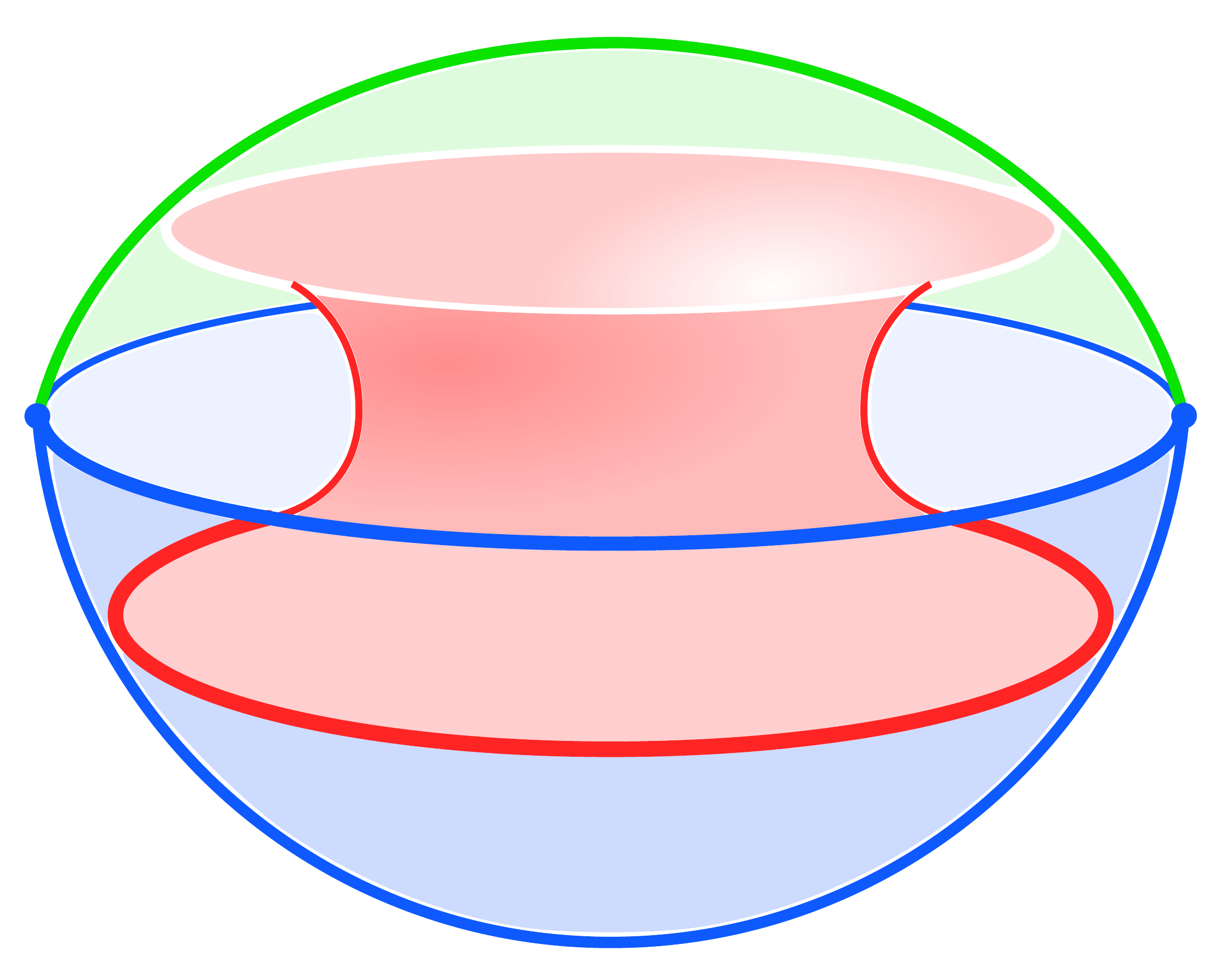}}%
    \put(0.71250948,0.42066535){\color[rgb]{0,0,0}\makebox(0,0)[lt]{\lineheight{1.25}\smash{\begin{tabular}[t]{l}$\Sigma_\xR$\end{tabular}}}}%
    \put(0.63804941,0.12333285){\color[rgb]{0,0,0}\makebox(0,0)[lt]{\lineheight{1.25}\smash{\begin{tabular}[t]{l}$\Sigma_\mt{CFT}$\end{tabular}}}}%
    \put(0,0){\includegraphics[width=\unitlength,page=2]{3d2.pdf}}%
    \put(0.63804997,0.60594907){\color[rgb]{0,0,0}\makebox(0,0)[lt]{\lineheight{1.25}\smash{\begin{tabular}[t]{l}$\sigma_\xR$\end{tabular}}}}%
  \end{picture}%
\endgroup%

		\caption{A timeslice of AdS$_{d+1}$ space. The entangling surface $\Sigma_\mt{CFT}$ lies on the CFT boundary and the RT surface $\Sigma_\xR$ intersects the brane at $\sigma_\xR$.
		}
		\label{fig:cutoffs}
	}
\end{figure}

In the remainder of this chaper, we will focus on a specific calculation of the entanglement entropy (EE): We consider the vacuum state of our boundary CFT on $\Rbb(\text{time})\times S^{d-1}(\text{space})$ with a conformal defect running along the equator of the $S^{d-1}$. As described in section \ref{sec:branegravity}, the bulk spacetime has locally an AdS$_{d+1}$ geometry and is bisected by a brane extending out to the defect position on the asymptotic boundary. Now we wish to evaluate the EE in the boundary CFT for a region $\xR$ comprised of the union of two polar spherical caps on the $S^{d-1}$ -- see figure \ref{EEprob}. We follow the usual holographic prescription to compute the EE. That is, we examine the bulk surfaces $\xV$ which are homologous to  $\xR$ and extremize the generalized entropy functional
\beq
 S_\EE(\xR) = {\rm min}\left\{\extr\,
 S_\gen(\xV)\right\}={\rm min}\left\{\extr
  \(
  \frac{A(\xV)}{4 G_\bulk} + \frac{A(\xV \cap {\rm brane})}{4 G_\brane}\)\right\}\,.
 \label{eq:sad}
\eeq
Of course, the first term above corresponds to the usual Ryu-Takayanagi (RT) term \cite{Ryu:2006bv,Ryu:2006ef} while, as discussed in appendix \ref{generalE}, we expect the second term to arise whenever the bulk surface crosses a DGP brane.\footnote{Implicitly, eq.~\eqref{eq:sad} assumes that the bulk and brane gravitational actions both correspond to the Einstein-Hilbert action (with a cosmological constant term), as in eqs.~\reef{act2} and \reef{newbran}.} Let us denote the extremal bulk surface as $\Sigma_\xR$, and the intersection with the brane $\sigma_\xR=\Sigma_\xR \cap {\rm brane}$, see figure \ref{fig:cutoffs}. Importantly, if there are multiple extrema, the EE is given by chosing the extremal surface yielding the smallest value for $S_\gen(\Sigma_\xR)$, as indicated above.

For the calculation described above, the candidate RT surfaces are anchored at the AdS boundary to the entangling surface $\SCFT=\partial \xR = \partial \RTsurf$, \ie the boundaries of these surfaces are comprised of two $(d-2)$-spheres, which are the boundaries of the two polar caps. We will find that there are two topologically distinct candidates for $\RTsurf$ which extremize the generalized entropy in eq.~\reef{eq:sad}, see figure \ref{fig:RTPhases}. The first consists of two disconnected disks on either side of the brane (in which case $\sigma_\xR=\{\varnothing\}$). The second candidate has a cylindrical geometry which pierces the brane. Hence it is only in this latter case that the second term contributes in eq.~\reef{eq:sad}. As noted above, the correct RT surface is chosen from these two candidates as the one which yields a smaller generalized entropy. Generally, we shall find that when the two polar caps are small, the disconnected discs are favoured, while the cylindrical surface can be the leading saddle for when the polar caps are large. We will denote the first situation as the `disconnected' phase and the latter as the `connected' phase. As we will describe in section \ref{sec:examples}, the details of the transition between these two phases also depends on other parameters in the holographic model, \eg the tension and gravitational coupling of the brane.

To understand the interpretation of these results in terms of quantum extremal islands, we turn to the `brane perspective' described in the previous section. This effective description gives the `island rule' proposed in \cite{Almheiri:2019hni} for the entanglement entropy,
\beqa
 S_\EE(\xR) &=& {\rm min}\left\{\extr\,
 S_\gen(R\cup \text{islands} )\right\} \labell{rule1}\\
&=&{\rm min}\left\{\extr
  \(  S_\EE(\braneR \cup \text{islands}) + \frac{A\(\partial(\islands)\)}{4 G_\brane}\)\right\}\,.
 \nonumber
\eeqa
As geometries $\CFTR=\braneR$, but we have used a different font on the right-hand side to emphasize the fact that the effective `brane perspective' does not give the same detailed description of the CFT state on $\CFTR$, as the boundary or bulk perspectives. Based on our discussion in the previous sections, one might have expected that the gravitational term in eq.~\reef{rule1} would involve $G_\eff$ rather than $G_\brane$.  This is implicit in \eqref{rule1}, as we will see below. In the presence of an island, the first term $S_\EE(\braneR \cup \text{islands})$ receives two large contributions, coming from the asymptotic AdS boundary and the region close to the brane. It is this second term, proportional to $1 / G_\text{RS}$, which combines with the last term in eq. \eqref{rule1} to yield the expected island contribution proportional to $1/G_\eff$, c.f. eq.~\reef{Newton33}.

It is now straightforward to interpret the previous holographic discussion in terms of the effective theory on the brane, eq.~\reef{rule1}. In the connected phase, the holographic RT surface crosses the brane and (if DGP couplings are turned on) we see an explicit brane contribution in eq.~\reef{eq:sad}. From the brane perspective, a quantum extremal island has formed in the gravitational region (\ie the region on the brane enclosed by $\sigma_\xR$)\footnote{Let us add that from the bulk perspective, entanglement wedge reconstruction \cite{EW1,EW2,EW3,Jafferis:2015del,Dong:2016eik,Faulkner:2017vdd,Cotler:2017erl} ensures that operators within this island can be reconstructed from boundary CFT data in $\xR$.} and the analogous gravitational term appears in the island rule \reef{rule1}. The bulk RT contribution in eq.~\reef{eq:sad} corresponds to $S_\EE(\braneR \cup \text{islands})$ in eq.~\reef{rule1}. As alluded to above, this makes clear that the gravitational contribution to the island is comprised of two components: the bare contribution $\sim 1/G_\brane$, which arises from a DGP coupling added to the brane, and the bulk contribution proportional to $L/G_\bulk$, which arises from the volume of the RT surface close to the brane. 
To see how the latter arises in the effective theory, notice that we can split $S_\EE$ into UV-finite and UV-`divergent' contributions close to the brane, where the latter are contributions proportional to inverse powers of $\s$. These are the analog of the UV divergent boundary contributions for the boundary CFT in the regions without gravity. As discussed in section \ref{face}, the brane position imposes a UV cutoff for the CFT on the brane, and hence the corresponding `divergent' contributions to the EE are in fact finite and instead yield contributions which match those expected for the gravitational entropy from the induced contributions to eq.~\reef{act3}. This makes contact with the usual notion of generalized entropy as the sum of the geometric gravitational entropy and the entropy of the quantum fields \cite{Faulkner:2013ana,Engelhardt:2014gca}.

In the disconnected phase, the EE only involves the modes enclosed within the two polar caps and there is no contribution from the CFT in the gravitational region, \ie on the AdS$_d$ brane. In passing, let us recall that the short wavelength modes in the vicinity of the entangling surface $\SCFT$ produce various UV divergent boundary contributions, such as the celebrated area law term \cite{Sorkin_1983,Bombelli:1986rw,Srednicki:1993im}. Of course, in both phases these contributions are regulated in the holographic calculation by introducing a cutoff surface near the asymptotic AdS boundary \cite{Rangamani:2016dms}.

\subsection{RT meets the Brane}
\label{sec:enzyme}
In this section, we shall introduce some technical details, which are useful to calculate the EE associated with the two polar caps in the boundary CFT. In particular, we examine the behaviour of the bulk RT surface $\Sigma_\xR$ as it crosses the brane, \ie how the intersection surface $\sigma_\xR$ is determined.
However, we begin by specifying our EE calculation more precisely and reviewing the metrics describing the bulk spacetime.

Let us describe the $\Rbb\times S^{d-1}$ geometry on which the boundary CFT lives with,
\beq\label{metricCFT}
ds^2=R^2\left[-dt^2+
d\theta^2+\sin^2\!\theta\,d\Omega_{d-2}^2 \right]\,,
\eeq
where $R$ is the radius of curvature of the $(d-1)$-sphere. The polar angle $\theta$ runs over $0\le\theta\le\pi$, and the conformal defect sits at the equator $\theta=\pi/2$. As illustrated in figure \ref{EEprob}, we wish to evaluate the EE in the boundary CFT for a region $\xR$ comprised of two polar caps on the $S^{d-1}$. More specifically, we choose the entangling surface $\SCFT$ to be two circles placed  symmetrically on either side of the defect at $\theta=\pi/2\pm\thb$. Hence we are evaluating the EE between these two balls and the complementary region, which corresponds to a `belt' of width $2\thb$ centered on the conformal defect.

Turning now to the bulk geometry, recall that in section \ref{BranGeo}, we discussed the background solution in terms a metric where the AdS$_{d+1}$ geometry was foliated by AdS$_d$ slices. Eq.~\reef{metric3} describes the local geometry on either side of the brane
located at $z=\s$ with
\beq\label{metric3a}
ds^2=\frac{L^2}{z^2}\left[dz^2 +  L^2\left(1 + \frac{z^2}{4\,L^2}\right)^2 \left(-\cosh^2\!\tdr\,dt^2+d\tdr^2+\sinh^2\!\tdr\,d\Omega_{d-2}^2 \right)\right]\,.
\eeq
While these coordinates are well suited to discuss the brane geometry, we also consider `global' coordinates for the AdS$_{d+1}$ geometry
\beq\label{metric2s}
ds^2=L^2\left[-\cosh^2\!r\,dt^2+dr^2+\sinh^2\!r\,\left(
d\theta^2+\sin^2\!\theta\,d\Omega_{d-2}^2\right) \right]\,,
\eeq
which are better adapted to discuss the boundary theory. That is, up to a Weyl rescaling, the geometry on fixed $r$ surfaces matches eq.~\reef{metricCFT}  in the asymptotic region, and the UV regulator surface needed to properly define the holographic EE can be simply chosen as some slice $r=r_\mt{UV}\gg L$.

However, while we refer to eq.~\reef{metric2s} as `global' coordinates, they do not cover the entire back-reacted bulk solution depicted in figure \ref{fig:brane2}. Rather we use the coordinates in eq.~\reef{metric2s} to cover two patches on either side of the brane and near the asymptotic AdS$_{d+1}$ boundary.\footnote{Of course, the same applies for the previous coordinates in eq.~\reef{metric3a}.} Comparing eqs.~\reef{metric3a} and \reef{metric2s}, it is straightforward to identify the transformation between the two coordinate systems as
\beq\label{transfor1}
\tanh\tdr=\tanh r\sin\theta\,,\quad \frac{z}L=-2\sinh r\,\cos\theta\pm2\sqrt{\sinh^2\! r\,\cos^2\!\theta+1}\,.
\eeq
With the + (--) sign, the brane at $z=z_\mt{B}\ll L$ resides near the boundary hemisphere with $0\le\theta\le\pi/2$ ($\pi/2\le\theta\le\pi$) and $r\to\infty$. Therefore letting $\theta$ run from 0 to $\pi$ on the boundary with the defect at $\theta=\pi/2$, we choose the -- (+) sign to cover the patch covering the asymptotic boundary hemisphere $0\le\theta\le\pi/2$ ($\pi/2\le\theta\le\pi$).

Using the AdS foliation \reef{metric3a}, the position of the brane was specified by $z=\s$. In terms of the global coordinates \reef{metric2s} , the brane position can be specified with
\beq\label{eq:foobar}
\sinh^2\! r\, \cos^2\!\theta= \frac{L^2}{\s^2}\(1-\frac{\s^2}{4L^2}\)^2\,.
\eeq
The specific sign of $\cos\theta$ depends on whether one considers the coordinate patch above or below the brane -- see comments below eq.~\reef{transfor1}. Further, we reach the asymptotic boundary on the brane by taking $\tdr\to\infty$, which in the global coordinates then corresponds to $r\to\infty$ and $\theta\to\pi/2$. Hence, we see that the brane intersects the asymptotic boundary at the position of the conformal defect, as expected.

To examine the behaviour of the bulk RT surface $\Sigma_\xR$ where it crosses the brane, it is useful to consider the problem of extremal surfaces using the metric~\reef{metric3a}. Because the bulk geometry is static, the RT surfaces will be confined to a constant time slice in the bulk. The entangling surfaces in the boundary are spherically symmetric and so we only need to consider bulk surfaces with the same rotational symmetry on the $S^{d-2}$, that is, we parametrize the surfaces as $\tdr=\tdr(z)$ and the bulk contribution to the holographic EE is then given by
\beq\label{area0}
S_\mt{bulk}= 2\,\frac{L^{d-1}\, \Omega_{d-2}}{4\,\Gbk} \int \frac{dz}z \,\[\frac{L}{z}\left(1 + \frac{z^2}{4\,L^2}\right)\sinh\tdr\]^{d-2} \sqrt{1+L^2\left(1 + \frac{z^2}{4\,L^2}\right)^2\(\frac{d\tdr}{dz}\)^2}
\eeq
where $\Omega_{d-2}$ is the area of a unit $(d-2)$-sphere.\footnote{Recall that the area of a unit $n$-sphere is given by $\Omega_{n} = 2\pi^{\frac{n+1}2}/\Gamma\(\frac{n+1}2\)$. \label{footsphere}} An overall factor of 2 is included here because we assume that the profile $\tdr(z)$ will be reflection symmetric about the brane, and hence $S_\mt{RT}$ recieves the same contribution from both sides.

Treating eq.~\reef{area0} as an action, we would derive an `Euler-Lagrange' equation for the profile whose solution corresponds to an extremal surface in the bulk, \ie away from the brane.\footnote{This equation is rather involved and the details are not important here.} However, this equation is second order and so the solutions are parameterized by two integration constants. One of these constants is fixed by the angle $\theta_\mt{CFT}$ on the asymptotic boundary (\ie the position of the entangling surface $\SCFT$ in the boundary theory), and the other, by the radius $\tdr_\mt{B}$ at which the RT surface intersects the brane, \ie $\tdr(z=\s)=\tdr_\mt{B}$. We are thus left with the question of fixing the boundary condition at the brane.

\begin{figure}[h]
	\def\svgwidth{\linewidth}
	\centering{
\begingroup%
  \makeatletter%
  \providecommand\color[2][]{%
    \errmessage{(Inkscape) Color is used for the text in Inkscape, but the package 'color.sty' is not loaded}%
    \renewcommand\color[2][]{}%
  }%
  \providecommand\transparent[1]{%
    \errmessage{(Inkscape) Transparency is used (non-zero) for the text in Inkscape, but the package 'transparent.sty' is not loaded}%
    \renewcommand\transparent[1]{}%
  }%
  \providecommand\rotatebox[2]{#2}%
  \newcommand*\fsize{\dimexpr\f@size pt\relax}%
  \newcommand*\lineheight[1]{\fontsize{\fsize}{#1\fsize}\selectfont}%
  \ifx\svgwidth\undefined%
    \setlength{\unitlength}{1500bp}%
    \ifx\svgscale\undefined%
      \relax%
    \else%
      \setlength{\unitlength}{\unitlength * \real{\svgscale}}%
    \fi%
  \else%
    \setlength{\unitlength}{\svgwidth}%
  \fi%
  \global\let\svgwidth\undefined%
  \global\let\svgscale\undefined%
  \makeatother%
  \begin{picture}(1,0.33333333)%
    \lineheight{1}%
    \setlength\tabcolsep{0pt}%
    \put(0,0){\includegraphics[width=\unitlength,page=1]{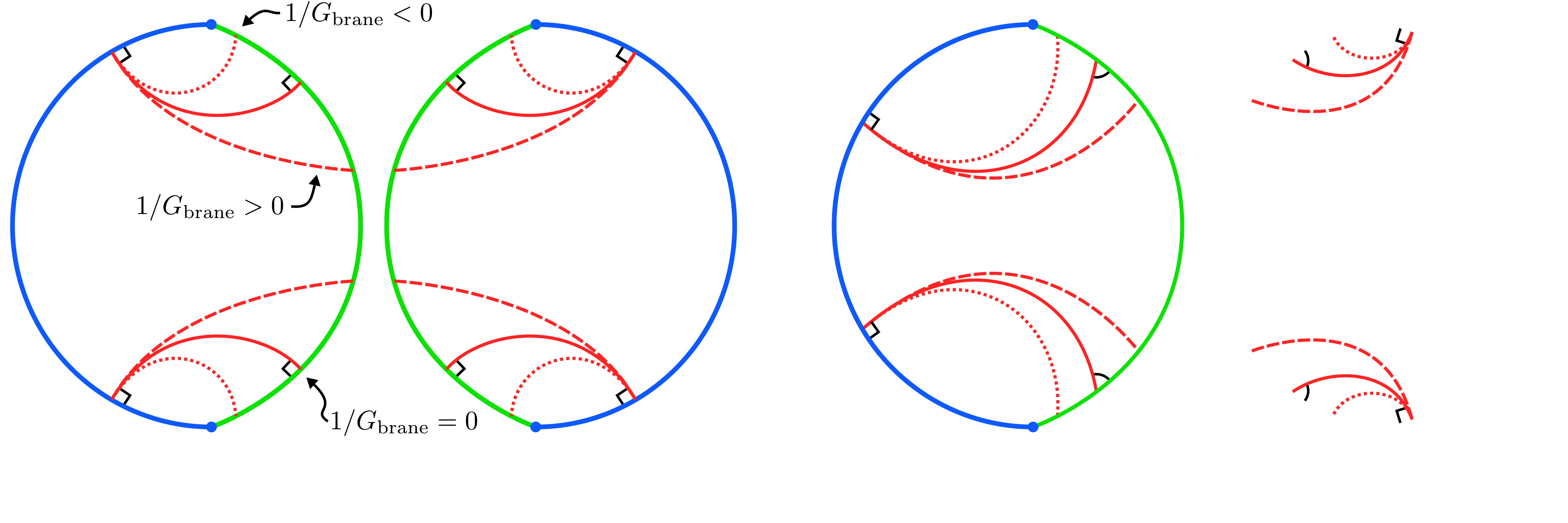}}%
    \put(0.16214927,0.0064484){\color[rgb]{0,0,0}\makebox(0,0)[lt]{\lineheight{1.25}\smash{\begin{tabular}[t]{l}$\mathbb{Z}_2$-symmetric\end{tabular}}}}%
    \put(0.67871974,0.00644822){\color[rgb]{0,0,0}\makebox(0,0)[lt]{\lineheight{1.25}\smash{\begin{tabular}[t]{l}$\mathbb{Z}_2$-asymmetric\end{tabular}}}}%
    \put(0,0){\includegraphics[width=\unitlength,page=2]{RTBraneIntersectionJS.pdf}}%
  \end{picture}%
\endgroup%

		\caption{
                  Families of extremal surfaces anchored at fixed positions on the asymptotic AdS boundary. The true RT surfaces are the members of these families which extremize area in the bulk, or equivalently, generalized entropy in the brane perspective. The RT surfaces in the case of zero, positive, and negative $1/\Gbr$ are respectively shown in solid, dashed, and dotted red. In the absence of a DGP Einstein-Hilbert action ($1/\Gbr=0$), the RT surfaces passe `straight' through the brane. The left (right) panel shows the computation of entanglement entropy for a region in the boundary CFT that is $\Zbb_2$-symmetric (-asymmetric) about the defect. 
		}
		\label{fig:rtintersectionangles}
	}
\end{figure}

There are two contributions that come into play at the brane. The first is the DGP contribution in eq.~\reef{eq:sad},
\beq\label{area1}
S_\mt{brane}= \frac{L^{d-2}\, \Omega_{d-2}}{4\,\Gbr} \,\[\frac{L}{\s}\left(1 + \frac{\s^2}{4\,L^2}\right)\sinh\tdr_\mt{B}\]^{d-2} \,.
\eeq
The second is a boundary term that comes from integrating by parts in the variation of the RT functional \reef{area0}. Combining these, one arrives at the following expression\footnote{Implicitly, we assume that we care considering the RT surfaces with a cylindrical topology, \ie in the connected phase. Examining these boundary terms carefully, one also finds that they are eliminated with $\tdr_\mt{B}=0$. This solution points towards the existence of the second phase of disconnected surfaces, which do not intersect the brane.}
\beq\label{ortho1}
L\,\frac{d\tdr}{dz}\bigg|_{z=\s}=\frac1{\Gbr}\,
\frac{\s/L}{1 + \frac{\s^2}{4\,L^2}}\[
\left(1 + \frac{\s^2}{4\,L^2}\right)^2\(\frac{2L}{\Gbk}\,\tanh\tdr_\mt{B}\)^2-\(\frac{\s/L}{\Gbr} \)^2\]^{-\frac12}\,.
\eeq
Hence, scanning through the family of RT surfaces parametrized by $\tdr_\mt{B}$, the solution which satisfies the above boundary condition is the one that properly extremizes the full entropy functional in eq.~\reef{eq:sad}. One observation is that without the DGP term, \ie $1/\Gbr=0$, the boundary condition simplifies to $L\,{d\tdr}/{dz}|_{z=\s}=0$. That is, the RT surface intersects the brane at a right angle. Turning on the gravitational action on the brane (with a positive coupling) produces $L\,{d\tdr}/{dz}|_{z=\s}>0$, which arises from pushing $\tdr_\mt{B}$ to a smaller value. The decrease in $\tdr_\mt{B}$ is natural here because the DGP contribution in eq.~\reef{area1} adds an additional penalty for large areas on the brane and the effect is to shrink the area of $\sigma_\xR$.\footnote{If $1/\Gbr<0$ as we consider in section \ref{sec:examples}, then the DGP entropy \reef{area1} facilitates a larger area for $\sigma_\xR$ and so we find that $\tdr_\mt{B}$ increases.} This is illustrated in the left panel of figure \ref{fig:rtintersectionangles}.

We observe that the above analysis has a simple interpretation in terms of the island rule \reef{rule1}. Recall that extremizing the RT functional \reef{area0} leads to a family of bulk solutions that are parametrized by $\tdr_\mt{B}$, their radius on the brane. Evaluating $S_\mt{bulk}+S_\mt{brane}$ for these different solutions is equivalent to evaluating the $S_\gen$ in eq.~\reef{rule1} with different candidates for the island geometry. The final step of extremizing with respect to variations of $\tdr_\mt{B}$ then matches the extremization in the island rule and identifies the quantum extremal surface $\sigma_\xR$ on the brane.

Next, we provide a more general geometric discussion of the boundary conditions.
The orthogonality between the RT surface and the RS brane is a special feature of the reflection symmetry of our setup. For a better geometric understanding of the boundary conditions, let us examine eq.~\reef{eq:sad} in more detail. Consider a $(d-1)$-dimensional surface $\xV$ parametrized by intrinsic coordinates $\xi^\alpha$, embedded in the $(d+1)$-dimensional bulk spacetime with coordinates $X^\mu$ and metric $g_{\mu\nu}$. Hence we describe the embedding of this surface in the bulk spacetime as $X^\mu = X^\mu(\xi^\alpha)$, and the induced metric then becomes
\beq
h_{\alpha\beta}= g_{\mu\nu}\,
  \frac{\partial X^\mu}{\partial \xi^\alpha}\,
  \frac{\partial X^\nu}{\partial \xi^\beta}\,.
\label{badnote}
\eeq
Hence the bulk contribution in eq.~\reef{eq:sad} becomes
\beq
  S_\mt{bulk}=\frac{A(\xV)}{4 G_\bulk}
  = \frac{1}{4G_\bulk}\int_\xV d^{d-1}\xi\; \sqrt{h}\,.
\label{eq:badNotation}
\eeq
Next to evaluate the brane contribution in eq.~\reef{eq:sad}, we introduce $(d-2)$ coordinates $y^a$ to parameterize the intersection of $\xV$ and the brane. The induced metric on this intersection surface then becomes
\beq\label{badnote2}
  \tilh_{ab}
  = h_{\alpha\beta}
  \frac{\partial \xi^\alpha}{\partial y^a}
  \frac{\partial \xi^\beta}{\partial y^b}\,,
\eeq
and the corresponding contribution to the generalized entropy is
\begin{align}
  S_\brane
  =& \frac{1}{4\Gbr}\int d^{d-2}y\; \sqrt{\tilh}\,.
  \label{eq:facundo}
\end{align}

Now following the prescription in eq.~\reef{eq:sad}, we wish to  extremize the sum of the two quantities above. So we begin with the variation of $S_\mt{bulk}$, which yields
\beq
\begin{split}
  \delta S_\mt{bulk}
  =& \frac{1}{4G_\bulk}\Bigg[ \int_{\xV\cap{\rm brane}}\!\!\!\!\!\!\!\! d^{d-2} y\,
  \sqrt{\tilh}\, g_{\mu\nu}\, (\partial_{n_\mt{R}} X^\mu+\partial_{n_\mt{L}}X^\mu) \,\delta X^\nu
  \\
  &\qquad+\int_\xV d^{d-1}\xi \sqrt{h}\,[\text{e.o.m.}]_\nu\, \delta X^\nu
  \Bigg]\,.
\end{split}
  \label{eq:recipe}
\eeq
Here we assume that the equations of motion along the bulk of $\xV$ can be satisfied and so the second term above vanishes. However, one must integrate by parts to arrive at these equations and so we are left with a boundary term where $\xV$ crosses the brane.\footnote{Dirichlet boundary conditions remove the analogous boundary contributions at the asymptotic AdS boundary.} Here we are assuming that the extremal surface is not necessarily smooth at the brane and so $n_\mt{R}^\alpha$ and $n_\mt{L}^\alpha$ are unit normals to the intersection surface directed along the extremal surface approaching the brane from either side.

In the absence of the DGP term \reef{newbran}, there is no brane contribution \reef{eq:facundo} and then the vanishing of the boundary term in eq.~\reef{eq:recipe} dictates $n_\mt{R}^\alpha+n_\mt{L}^\alpha=0$.\footnote{Actually, the requirement is $\tg_{i\alpha}(n_\mt{R}^\alpha+n_\mt{L}^\alpha)=0$, \ie the projection into the brane of the sum of the two normals vanishes -- see the discussion after eq.~\reef{ortho7}. However, the vanishing of the full vector sum follows from this restriction.} That is, with $1/\Gbr=0$, the boundary condition is that the RT surface should pass smoothly through the brane --- this is illustrated by the solid red RT surfaces in figure \ref{fig:rtintersectionangles}. In the reflection symmetric setup considered above, this can only be accomplished if the RT surface is orthogonal to the brane, \ie both $n_\mt{R}^\alpha$ and $n_\mt{L}^\alpha$ are orthogonal to the brane.

Of course, with a DGP brane, we must also consider the variation of $S_\brane$ in eq.~\reef{eq:facundo}, which yields
\begin{align}
  \delta S_\brane
  =& \frac{1}{4\Gbr}
  \int d^{d-2}y\; \sqrt{\tilh}\,
  \inducedK_i \,\frac{\partial x^i}{\partial X^\nu}\, \delta X^\nu,
  \label{eq:puerto}
\end{align}
where $\inducedK_i$ denotes the trace of the extrinsic curvature of the intersection surface on the brane, as viewed from the brane geometry (with the $d$ coordinates $x^i$).\footnote{In deriving eq.~\eqref{eq:puerto}, we used that $\inducedK_i$ gives the expansion of the area element $\sqrt{\tilh}$ under the map produced by geodesics shooting out normal to the intersection surface, ${\rm RT}\cap\Brane$. }
Requiring the sum of eqs.~\eqref{eq:recipe} and \eqref{eq:puerto} to vanish then yields the boundary condition
\beq\label{ortho7}
 0  =  \tg_j{}^\nu\(g_{\mu\nu} (\partial_{n_\mt{R}} X^\mu+\partial_{n_\mt{L}}X^\mu)
  + \frac{G_\bulk}{G_\brane}\,\inducedK_i \,\partial_\nu x^i\)\,.
\eeq
Here, we think of the induced metric on the brane as the bulk tensor $\tg_{\mu\nu}=g_{\mu\nu}-N_\mu N_\nu$, where $N_\mu$ is the unit normal orthogonal to the brane. Then, the initial factor $\tg_j{}^\nu$ above projects the vector expression in the brackets on to the brane. This projection is required because  $\delta X^\nu$ in eqs.~\eqref{eq:recipe} and \reef{eq:puerto} is restricted to be parallel to the brane.\footnote{In writing eq.~\eqref{eq:recipe}, we have assumed that the same domain for the coordinates $\xi^a$ mapped to the portion of the RT surface on either side of the brane under both $X^\mu(y^a)$ and $X^\mu(y^a)+\delta X^\mu(y^a)$. Said another way, $S_\bulk$ has the same integration limits in $y^a$ both before and after the variation.}
Hence the brane contribution \reef{eq:facundo} leads to a discontinuity in the first derivative of the RT surface at the brane, as was implicitly found in eq.~\reef{ortho1} above.

\subsection{Wald-Dong entropy} \label{sec:genera}

As alluded to above, one of the striking features of EE for subregions in quantum field theory is that the result is dominated by short wavelength modes in the vicinity of the entangling surface and the EE is UV divergent. Of course, the leading contribution is the famous area law term \cite{Sorkin_1983,Bombelli:1986rw,Srednicki:1993im} and in higher dimensions, there are subleading UV divergences which are also determined by the geometry of the entangling surface (as well as the dimensionful couplings of the underlying theory). In the holographic context, these divergences arise because the RT surface in the bulk extends out to the asymptotic boundary and hence the unregulated area is infinite \cite{Ryu:2006bv,Ryu:2006ef,Rangamani:2016dms}. In the context of braneworld gravity, like the construction in the previous section with $\s\ll L$, one expects large UV contributions when the RT surface crosses the brane. However, in this instance, the corresponding UV cutoff remains finite and set by the position of the brane, as discussed above. We show below that the corresponding UV contributions to the holographic EE can be interpreted as the Wald-Dong entropy \cite{Wald:1993nt,Iyer:1994ys,Jacobson:1993vj,Dong:2013qoa}\footnote{Our calculations will include the subleading contributions arising from the curvature-squared terms in eq.~\reef{act3}. Because the corresponding quantum extremal surfaces have nonvanishing extrinsic curvature, we will need the full expression for the gravitational entropy derived by Dong \cite{Dong:2013qoa}.} of the induced gravity on the brane \cite{Emparan:2006ni,Myers:2013lva}. Of course, the leading UV contributions studied here do not probe the full bulk profile of the RT surface, and we leave the full calculation of holographic EE to Section \ref{sec:examples}.

To evaluate the leading contributions to the holographic entropy where the RT surface crosses the brane, first recall the bulk metric \reef{metric3} with bulk coordinates $X^\mu=(z,x^i)$ and the brane positioned at $z=\s$. Now, for the RT area functional \eqref{eq:badNotation}, we choose the coordinates on the RT surface as $\xi^\alpha=(z,y^a)$ where $z$ is the same radial coordinate as in the bulk and $y^a$ are the $d-2$ spatial coordinates describing the profile of the RT surface in slices of constant $z$ (and time). Now following \cite{Hung:2011ta},  we can construct a Fefferman-Graham expansion for the transverse profile $x^i(\xi)$ of the RT surface for small $z$ (\ie in the vicinity of the brane) to find\footnote{We will assume that the RT surface is $\Zbb_2$ symmetric across the brane. However, in principle, there are two independent profiles on either side of the brane, \ie
$x_\mt{R}^i(z,y^a)$ and $x_\mt{L}^i(z,y^a)$. Of course, the profiles agree where they meet on the brane, $x_\mt{R}^i(z=\s,y^a)=x_\mt{L}^i(z=\s,y^a)$ and satisfy the boundary condition \reef{ortho7}. At this point, let us also recall that the profile in the time direction is trivial here, \ie $x^t(z,y^a) =\overset{\scriptscriptstyle{(0)}}{x}{}^{\,t}(y^a) =$ constant.}
\beq\label{run11}
x^i(z,y^a) = \overset{\scriptscriptstyle{(0)}}{x}{}^{\,i}(y^a) + \frac{z^2}{L^2}\,\overset{\scriptscriptstyle{(1)}}{x}{}^{\,i}(y^a) + \frac{z^4}{L^4} \overset{\scriptscriptstyle{(2)}}{x}{}^{\,i}(y^a) + \cdots\,.
\eeq
In principle, the functions $\overset{\scriptscriptstyle{(n)}}{x}{}^{\,i}(y^a)$ are determined recursively through extremization of the RT area functional \eqref{eq:badNotation}, however, we will simply quote the next-to-leading result found in \cite{Hung:2011ta}:
\beq\label{ramble}
\overset{\scriptscriptstyle{(1)}}{x}{}^{\,i}
= -\frac{L^2\, \overscript{\curvK}{0}^i}{2(d-2)}
= -\frac{L^4\, \inducedK^i}{2(d-2) \s^2} + \mO\left(\frac{\s^2}{L}\right)\,,
\eeq
where $\overscript{\curvK}{0}^i$ is the trace of the extrinsic curvature of the surface $\overscript{x}{0}^i(y^a)$ (at $z=0$) with the boundary metric,  $\overscript{g}{0}_{ij}=g_{ij}^{\mt{AdS}_d}$ as given in eq.~\eqref{metric2}. As the latter is an unphysical surface in the present context, we introduced the second expression with $\inducedK^i$, the trace of extrinsic curvature of intersection surface $\sigma_\xR$ on the brane, \ie $x^i(z=\s,y^a)$, evaluated with induced metric $\inducedg_{ij}$. This expression follows using the relation \eqref{relate} between the boundary metric and the induced metric on the brane,\footnote{Note that in contrast to \cite{Hung:2011ta}, the indices $i$ on the extrinsic curvatures in eq.~\reef{ramble} are coordinate indices, rather than orthonormal frame indices. This introduces an extra factor of $L/\s$ in the leading term on the right-hand side of eq.~\eqref{ramble}. We also note that the sign of our extrinsic curvatures differs from that in \cite{Hung:2011ta}, \ie the extrinsic curvature of a sphere embedded in flat space is positive here.} and the relation \eqref{run11} between $\overscript{x}{0}^i(y^a)$ and $x^i(\s,y^a)$. Note that the leading term on the right-hand side of eq.~\eqref{ramble} scales as $\s^0$ since $\inducedK^i\sim \overscript{\curvK}{0}^i \s^2/L^2$ and by this counting, the first correction is $\mO(\s^2/L)$.

Using eq.~\eqref{badnote}, we now evaluate the non-vanishing components of the induced metric on the RT surface. First, the $h_{zz}$ component is given by
\beqa
&&h_{zz}=\frac{L^2}{z^2}\left[
1 + \frac{z^2}{L^2}\, \overset{\scriptscriptstyle{(1)}}{h}_{zz}
 + \mO\left(\frac{z^4}{L^4}\right)
\right]\,,
\labell{eq:zirkova}\\
{\rm where}\qquad&& \overset{\scriptscriptstyle{(1)}}{h}_{zz}= \frac{4}{L^2}\, \AdSdmetric_{ij} \overset{\scriptscriptstyle{(1)}}{x}{}^{\,i} \overset{\scriptscriptstyle{(1)}}{x}{}^{\,j}
  = \frac{L^4}{(d-2)^2 \s^2}\, \inducedK_i\, \inducedK^i+ \mO\left(\frac{\s^2}{L^2}\right)\,.
\nonumber
\eeqa
In the final expression and throughout the following, the indices on $\inducedK^i$ are contracted using the induced metric $\tg_{ij}$. The remaining nonvanishing components are
\beq
h_{ab}
= \frac{L^2}{z^2}\left(1+\frac{z^2}{4L^2}\right)^2 \QESmetricNoz_{ab}\,,
\quad{\rm with}\ \
\QESmetricNoz_{ab}
\equiv \AdSdmetric_{ij}
\frac{\partial x^i}{\partial y^a}
\frac{\partial x^j}{\partial y^b}
= \overset{\scriptscriptstyle{(0)}}{\QESmetricNoz}_{ab}
+ \frac{z^2}{L^2} \overset{\scriptscriptstyle{(1)}}{\QESmetricNoz}_{ab} + \mO\left(\frac{z^4}{L^4}\right)\,.
\label{jackDaniels}
\eeq
The leading term in $\QESmetricNoz_{ab}$ is simply given by
\begin{align}
  \overset{\scriptscriptstyle{(0)}}{\QESmetricNoz}_{ab}
  =& \AdSdmetric_{ij}\,
  \frac{\partial \overset{\scriptscriptstyle{(0)}}{x}{}^{\,i}}{\partial y^a}\,   \frac{\partial \overset{\scriptscriptstyle{(0)}}{x}{}^{\,j}}{\partial y^b}\,,
\end{align}
and while the individual components $\overset{\scriptscriptstyle{(1)}}{\QESmetricNoz}_{ab}$ will not be needed, we will use the next-to-leading order expansion of the measure
\beq
  \sqrt{\QESmetricNoz}
  = \sqrt{\overset{\scriptscriptstyle{(0)}}{\QESmetricNoz}}\left\{
  1 - \frac{L^2 z^2}{2(d-2)\s^2}\, \inducedK_i \,\inducedK^i\left[
	1+ \mO\left(\frac{\s^2}{L^2}\right)
	\right]
  + \mO\left(\frac{z^4}{L^4}\right)
  \right\}\,.
  \label{eq:absolut}
\eeq
The latter is obtained by interpreting $\overscript{\curvK}{0}^i \sim L^2 \inducedK^i/\s^2$ as giving an expansion of the area element $\sqrt{\QESmetricNoz}$. 

Combining these expressions, the area functional \eqref{eq:badNotation} for the RT surface $\Sigma_\xR$ in the vicinity of the brane becomes
\begin{align}\label{ramble2}
  \begin{split}
  \frac{\area(\Sigma_\xR)}{4 G_\bulk}
  \simeq& \frac{1}{2 G_\bulk} \int_{\s} dz\;\Bigg\{
  \left(\frac{L}{z}\right)^{d-1}
  \left(1+\frac{z^2}{4L^2}\right)^{d-2} \\
  &\quad\times\ \ \int d^{d-2}y\; \sqrt{\QESmetricNoz}
  \left[
  1 + \frac{z^2}{2L^2} \overscript{\RTmetric}{1}_{zz}
  + \Ocal\left(\frac{z^4}{L^4}\right)
  \right]   \Bigg\} \\
	=& \frac{L^{d-1}}{2 G_\bulk \s^{d-2}}
	\int d^{d-2}y\; \sqrt{\overscript{\QESmetricNoz}{0}}
	\Bigg[
	\frac{1}{d-2}
	+ \frac{d-2}{4(d-4)} \left(\frac{\s}{L}\right)^2
	\\
	&\quad- \frac{d-3}{2(d-2)^2(d-4)} L^2 \inducedK_i \inducedK^i
	+ \Ocal\left(\frac{\s^4}{L^4}\right)
	\Bigg]
\end{split}
\end{align}
where an overall factor of $2$ was included to account for the contributions coming from both sides of the brane.\footnote{Recall that we are assuming that the RT surface is symmetric under  reflection across the brane.}
Next, we evaluate the area of the intersection surface $\sigma_\xR=\Sigma_\xR\cap\Brane$ using the metric induced on this surface, \ie $\tilh_{ab}=h_{ab}|_{z=\s}$ where $h_{ab}$ appears in eq.~\reef{jackDaniels}:
\begin{align}
  A(\sigma_\xR)
  =& \int_{\sigma_\xR}\!\!\! d^{d-2}y\; \sqrt{\tilde{h}} 	
  \nonumber\\
	=& \left(\frac{L}{\s}\right)^{d-2}
	\int_{\sigma_\xR}\!\!\! d^{d-2}y\; \sqrt{\overscript{\QESmetricNoz}{0}}
	\left[
	1
	+\frac{d-2}{4}\left(\frac{\s}{L}\right)^2
	-\frac{L^2 \inducedK^i \inducedK_i}{2(d-2)}
	+ \Ocal\left(\frac{\s^4}{L^4}\right)
	\right].\label{numnum}
\end{align}
Hence we may rewrite the result in eq.~\reef{ramble2} as 
\begin{align}
	\begin{split}
  \frac{\area(\Sigma_\xR)}{4 G_\bulk}
  \simeq&
  \frac{L\,\area(\sigma_\xR)}{2(d-2) G_\bulk}+\frac{L}{4(d-4) G_\bulk}\int_{\sigma_\xR}\!\!\! d^{d-2}y\; \sqrt{\tilh}\bigg[ 
    \frac{\s^2}{L^2}
- \frac{L^2\,\inducedK_i \inducedK^i}{(d-2)^2}
	\bigg]
  + \mO\left(\frac{L^{d-6}}{ \s^{d-6}}\right)\,.
	\end{split}
  \label{eq:bazinga}
\end{align}
Of course, if the brane action also includes a DGP contribution \reef{newbran}, one would add the corresponding Bekenstein-Hawking term, as in eq.~\reef{eq:sad}, to produce
\begin{align}
	\begin{split}
  \frac{\area(\Sigma_\xR)}{4 G_\bulk}+ \frac{\area(\sigma_\xR)}{4 G_\brane}
  \simeq&
  \frac{\area(\sigma_\xR)}{4G_\mt{eff}}+\frac{L}{4(d-4) G_\bulk}\int_{\sigma_\xR}\!\!\! d^{d-2}y\; \sqrt{\tilh}\bigg[ 
    \frac{\s^2}{L^2}
- \frac{L^2\,\inducedK_i \inducedK^i}{(d-2)^2}
	\bigg]
  + \mO\left(\frac{L^{d-6}}{ \s^{d-6}}\right)\,,
	\end{split}
  \label{eq:bazinga2}
\end{align}
where the two leading contributions proportional to $\area(\sigma_\xR)$ were combined using eq.~\reef{Newton33}.

It is clear that the first term in eq.~\reef{eq:bazinga2} corresponds to the Bekenstein-Hawking entropy of the surface $\sigma_\xR$ for the gravity action \reef{act3} induced on the brane. 
We now show that leading corrections in eq.~\eqref{eq:bazinga2} match the contributions to the Wald-Dong entropy \cite{Dong:2013qoa} coming from the curvature-squared terms. That is, given the gravity action \reef{act3}, the corresponding Wald-Dong entropy is given by
\begin{align}\label{dong}
	\begin{split}
S_\mt{WD} =& \frac{\area(\sigma_\xR)}{4G_\mt{eff}}+
\frac{L^3}{4(d-2)^2 (d-4)G_\bulk}\int_{\sigma_\xR}\!\!\! d^{d-2}y \sqrt{\tilh}
\left(2 \inducedR_{ij} n^{im} \tensor{n}{^j_m}
-\frac{d}{d-1} \tilde{R}-\inducedK_i \inducedK^i
\right)\,,
\end{split}
\end{align}
where $n^i_m$ are two unit normals to the entangling surface $\sigma_\xR$ embedded in the $d$-dimensional brane geometry, and as in section \ref{sec:branegravity}, $\inducedR_{ij}$ and $\inducedR$ are the Ricci tensor and scalar curvatures, respectively, evaluated with $\inducedg_{ij}$. Comparing eqs.~\reef{eq:bazinga2} and \reef{dong}, we immediately see that the coefficients precisely match for the term proportional to $\inducedK_i \inducedK^i$.
Then using eqs.~\eqref{curve1} and \eqref{Ricky2}, we can evaluate the remaining two curvature terms in eq.~\eqref{dong},
\begin{multline}\label{gam}
	\frac{L^3}{4(d-2)^2(d-4)G_\bulk} \int d^{d-2}y\; \sqrt{\tilh}
  \left( 2\inducedR_{ij} n^{im} \tensor{n}{^j_m}
  -\frac{d}{d-1} \tilde{R}
  \right)
	\\
  = \frac{\s^2}{4(d-4)G_\bulk L} \int d^{d-2}y\; \sqrt{\tilh}\,,
\end{multline}
which matches the $\mO(\s^2/L^2)$ term in eq.~\eqref{eq:bazinga2}. Hence, as expected \cite{Emparan:2006ni,Myers:2013lva}, in the regime $\s\ll L$, one finds that the leading contributions to the holographic entanglement entropy \reef{eq:sad}  where the RT surface crosses the brane reproduce the Wald-Dong entropy of the intersection surface derived for the gravity action \reef{act3}.


To close this section, we briefly remark on the case of $d=2$, which is somewhat special in that the intersection between the RT surface and the brane is a point. Consequently, the leading UV contribution to entropy is not a standard area term, but rather a logarithmic term. Integrating the RT area (in this case, length) across gives
\beq
S
\simeq \frac{L}{2G_\bulk} \log\left(\frac{\ell_\IR}{\s}\right)\,.
\label{eq:kaboom}
\eeq
where an IR length scale $\ell_\IR$ must appear to make the argument of the logarithm dimensionless.\footnote{As in eq.~\reef{ramble2}, a factor of two has been included to account for both sides of the brane.} 

Following \cite{Myers:1994sg}, we can find (the leading contribution to) the gravitational entropy for the brane theory evaluating the Wald entropy formula \cite{Wald:1993nt} to the Polyakov-Liouville action \eqref{PolyAct2}, and then substituting the on-shell solution \reef{sol4} for the scalar $\phi$,\footnote{Note that the action \eqref{PolyAct2} is multiplied by a factor of two for the full induced brane action,  \ie $I_\mt{induced} = 2\,I_\mt{diver} + I_\mt{brane}$.}
\beq\label{arc}
S=\frac{L}{4 \Gbk}\,\phi_0=-\frac{L}{4 \Gbk}\,\log\!\(-\frac{L^2\tilde R}{2}\)\,.
\eeq
Now substituting $\tilde R\simeq -2\s^2/L^4$ reproduces the leading singular behaviour in the holographic result \reef{eq:kaboom}. The same answer can be obtained by evaluating the Wald-Dong entropy formula \cite{Wald:1993nt,Dong:2013qoa} directly on the induced gravity action \eqref{induct}. Hence, once again in this special case, the holographic entanglement entropy \reef{eq:sad} reproduces the Wald-Dong entropy for the corresponding gravity action on the brane.



\subsection{Explicit Calculations}\label{sec:examples}

In this section, we explicitly evaluate the holographic EE and examine the transition between the two classes of RT surfaces. While we set up the calculations for general $d>2$, our explicit results are given for $d=3$ in which case the bulk spacetime locally has the geometry of AdS$_{4}$. We add some comments about $d=2$, and the addition of Jackiw-Teitelboim gravity \reef{JTee} on the brane, in the discussion section.

\subsubsection*{Setting up the calculation for general dimension}

In section \ref{sec:enzyme}, we reviewed two different coordinate systems in AdS$_{d+1}$. The AdS$_d$ foliation \reef{metric3a} was well suited to discuss the brane geometry, while the global coordinates are adapted to discuss the background geometry of the boundary CFT. However, our explicit calculations of the holographic EE are best performed in a new `cylindrical' coordinate system. In particular, following \cite{Krtous:2014pva}, we introduce cylindrical coordinates $P,\,\zeta$ where $\zeta$ specifies the position along the axis of the cylinder while $P$ measure the distance from the axis. These are related to the global coordinates in eq.~\reef{metric2s} by
\begin{align}\label{cylie}
\cosh r&=\sqrt{P^2+1}\,\cosh\zeta\,,\\
\tan\theta&=\frac{P}{\sqrt{1+P^2}}\,\frac{1}{\sinh\zeta}\,,
\end{align}
while the rest of the spherical angles remain unchanged. With this transformation, the metric becomes
\beq\label{cylindd}
ds^2=L^2\[ -(P^2+1)\cosh^2\zeta\, dt^2+\frac{dP^2}{1+P^2}+\left( 1+P^2 \right)d\zeta^2+P^2\,d\Omega_{d-2}^2\]\,.
\eeq
The range of these coordinates is $P\in (0,\infty)$ and $\zeta\in(-\infty,\infty)$. The conformal boundary is reached with $P\to \infty$ (or $\zeta\to\pm\infty$ with fixed $P$).  The upper ($0\le\theta\le\pi/2$) and lower ($\pi/2\le\theta\le\pi$) hemispheres are mapped to the upper ($\zeta\ge0$) and lower ($\zeta\le0$) halves of the cylindrical system. The conformal defect is positioned at $\zeta=0$. As noted above, the RT surfaces will be restricted to a constant time surface and hence the convenience of the cylindrical coordinates becomes evident, \ie  $\zeta$ becomes an extra Killing coordinate in the corresponding spatial geometry.

A few more technical details are needed  for our calculations:
in cylindrical coordinates \reef{cylindd}, the boundary entangling surface corresponds to the two circles $\zeta=\pm\zeb$, where
\beq\label{zeta0}
\sinh\zeb=\tan\thb\,,
\eeq
seen in the limit $P\to\infty$ of the second line in eq.~\reef{cylie}. Using the AdS foliation of eq.~\reef{metric3a}, the position of the brane was $z=\s$. Using eq.~\reef{eq:foobar}, the brane position can be specified in cylindrical coordinates \reef{cylindd} according to 
\beq\label{eq:foobar2}
 \(1+P^2\)\sinh^2\!\zeta=\frac{L^2}{\s^2}\(1-\frac{\s^2}{4L^2}\)^2\,.
\eeq
Recall that the brane intersects the asymptotic boundary at the position of the conformal defect, \ie at $\theta=\pi/2$ with $r\to\infty$, which corresponds to $\zeta= 0$ with $P\to \infty$ in cylindrical coordinates. Further recall that RT surface areas are UV divergent since they extend to the asymptotic boundaries.  Hence we introduced a UV regulator surface at $r=r_\mt{UV}$, which in cylindrical coordinates becomes
\beq\label{regular}
(P^2+\tanh^2\zeta)\cosh^2\zeta=\sinh^2\! r_\mt{UV}\,.
\eeq
We will be mainly interested in comparing the areas of different surfaces for fixed $\zeb$, as discussed above. Since the UV divergent terms only depend of the geometry of the entangling surface, they will cancel in the difference of the two areas. Hence, we can then safely take the UV cutoff to infinity.

As noted, the RT surfaces all lie in a fixed time slice and thus we only need consider configurations with cylindrical symmetry (\ie rotational symmetry on the $S^{d-2}$). Hence it is convenient to use the cylindrical coordinates \reef{cylindd} and  parametrize the profile of the bulk surfaces as $\zeta=\zeta(P)$. The bulk contribution to the holographic EE is given by
\begin{align}\label{area}
S_\mt{bulk}= \frac{L^{d-1}\, \Omega_{d-2}}{2\,\Gbk} \int dP P^{d-2} \sqrt{\frac{1}{1+P^2}+(1+P^2)\,\zeta'^2}
\end{align}
where again $\Omega_{d-2}$ is the area of the unit $(d-2)$-sphere -- see footnote \ref{footsphere}. As in eq.~\reef{area0}, an overall factor of 2 is included here to account for the reflection symmetry of the profile $\zeta(P)$ about the brane. Since this expression does not contain an explicit $\zeta$ dependence, it is straightforward to derive
\begin{align}\label{zetap}
\zeta'(P)=\pm \frac{1}{1+P^2}\,\sqrt{\frac{P_0^{2(d-2)}\(1+P_0^2\)}{P^{2(d-2)}\(1+P^2\)-P_0^{2(d-2)}\(1+P_0^2\)}}
\end{align}
where the two branches correspond to two identical surfaces related by a reflection with respect to $\zeta=0$. $P_0$ corresponds to the turning point, where the surface makes its closest approach to the symmetry axis.

\begin{figure}
	\def\svgwidth{0.8\linewidth}
	\centering{
		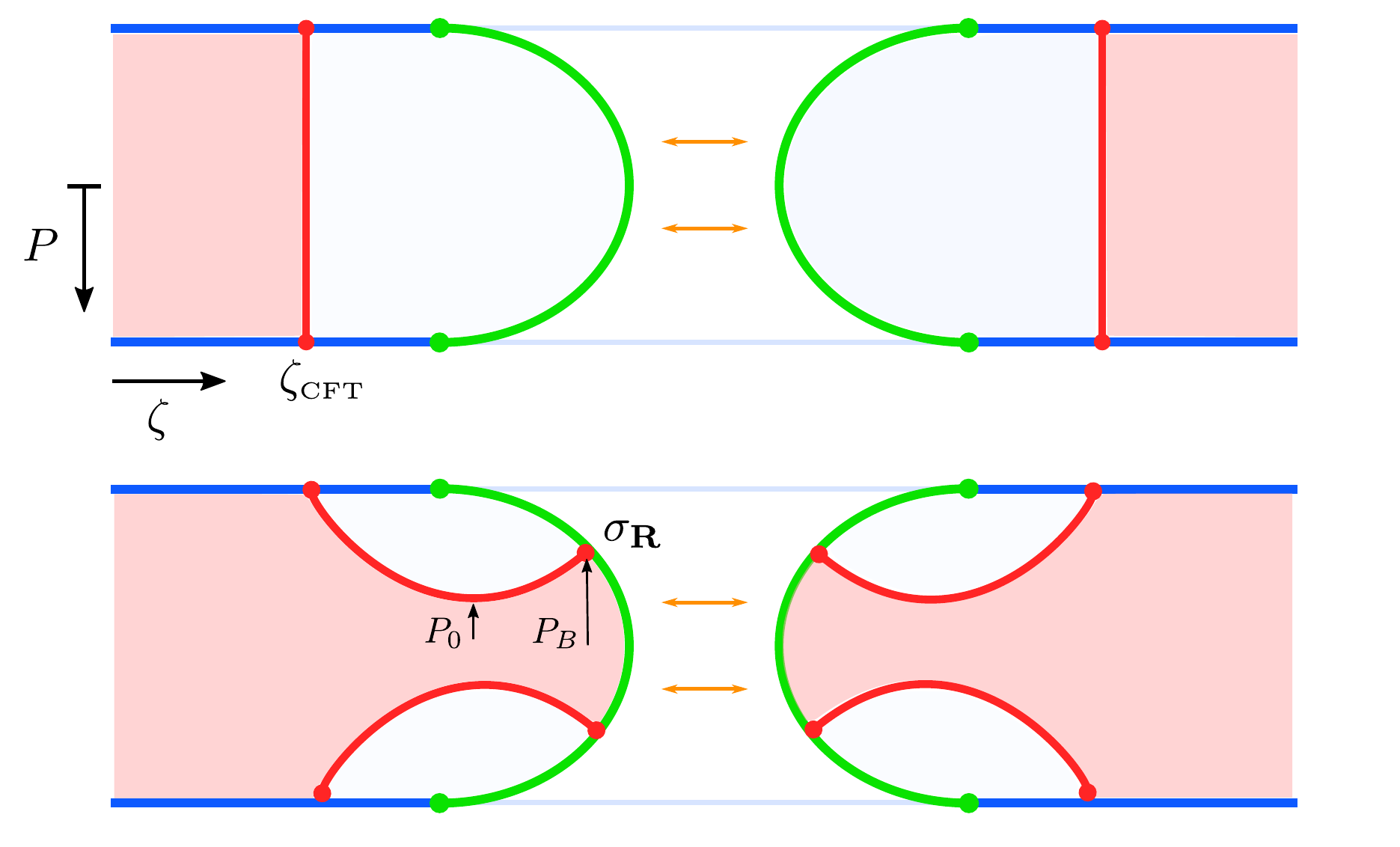
		\caption{Sketch of fixed time slices of our symmetric setup, showing the two possible configurations. The shaded red region corresponds to the entanglement wedge. The connected solution contains an island on the brane, where gravity is dynamical.}
		\label{fig:RTPhases}
	}
\end{figure}

We now discuss the disconnected phase described at the beginning of this section. It corresponds to the `trivial' solution with $P_0=0$. We find $\zeta(P)=\pm\zeb$, which in cylindrical coordinates looks simply as a pair of disks anchored at the boundary entangling surface. Substituting $\zeta'=0$ into eq.~\reef{area}, the area of the two discs can be integrated up to some cutoff radius $P_\mt{UV}$, and the corresponding holographic EE is
\begin{align}\label{A_disc}
S_\mt{disc}=\frac{L^{d-1}\, \Omega_{d-2}}{2(d-1)\,\Gbk} \ \puv^{d-1}\, {}_{2}F_1\left[ \frac{1}{2},\frac{d-1}{2},\frac{d+1}{2},-\puv^2 \right]\,.
\end{align}
In this case, the entanglement wedge corresponds to two identical disconnected pieces contained between each component of the RT surface and the asymptotic boundary, \ie the regions $\zeta\ge+\zeb$ and $\zeta\le-\zeb$, as sketched in the upper panel of figure \ref{fig:RTPhases}. 
%

The connected phase corresponds to $P_0>0$, which leads to a cylindrical RT surface. Integrating eq.~\reef{zetap} yields a family of bulk surfaces, which are symmetric about the brane and which are anchored on the asymptotic boundary at $\zeta=\pm\zeb$. Recalling the discussion below eq.~\reef{area0}, we observe that in this configuration, $P_0$ is the second integration constant which must be tuned in order to satisfy the appropriate boundary condition \reef{ortho1} at the brane, see the lower panel of figure \ref{fig:RTPhases}.

Before we calculate the entropy in the most general setting, let us consider the case of a zero-tension brane with $1/\Gbr=0$, \ie empty AdS$_{d+1}$. In this case, the brane is positioned at $\s=2L$ or simply, $\zeta=0$. Now, the `plus' branch of eq.~\eqref{zetap} can be integrated to produce a profile extending from $P=\puv$ at $\zeta=+\zeb$ to the maximal depth $P=P_0$ at some $\zeta=\zeta_0(\zeb,P_0)<\zeb$. Since eq.~\reef{ortho1} indicates that the RT surface must intersect the brane orthogonally, we must tune $P_0$ (with fixed $\zeb$) such that $\zeta_0=0$, \ie the RT surface reaches its maximal depth at the brane position. Now, substituting eq.~\reef{zetap} into eq.~\reef{area}, the holographic EE (for empty AdS$_{d+1}$) becomes
\begin{align}\label{A_conn}
S_\mt{conn}(T_o=0)
&=\frac{L^{d-1}\, \Omega_{d-2}}{2\,\Gbk} \int_{P_0}^{\puv}\!\!\! dP\,  \frac{P^{2(d-2)}}{\sqrt{P^{2(d-2)}(1+P^2)-P_0^{2(d-2)}(1+P_0^2)}}\,.
\end{align}

In the general case, this exercise is slightly more complicated for the case of interest with a finite-tension DGP brane at some $z=\s\ll L$, and the geometry of the corresponding RT surface is illustrated in the lower panel of figure \ref{fig:RTPhases}. The RT surface is again symmetric about the brane and so as above, we focus on the portion starting at $\zeta=+\zeb$ at the asymptotic boundary (\ie at $P=\puv$). As before, the `plus' branch of eq.~\eqref{zetap} produces a surface reaching its maximal depth $P=P_0$ at some $\zeta=\zeta_0(\zeb,P_0)<\zeb$.\footnote{In fact, $\zeta_0(\zeb,P_0)$ is precisely the same function introduced above, since the turning point of the RT surfaces are completely independent of the brane properties.}  Now one continues from this point using the `minus' branch of eq.~\reef{zetap}, which then meets the brane as some $P=P_\mt{B}(\zeb,P_0)$ and $\zeta=\zeta_\mt{B}(\zeb,P_0)$.\footnote{Of course, $P_\mt{B}$ and $\zeta_\mt{B}$ are related as in eq.~\reef{eq:foobar2}.} One would again tune $P_0$ (for fixed $\zeb$) to ensure the appropriate boundary condition \reef{ortho1} is satisfied at the brane.
The bulk contribution to the holographic EE then becomes
\beqa
S_\mt{conn}(T_o>0)&=&\frac{L^{d-1}\, \Omega_{d-2}}{2\,\Gbk}\[ \int_{P_0}^{\puv}\!\!\!  dP\,  \frac{P^{2(d-2)}}{\sqrt{P^{2(d-2)}(1+P^2)-P_0^{2(d-2)}(1+P_0^2)}}\right.
\labell{Acon2}\\
&&\qquad\qquad\qquad+\left. \int_{P_0}^{\pb}\!\!  dP\,  \frac{P^{2(d-2)}}{\sqrt{P^{2(d-2)}(1+P^2)-P_0^{2(d-2)}(1+P_0^2)}}\]\,.
\nonumber
\eeqa
Of course, if there is no gravitational term on the brane (\eg as in eq.~\reef{newbran}), then this expression yields the entire generalized entropy \reef{eq:sgen_intro} for the connected phase. Now
rather than explicitly examining the brane boundary condition \reef{ortho1} in cylindrical coordinates, we will simply evaluate the generalized entropy and find the minimum numerically in the following. Hence to proceed further we will have to choose a specific value for the boundary dimension $d$.

\subsubsection*{Explicit results for $d=3$}
In this section, we consider the above discussion for $d=3$, in which case the boundary geometry becomes $\Rbb\times S^{2}$, the bulk spacetime is locally AdS$_4$, and the branes have an AdS$_3$ geometry. We will also consider supplementing the the four-dimensional bulk action \reef{act2} with a Gauss-Bonnet term, 
 \beq
I_\mt{top} = \frac{\lgb}{16\pi^2}  \int \mathrm{d}^4x \, \sqrt{-g}\, \left[ R_{abcd}R^{abcd}-4\,R_{ab}R^{ab}+R^2 \right]\,.
 \labell{top2}
 \eeq
Note that we have ignored the necessary boundary terms which ensure that this interaction is proportional to the Euler density, \eg see \cite{Myers:1987yn}. Although this curvature-squared term does not effect the bulk equations of motion, it will contribute to the generalized entropy \cite{Dong:2013qoa,Hung:2011xb}\footnote{One may worry that the topological nature of $I_\mt{top}$ undercuts the usual derivations of the generalized entropy. However, individually the three terms in eq.~\reef{top2} are dynamical and one can apply the results of \cite{Dong:2013qoa} for each separately and then take the sum of the corresponding contributions to the holographic entropy, which one finds matches the result in eq.~\reef{Euler3}.}
 \beq
S_\mt{JM} = \frac{\lgb}{4\pi} \int_{\Sigma_\xR} d^2x\sqrt{h}\,\mR +\frac{\lgb}{2\pi} \int_{\partial \Sigma_\xR}
dx\sqrt{h}\,\mK_g \,,
 \labell{Euler3}
 \eeq
where $\mR$ denotes the Ricci scalar for the intrinsic geometry on the RT surface $\Sigma_\xR$. Similarly, $\mK_g$ denotes the geodesic curvature of the boundary $\partial \Sigma_\xR$. Of course, eq.~\reef{Euler3} gives a topological contribution proportional to the Euler character of the two-dimensional extremal surfaces\footnote{The normalization is chosen so that for an RT surface with two-sphere topology, $S_\mt{JM} = 2 \lgb$.} and so their geometry remains unaffected by this term. However, in the following, this additional contribution will  provide an extra parameter which allows us to adjust the transition between the connected and disconnected phases.

For $d=3$, some analytic expressions for the extremal surfaces can be obtained \cite{Krtous:2014pva}. For example,
integrating eq.~\eqref{zetap} yields the following profile for the extremal surface in empty AdS$_4$ \cite{Krtous:2014pva}
\beqa
&&\zeta_\pm(P;P_0,\zeta_0)=\zeta_0\pm \frac{P_0}{\sqrt{(1+P_0^2)(1+2P_0^2)}} \labell{zetasol} \\
&&\times\left[ (1+P_0^2)\, F\!\left( \mbox{Arcos} \frac{P_0}{P},\sqrt{\frac{1+P_0^2}{1+2P_0^2}} \right)-P_0^2\, \Pi\!\left( \mbox{Arccos}\frac{P_0}{P},\frac{1}{1+P_0^2},\sqrt{\frac{1+P_0^2}{1+2P_0^2}} \right) \right]
\nonumber
\eeqa
where $F$ and $\Pi$ correspond to incomplete elliptic integrals of the first and third kind, respectively.\footnote{Our notation for the elliptic integrals matches that in \cite{Gradshteyn:1702455}, section 8.1.} Again, the $\pm$ branches correspond to the two portions of the surface, symmetric with respect to $\zeta_0=0$. Of course, we need to know where this surface is anchored at the boundary. Hence we define
\beqa
\zeta_\infty&\equiv&\zeta_+(P\to \infty;P_0,\zeta_0)-\zeta_0
\labell{alphadog}\\
&=&\frac{P_0\left[ (1+P_0^2)\,K\!\left( \sqrt{ \frac{1+P_0^2}{1+2P_0^2} }\right) - P_0^2\, \Pi\!\left( \frac{1}{1+P_0^2},\sqrt{\frac{1+P_0^2}{1+2P_0^2}} \right) \right]}{\sqrt{(1+P_0^2)(1+2P_0^2)}}
\nonumber
\eeqa
and the surface reaches the asymptotic boundary at $\zeta_\pm(P\to \infty)=\zeta_0\pm \zeta_\infty$. Hence the two components of the entangling surface in the boundary theory are separated by $2\zeta_\infty$, in the cylindrical coordinates.

\begin{figure}[h]
	\def\svgwidth{0.5\linewidth}
	\centering{
		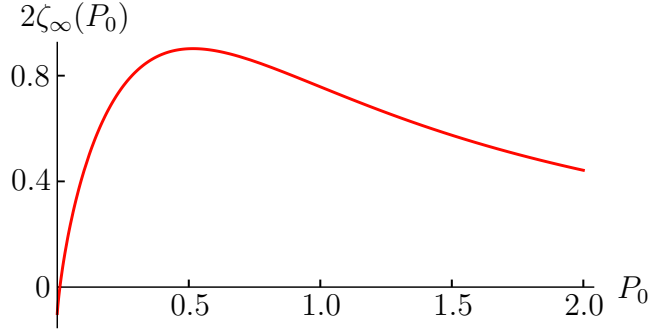
		\caption{Plot of the `height' of the RT surface in cylindrical coordinates, as a function of the turning point $P_0$ characterising the surface. For $\zeta_\infty<\zeta_\infty^{\mt{crit}}$, there are two minimal surfaces anchored at the same regions; otherwise there exists none. 
		}
		\label{figzetainfty}
	}
\end{figure}

Figure \ref{figzetainfty} plots $\zeta_\infty$ as a function of $P_0$. The maximum is obtained at $P_0=P_0^{\mt{crit}}\approx 0.51633$ with $\zeta_\infty=\zeta_\infty^{\mt{crit}}\approx 0.5011$. An interesting observation in \cite{Krtous:2014pva} was that, for $P_0<P_0^{\mt{crit}}$, there exist \textit{two} values of $P_0$ with the same $\zeta_\infty$. That is, if the two components of the entangling surface are sufficiently `close' on the boundary sphere, there actually exist \textit{two} extremal RT surfaces that connect them in the bulk. However, one branch (with the smaller value of $P_0$) is always subdominant, and therefore will be of little interest in our analysis. On the other hand, if the separation of the two entangling spheres  is larger than the critical value $2\zeta_\infty^{\text{max}}$ (in cylindrical coordinates), there is no connected extremal surface that joins them.
%

\begin{figure}[th]
	\def\svgwidth{0.8\linewidth}
	\centering{
		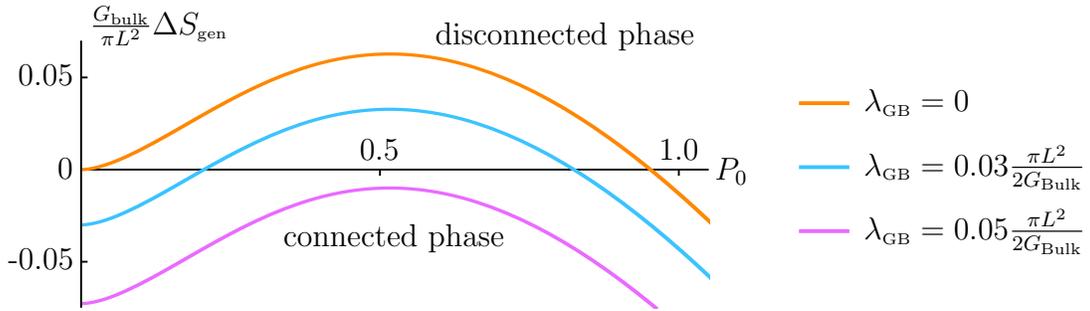
		\caption{Renormalised entropy from eq. \eqref{dAP01}. The connected (disconnected) surface dominates when $\Delta S<0\,(\Delta S>0)$. When $\lgb$ becomes very large, $\lgb\sim c_T$, the connected solution becomes favoured.}
		\label{figdeltaA}
	}
\end{figure}

Let us now describe the solutions corresponding to different values of the tension and DGP term:
\paragraph{a) $T_o=0;\ 1/\Gbr=0$\ :} First we consider the holographic EE in empty AdS$_4$ as a lead-in to the case with a brane. As emphasized above, the area of these surfaces is divergent, and so one introduces a UV regulator surface,  integrating of the area from $P_0$ to some $\puv\gg1$ \cite{Krtous:2014pva}. For the disconnected solution (\ie a pair of disks), eq.~\eqref{A_disc} with $d=3$ gives
\begin{align}\label{Sdisc}
S_\mt{disc}(\puv)
=& \frac{\pi L^2}{\Gbk}\,\left( \sqrt{1+\puv^2}-1 \right)+2 \lgb
\\
=& \frac{A(S^1_{\puv})}{4G_\mt{eff}}
- \frac{\pi L^2}{G_\bulk} + 2\lgb
+ \Ocal(\puv^{-1})\,.
\label{eq:lonely}
\end{align}
where
\beq\label{sample}
\frac{A(S^1_P)}{4G_\mt{eff}}=\frac{\pi L^2}{\Gbk}\,P\,,
\eeq
is the length of $S^1_P$, a circle with radius $P$, and we used eq.~\reef{Newton2} to write $\frac{1}{G_\mt{eff}}=\frac{2\,L}{\Gbk}$.
We have included in eq.~\eqref{Sdisc} the topological contribution in eq.~\reef{Euler3}. On the other hand, for the connected surfaces the area formula \eqref{A_conn} yields
\begin{align}\label{eq:dietCoke}
\begin{split}
\MoveEqLeft[3]
S_\mt{conn}(\puv,P_0)
\\
=& \frac{\pi L^2}{\Gbk}\,\frac{P_0^2}{\sqrt{1+2P_0^2}}\, \Pi\!\left( \mbox{Arccos}\frac{P_0}{\puv},1,\sqrt{\frac{1+P_0^2}{1+2P_0^2}} \right)
\end{split}
\\
\begin{split}
=& \frac{A(S^1_{\puv})}{4G_\mt{eff}}
+ \frac{\pi L^2}{G_\bulk}\left[
-\sqrt{1+2P_0^2} E\left(\sqrt{\frac{1+P_0^2}{1+2P_0^2}}\right)
+ \frac{P_0^2}{\sqrt{1+2P_0^2}} K\left(\sqrt{\frac{1+P_0^2}{1+2P_0^2}}\right)
\right]
\\
&+ \Ocal(\puv^{-1})\,,
\end{split}
\label{eq:friendless}
\end{align}
where $E$ is the elliptic integral of the second kind. We emphasize that this result only applies for  vanishing $T_o$ and vanishing $1/\Gbr$, \ie  for the AdS$_4$ vacuum. Note that the Euler character of the cylindrical RT surface is zero and hence there is no contribution proportional to $\lgb$. As expected, the divergence in the $\puv\to\infty$ limit matches for the areas of the connected and disconnected surfaces. Hence we can safely take the limit when considering the difference
\begin{align}\label{dAP01}
\Delta S(P_0)&=\lim_{\puv\to \infty}\left( S_\mt{conn}(\puv,P_0)-S_\mt{disc} (\puv)\right)\,,
\end{align}
given by the difference in $O\left( \left( P_0/\puv \right)^0 \right)$ terms in eq.~\eqref{eq:friendless} and eq.~\eqref{eq:lonely}.
A plot of $\Delta S$ is shown in figure \ref{figdeltaA}. When $\Delta S>0$, the disconnected RT surface is the dominant saddle, while for $\Delta S<0$, the connected solution dominates. Notice that with a larger (positive) topolgical coupling $\lgb$, the entropy in eq.~\reef{Sdisc} increases while eq.~\reef{eq:dietCoke} is unaffected, and hence the range of the disconnected phase is decreased in figure \ref{figdeltaA}.

%

\paragraph{b) $T_o\ne 0;\ 1/\Gbr=0$\ :} The next step is to introduce the brane, however, we do not include a gravitational term in the brane action yet, \ie $1/\Gbr=0$. In this case, we saw  in eq.~\reef{Acon2} that there is an additional contribution as the RT surface extends from the maximal depth $P_0$ back out to meet the brane at $P_\mt{B}$. Both contributions in eq.~\reef{Acon2} take the same form except for the limits of integration, hence the $d=3$ result in eq.~\reef{eq:dietCoke} is replaced by
\beqa
S_\mt{conn}(\puv,P_0)&=& \frac{\pi L^2}{\Gbk}\,\frac{P_0^2}{\sqrt{1+2P_0^2}}\,\[ \Pi\!\left( \mbox{Arccos}\frac{P_0}{\puv},1,\sqrt{\frac{1+P_0^2}{1+2P_0^2}} \right)\right.
\labell{CokeZero}\\
&& \left.\qquad\qquad+\Pi\!\left( \mbox{Arccos}\frac{P_0}{\pb},1,\sqrt{\frac{1+P_0^2}{1+2P_0^2}} \right)\]\,.
\nonumber
\eeqa

Of course, the entropy for the disconnected phase remains the same as in eq.~\reef{Sdisc} and we can consider the difference of the generalized entropy evaluated on the connected and disconnected extremal surfaces, as in eq.~\reef{dAP01}. Just as we saw a leading divergent contribution in eq.~\reef{eq:dietCoke} for $\puv\to\infty$, we expect that eq.~\reef{CokeZero} will contain an analogous large contribution for $\pb\gg P_0$. However, this term will not be cancelled in $\Delta S$. In fact, in this regime, we can expand the difference as
\begin{align}\label{radishes}
\begin{split}
\MoveEqLeft[2]\Delta S(P_0)
\\
=& \frac{A(\sigma_\xR)}{4G_\mt{eff}}
+ \frac{\pi L^2}{G_\bulk}\left[
1-2\sqrt{1+2P_0^2} E\left(\sqrt{\frac{1+P_0^2}{1+2P_0^2}}\right)
+ \frac{2P_0^2}{\sqrt{1+2P_0^2}} K\left(\sqrt{\frac{1+P_0^2}{1+2P_0^2}}\right)
\right]
\\
&-2 \lgb
+ \Ocal(\pb^{-1})\,.
\end{split}
\end{align}
Here, the intersection $\sigma_\xR$ of the RT surface and the brane is a circle of radius $\pb$ with area $A(\sigma_\xR)=2\pi L\,\pb$ given by eq.~\eqref{cylindd}. The fact that the leading term can be expressed as the gravitational entropy for the induced gravity action \reef{act3} on the brane is in perfect agreement with our discussion in the previous section. As we will see below, the finite terms will play a role once we turn on the DGP term, allowing for the appearance of a different island on the brane. 

From the above expansion, we see that there is a strong penalty for having a large $\sigma_\xR$ in the connected phase. From the brane perspective, the gravitational entropy results in a large penalty against forming an island on the brane. In fact, generally we expect that $\Delta S>0$ in this regime and hence the disconnected solution provides the dominant saddle point. However, if we tune the topological coupling $\lgb$ to be large\footnote{We note that this requires $\lgb\sim {L^2}/{\Gbk}\sim \cT$, the central charge of the boundary CFT -- see further discussion in section \reef{sec:discussion}.} (and positive), this contribution can compensate for the leading gravitational entropy term, at least for $\sigma_\xR$ up to a certain size.

\begin{figure}[h]
	\def\svgwidth{0.9\linewidth}
	\centering{
		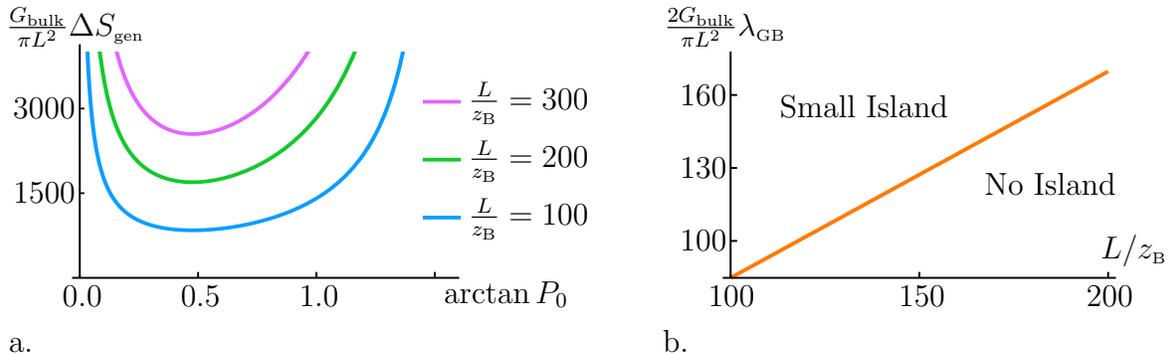
		\caption{Panel a. illustrates the renormalised area from eq. \eqref{radishes} of connected RT surfaces, anchored at $\zeb=0$, with $\lgb=0$. Panel b. is a plot of the critical value of $\lgb$ such that $\mbox{min}( \Delta S)<0$.
		}
		\label{fig:dS0}
	}
\end{figure}

On the other hand, we must note that $\pb$  is not an independent parameter. Rather it is implicitly determined by $\zeb$ and the brane tension $T_o$, as well as the value of $P_0$ that minimises the area functional in eq. \eqref{radishes}. $\pb$ can be determined in the following way (see figure \ref{fig:RTPhases}). One begins by solving for $\zeta_0$ using
$\zeta_0+\zeta_\infty(P_0)=\zeb$ where $\zeta_\infty(P_0)$ is given in eq.~\reef{alphadog}. Then one finds `sample' values of $\pb,\zeta_\mt{B}$ where the extremal surface meets the brane by combining eqs.~\reef{eq:foobar2} and \reef{zetasol} and simultaneously solving
\beqa
 \(1+\pb^2\)\sinh^2\!\zeta_\mt{B}&=&\frac{L^2}{\s^2}\(1-\frac{\s^2}{4L^2}\)^2\,,\nonumber\\
 \zeta_-(\pb;P_0,\zeta_0)&=&\zeta_\mt{B}\,.
 \label{solver3}
\eeqa
This yields $P_B$ as a function of $P_0,\zeta_{\mt{CFT}}$ and $T_o$, and substituting $\pb$ into eq.~\reef{CokeZero} gives the area of the associated extremal surface. Below, we perform this calculation numerically. However, we have not yet considered the boundary conditions \reef{ortho1} in this analysis. Rather than explicitly examining the latter, we simply evaluate the area (or rather the difference $\Delta S$) over the range of possible $P_0$ (with fixed $\zeb,T_o$), as shown in figure \ref{fig:dS0}a. The correct RT surfaces are then identified as the minima in these plots. Further, the examples in the figure illustrate that without the topological contribution, $\Delta S>0$ for all minima and so the disconnected phase dominates, as generally expected. That is, no quantum extremal islands form on the brane in this case. However, as shown in figure \ref{fig:dS0}b, we see that with a sufficiently large topological coupling $\lgb$ one can achieve $\Delta S<0$, where a first order transition leads to the formation of an island. 

Although the above recipe is valid for arbitrary brane tensions, in the limit of very large tension we can approximate the solution analytically. Since, as stated above, the leading contribution to the entropy \eqref{sample} scales as $A(\sigma_\xR)\sim \pb$, the RT surface corresponds to that which has the minimal value of $\pb$. Moreover, since the function $\zeta_{\mt{B}}(P)$ defining embedding of the brane in \eqref{solver3} is monotonically decreasing with $P$, the surface must maximise its hight $\zeta_\infty(P_0)$, which is achieved for $P_0=P_0^{\mt{crit}}$, by definition (see discussion around figure \ref{figzetainfty}). This can be readily checked in figure \ref{fig:dS0}a, where the curves attain a minimum around $\mbox{arctan}(P_0^{\mt{crit}})\approx 0.47$, with a small correction due to the finite terms in \eqref{radishes}, which becomes smaller and smaller as we increase the tension. We shall refer to this solution with $P_0\approx P_0^{\mt{crit}}$ as the \textit{small island}, in order to distinguish it from a second island appearing below which corresponds to a circle with a larger radius.

\paragraph{c) $T_o\ne 0;\ 1/\Gbr\ne 0$\ :} Finally, we examine the holographic EE in the presence of a DGP brane. The only difference in this analysis is the additional contribution coming at the intersection of the RT surface with the brane in eq.~\reef{eq:sad}.
In the present setting, this means that we add the following,
\beq\label{sample2}
S_\mt{brane}=\frac{A(\sigma_\xR)}{4\Gbr}=\frac{\pi L}{2\Gbr}\,\pb\,,
\eeq
to the bulk contribution in eq.~\reef{CokeZero}. In fact, the  expansion of $\Delta S$ for $\pb\gg P_0$ takes precisely the same form as in eq.~\reef{radishes}. The only difference is that the induced Newton's constant on the brane is now given by eq.~\reef{Newton33}, \ie $\frac{1}{G_\mt{eff}}=\frac{2\,L}{\Gbk}+\frac{1}{\Gbr}$.

Generally, we might think of $1/\Gbr$ as a positive quantity, and so the DGP contribution \reef{sample2} would simply increase the penalty for having a large $\sigma_\xR$ in the connected phase, and enhance the dominance of the disconnected phase. However, there is no apriori reason why we should not also consider a negative gravitational coupling on the brane,\footnote{For example, integrating out quantum fields on the brane could produce either a positive or negative shift in Newton's constant. In particular, it can be negative for gauge fields or nonminimally coupled scalar fields, as discussed in the context of EE in \cite{Larsen:1995ax,Kabat:1995eq} -- see further discussion in section \ref{sec:discussion} and appendix \ref{bubble}.} in which case the DGP term serves as another mechanism to reduce the penalty for forming an island on the brane. It is this scenario that we will examine further here -- as well as in appendix \ref{bubble}.

It will prove convenient to work with the ratio $\lamb$ introduced in eq.~\reef{newdefs}. Let us recall what parameters are in play. The tension of the brane is controlled by $\s$, which we keep small but finite. The dimensionless ratio between the bulk and brane gravitational constants is controlled by $\lamb$. As discussed above, interesting things happen when $\lamb<0$, which is when $\Gbr<0$ while $\Gbk>0$.

\begin{figure}[h]
	\def\svgwidth{\linewidth}
	\centering{
		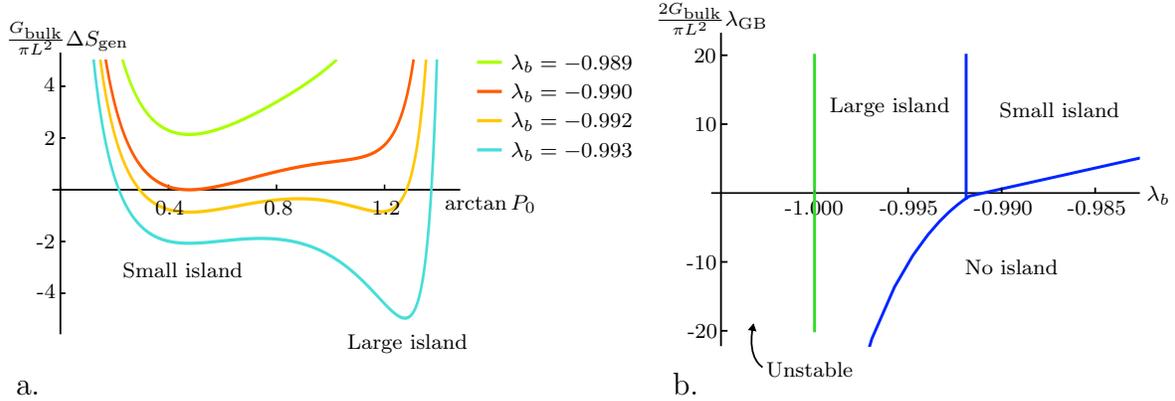
		\caption{ Panel a.: Generalised (renormalised) area from eq. \eqref{radishes} as function of $P_0$, for different values of the DGP coupling $\lamb$. Notice the appearance of a `large' island when $\lamb$ approaches $-1$, due to the partial cancellation of the induced and DGP area terms. Panel b.: Phase diagram: the black lines correspond to first order phase transitions, while the blue one at $\lamb=-1$ indicates the region where gravity becomes unstable. Both plots are done for fixed $L/\s=100$. 
		}
		\label{fig:dS}
	}
\end{figure}
Using the same approach described above, we can explore the transition between the connected and disconnected phases numerically. In figure \ref{fig:dS}a, we plot $\Delta S$ as function of $P_0$ for a fixed $\zeb=0.095,L/\s=100$ and $\lgb=0$, for different values of $\lamb$. These plots are analogous to those presented in figure \ref{fig:dS}a where $\lamb=0$ (but $L/\s$ is varied). Again, these plots are made in lieu of a detailed examination of the boundary conditions where the RT surfaces meet the brane, rather the correct boundary conditions \reef{ortho1} will be achieved where $P_0$ is tuned to produced an minimum in these plots. For small $\lamb$ the curves show a single minimum but $\Delta S>0$, indicating that the disconnected solution dominates in this case. As $\lamb$ becomes more negative, the curves are pulled down and eventually $\Delta S$ enters the negative region so that the connected solution becomes the dominant saddle point. This behaviour is as expected but we note that $\lamb$ is very close to $-1$ in this regime, which according to eq.~\reef{newdefs} means there is almost a complete cancelation between the induced gravitation coupling $1/G_\mt{RS}$ and the DGP term $1/\Gbr$. Of course, this near cancellation is alleviated by turning on the topological coupling $\lgb$, as shown in figure \ref{fig:dS}b. 

Another interesting feature shown in figure \ref{fig:dS}a is the appearance of a second minimum in the curves. This second solution occurs at a larger value of $P_0$ and also of $P_B$, and corresponds to a larger circle $\sigma_\xR$ on the brane, and therefore we refer to it as a \textit{large island}. The existence of this second island is due to the finite terms in \eqref{radishes}. Indeed, these terms are essentially what is plotted in figure \ref{figdeltaA}, and they are unbounded from below for large $P_0$. Therefore, when $\lamb$ becomes sufficiently negative as to produce a significant cancellation between the induced and DGP gravitational entropies, there is a new competition, now between $A(\sigma_\xR)/4G_\mt{eff}$ and the finite terms, producing the large island. As $\lamb\to-1$, the minimum rolls down to infinity ($P\to \infty,\Delta S_{\mt{gen}}\to-\infty$), indicating an instability at this point, which we explore further in appendix \ref{bubble}.

Figure \ref{fig:dS}b summarises the phase diagram of the system, for a fixed value of the tension $L/\s=100$, as we vary both the DGP coupling $\lamb$ and the topological coupling $\lambda_{\mt{GB}}$. The lines between no/small/large islands correspond to first order phase transitions, while the blue line at $\lamb$ indicates the region where the theory becomes unstable.

\section{Discussion}\label{sec:discussion}
%

We have described a holographic framework where quantum extremal surfaces and the island rule \reef{wonderA} can be examined in higher dimensions, \ie for gravity theories in $d\ge2$. In particular, the background is simple enough that the construction given in section \ref{sec:branegravity} is straightforward and purely analytic, in contrast to the numerical approach of \cite{Almheiri:2019psy}. In section \ref{face}, we were also able to describe the system from three different perspectives, analogous to the three descriptions of the two-dimensional system examined in \cite{Almheiri:2019hni}. In particular, we have the boundary perspective, where the system is described as a $d$-dimensional CFT coupled to a ($d-1$)-dimensional conformal defect;
the bulk gravity perspective, where ($d+1$)-dimensional gravity with a negative cosmological constant is coupled to a codimension-one brane; and the brane perspective, where the boundary CFT is coupled to an AdS$_d$ region which supports Einstein gravity and two copies of the same CFT, which are weakly coupled to each other. As we emphasized, this last perspective is an effective theory, as is made clear by the cut-off arising in this Randall-Sundrum braneworld scenario. As discussed and examined in some detail in section \ref{HEE}, this effective gravity theory lends itself to the appearance of quantum extremal islands in the brane perspective, although these have a conventional interpretation from the bulk gravity perspective, in terms of RT surfaces which cross the brane for certain of choices of the entangling geometry on the boundary.\\

\hd{Unconventional features:} Of course, the analysis presented in our paper is somewhat unusual in that we are finding quantum extremal islands but there are no black holes, no horizons and no Hawking radiation involved. Rather we simply considered the entanglement entropy of various entangling regions in the vacuum state of the boundary system. However, to favour the formation of these quantum extremal islands, and at the same time have the brane in the `Einstein gravity regime,' \ie $L/\leff\ll1$, we had to introduce somewhat unconventional couplings. That is, we considered a negative Newton's constant on the brane $\lamb<0$ and nonzero Gauss-Bonnet coupling $\lgb$ for a four-dimensional bulk. Both of these choices were enhancing the connected RT surfaces over the disconnected RT surfaces in calculating the holographic EE. Of course, an interesting question is the interpretation of these `exotic' bulk couplings in terms of data describing the boundary CFT (and the conformal defect). While we do not have a precise interpretation, some qualitative results can be stated.

As observed in section \ref{face}, using standard holographic techniques, one finds that the gravitational coupling in the DGP brane action \reef{newbran} affects the spectrum of defect operators in the boundary theory \cite{domino}. Now let us reiterate that there is no apriori reason not to consider $\lamb<0$. For example, integrating out quantum fields on the brane could produce either a positive or negative shift of Newton's constant. In particular, the shift can be negative for gauge fields or nonminimally coupled scalar fields, as was discussed in the context of EE in \cite{Larsen:1995ax,Kabat:1995eq} -- see also discussion is appendix \ref{bubble}. However, this scenario is not the one we are describing here. In particular, additional brane fields such as these would make significant contributions to the EE which are not accounted for in our calculations. Hence, implicitly, we simply assume that the gravitational coupling $1/\Gbr$ (either positive or negative) is induced by some unknown UV physics.

Introducing the Gauss-Bonnet term \reef{top2} does not modify the gravitational dynamics in the four-dimensional bulk, considered in section \ref{sec:examples}, and hence the correlators of the stress tensor are not modified in the dual three-dimensional boundary theory.\footnote{Of course, such modifications arise for holographic constructions in higher dimensions \cite{Buchel:2009sk}.} However, the topological coupling $\lgb$ affects the entanglement structure of the boundary CFT states. To see this, consider calculating the entanglement entropy holographically for two nearby regions in the boundary. The phase transition between connected and disconnected phase of the RT surfaces is sensitive to a Gauss-Bonnet term. For positive $\lgb$, the transition from disconnected to connected phase takes place earlier (and vice versa for negative $\lgb$). This means that with $\lgb>0$, the mutual information between these two regions remains of order $c_\mt{T}$ for larger separations, \eg \cite{Headrick:2010zt}. Note, however, that choosing positive $\lgb$ favours higher genus surfaces. A concern with this choice might be if higher genus extremal surfaces exist, they may produce unusual results. Finally, we note that the topological coupling appears directly in the expressions for the holographic EE, \eg see eq.~\reef{Sdisc}. Therefore to have an appreciable effect, we must choose this coupling to be of the order of the central charge of the boundary theory, \ie $\lgb\sim L^2/\Gbk\sim \cT$.

Let us add that in section \ref{sec:examples}, we focused on the example of $d=3$ with a four-dimensional bulk. In this case, the natural topological term to add to the bulk gravity is the Gauss-Bonnet term \reef{top2}. Of course, the scenario extends straightforwardly to any $d=2n-1$ for which there is a corresponding topological term which can be added to the bulk gravity action, \ie the Euler character for $2n$-dimensional manifolds, \eg see \cite{Hung:2011xb}. Similarly, for even boundary dimensions ($d=2n$), the analogous topological terms could be added to the brane action, where they would not modify the dynamics of gravity on the brane but they would modify the gravitational entropy associated with the boundary of the quantum extremal islands. 

In light of these unconventional features, a natural question therefore is whether we find quantum extremal islands in our analysis with both $\lamb =0= \lgb$. The answer is affirmative, however, one must reduce to the tension of the brane to reduce its backreaction and the extent of the additional geometry in the vicinity of the  brane's location. As a result, the connected RT surfaces will have a smaller (bulk) area contribution as they cross the brane. However, in this case, the curvature of the AdS geometry on the brane is also smaller, and hence the effective description of the brane theory in terms of Einstein gravity breaks down. That is, with $\leff\sim L$, the contributions of the higher curvature corrections in the induced action \reef{act3} are no longer suppressed relative to the Einstein term and these new interactions play an important role in the dynamics of gravity in the brane perspective. Furthermore, the cutoff of the corresponding CFT on the brane will be much lower. Alternatively, one could think about computing the EE in settings beyond the vacuum state that we studied here. In fact, in \cite{QEI}, we will explicitly show without additional Gauss-Bonnet or DGP couplings that quantum extremal islands appear for (nonextremal) eternal black holes in equilibrium with an external heat bath, \ie in a higher dimensional analog of the analysis in \cite{Almheiri:2019yqk}.

Let us conclude here by comparing our approach with the recent work \cite{Geng:2020qvw}, which appeared while the present paper was prepared for submission. The latter examines essentially the same model (with no DGP term) but concentrates on a very different regime. The authors of \cite{Geng:2020qvw} focused on the formation of islands for the case of a tensionless brane, where the brane gravity becomes very nonstandard, as explained above. Further, in the limit where the graviton becomes massless, \ie $\ell_\mt{eff}\to \infty$, they  observe that no islands form \cite{Geng:2020qvw}. On the other hand, the present work focuses the regime of large brane tension, where the theory on the brane can be well approximated by Einstein gravity (\ie the graviton mass and higher curvature interactions are negligible). We moreover show that by allowing either a topological term or a negative $\Gbr$, islands can appear even in the absence of horizons.\\ 

\hd{Resolving Puzzles:} Our construction clarifies certain conceptual puzzles that arose in early discussions of quantum extremal islands in a holographic framework, \eg for the two-dimensional gravity models introduced in \cite{Almheiri:2019hni} and studied in \cite{Almheiri:2019yqk, Chen:2019uhq}. For example in these models the Planck brane, which supports the JT gravity theory, appears at the boundary of the three-dimensional bulk spacetime. Hence one might have wondered if the brane degrees of freedom (including the JT gravity) are a part of the boundary theory or part of the bulk theory. In our construction, the Planck brane is in the middle of the spacetime geometry and so this question does not arise -- these degrees of freedom belong to the bulk. An important corrolary of this observation is that when a quantum extremal island appears on the brane, \eg see the lower panel in figure \ref{fig:RTPhases}, we are able to recover information about the island with data from the boundary CFT in the corresponding boundary subregion, by applying standard entanglement wedge reconstruction \cite{EW1,EW2,EW3,Jafferis:2015del,Dong:2016eik,Faulkner:2017vdd,Cotler:2017erl}. Of course, the latter would not apply if the brane degrees of freedom were a part of the boundary theory.

Further, our construction circumvents the question of whether RT surfaces are allowed to end on the Planck brane. Rather in our paper, the extremal surfaces just pass through the bulk and only end on the asymptotic boundary as usual. It is simply that in certain situations, the RT surfaces will pass through the brane, which of course, corresponds to the formation of a quantum extremal island.

Another `novel' feature of the two-dimensional JT gravity model of \cite{Almheiri:2019hni} was that the holographic entanglement entropy included an extra boundary term, \ie the gravitational entropy of the JT model, where the RT surface terminated on the Planck brane. That is, the holographic entanglement entropy was given by extremizing the sum of the bulk area of the RT surface and this additional boundary term. An analogous gravitational entropy term on the brane arises in our construction with a DGP brane -- see eq.~\reef{eq:sad}. In fact, our derivation in appendix \ref{generalE} suggests that if the brane supports intrinsic gravitational interactions then the corresponding Wald-Dong entropy on the brane is part of the holographic entanglement entropy formula, as shown in eq.~\reef{fish9}. Hence this general result agrees with the boundary term introduced in the two-dimensional JT gravity models, mentioned above. A shortcoming of the derivation in appendix \ref{generalE} is that the geometric configuration involved a high degree of symmetry, which precluded  finding the expected extrinsic curvature terms \cite{Dong:2013qoa}. Therefore it would be interesting to extend our construction there to more general configurations  along the lines of \cite{Lewkowycz:2013nqa,Dong:2016hjy}.

We want to emphasize the above discussion is distinct from finding in section \ref{sec:enzyme} that the leading contribution to the holographic EE where the RT surface crosses the brane matches the Wald-Dong entropy of the induced gravitational action on the brane\reef{act3}.\footnote{Recall that this analysis was general enough to see the extrinsic curvature contributions coming from the higher curvature interactions in eq.~\reef{act3}.} For example, the leading contribution is $\area(\sigma_\xR)/{4G_\mt{eff}}$, where $\sigma_\xR$ is the cross-section of the RT surface on the brane. As shown in eq.~\reef{eq:bazinga2}, the DGP term is one important contribution to this result, but the bulk area of the RT surface in the vicinty of the brane is also necessary. Of course, we still find the leading contributions reproduce the gravitational entropy of the induced gravity theory on the brane even without the DGP term, \ie with $1/\Gbr=0$. This must be closely related to the fact that the bulk Einstein equations combined with the Israel junction conditions are equivalent to the gravity equations of motion on the brane in the Randall-Sundrum scenario \cite{deHaro:2000wj}.

In passing we note here that $d=2$ is distinguished in the above discussion. In this case, the leading contribution corresponds to the Wald-Dong entropy for the the Polyakov-Liouville action \eqref{PolyAct2} and takes the form given in eq.~\reef{arc}. However, since it only depends on the curvature scalar which is constant across the AdS$_2$ geometry of the brane, this contribution takes the same value no matter where the RT surface  crosses the brane. This contrasts with the higher dimensional result $\area(\sigma_\xR)/{4G_\mt{eff}}$, which rapidly grows as the position of $\sigma_\xR$ moves to larger radii on the brane. That is, there is an enormous penalty against forming large quantum extremal islands for $d\ge3$. In contrast, no such penalty arises for $d=2$ facilitating the formation of islands, as discussed in detail in
\cite{Rozali:2019day}. Of course, if one adds JT gravity \reef{JTee} to the two-dimensional brane action, as in eq.~\reef{braneact2}, then the gravitational entropy on the brane includes $\(\Phi_0+
\Phi(x)\)/4\Gbr$, which will favour smaller quantum extremal islands because the dilaton profile grows with the radius on the brane \cite{Maldacena:2016upp}.

Of course, we can modify our higher dimensional construction to make it more analogous to the two-dimensional model introduced in \cite{Almheiri:2019hni} by taking a $\mathbb Z_2$ orbifold quotient across the brane. With this orbifold, the brane appears as the edge of the bulk geometry but clearly the association with the bulk degrees of freedom has not changed. The brane now only supports a a single copy of the boundary CFT and there are factors of 1/2 appearing in various expressions, \eg we make the following replacement in eq.~\reef{Newton2}: ${1}/{G_\mt{eff}}=L/((d-2)\Gbk)$. Similarly, the RT surfaces will now end on the orbifolded brane while satisfying the boundary condition,
\beq\label{ortho8}
 0  =  \tg_j{}^\nu\(g_{\mu\nu}\,\partial_{n}X^\mu
  + \frac{G_\bulk}{G_\brane}\,\inducedK_i \,\partial_\nu x^i\)\,,
\eeq
which replaces eq.~\reef{ortho7}. Further, the conformal defect becomes a conformal boundary in the orbifolded theory, \ie the spatial geometry on which the CFT lives is now a ($d-1$)-dimensional hemisphere with the conformal boundary being the $S^{d-2}$ at the edge of the hemisphere. 

Other questions that may have arisen from the early discussions of quantum extremal islands which focussed on JT gravity might include the importance of having a low spacetime dimension, \ie $d=2$, or of the JT model itself. The early work of \cite{Penington:2019npb} considered black hole evaporation with Einstein gravity in higher dimensions, and the holographic model of \cite{Almheiri:2019hni} was extended to a holographic framework with $d=4$ in \cite{Almheiri:2019psy} using numerical calculations. Hence our paper reinforces these results by describing quantum extremal islands in a new setting, in particular, in higher dimensions and with Einstein gravity. Our construction is also simple enough that further investigations of the role of quantum extremal islands in higher dimensions are straightforward, \eg see \cite{QEI}. Let us add that JT gravity can be seen as the gravitational dual of the so-called SYK model \cite{Maldacena:2016hyu,Sachdev:1992fk,Sachdev:2010um,Ktalks}. This duality involves an ensemble average over the couplings in the boundary quantum mechanics and so one may expect that this averaging plays a role in the appearance of quantum extremal islands. However, it seems that this is not the case as our construction relies on the standard holographic rules of the AdS/CFT correspondence where there is no such averaging of the couplings in the boundary theory.

One other perplexing issue with the island rule \reef{rule1} is the appearance of the entanglement of the CFT degrees of freedom in the region $\CFTR$ on both sides of the equation \cite{Almheiri:2019hni}. As explained in \cite{Almheiri:2019yqk}, we should distinguish the ``full quantum description'' of, \eg the Hawking radiation in the presence of black holes on the left-hand side from the ``semiclassical description'' which includes the outgoing radiation and purifying partners on the quantum extremal island on the right-hand side. Our holographic construction makes clear that the description of quantum states with islands in the brane picture is on a different footing than that solely in terms of the boundary theory. In particular, referring to the three perspectives discussed in section \ref{face}, it is clear that the boundary perspective (with the boundary CFT coupled to a conformal defect) gives a complete description of quantum state.  By the standard rules of the AdS/CFT correspondence, the bulk perspective (where Einstein gravity with a negative cosmological constant is coupled to a codimension-one brane) gives an equivalent description.\footnote{In this paper, we modeled the CFT defect with a simple brane in the bulk. This bottom-up approach is neither sufficient, nor completely correct. For example, in the case of $\mathcal N=4$ SYM theory on $S^4$, the presence of an interface breaks at least half of the supersymmetry generators and the $R$ symmetry. In a complete description, this will result in a deformation of the bulk $S^5$. For top-down models, see \cite{Karch:2001cw,DeWolfe:2001pq, DHoker:2007hhe, DHoker:2008rje, Chiodaroli:2009yw, Chiodaroli:2011nr, Chiodaroli:2012vc}. }
However, the brane perspective has a different character. In particular, the description in terms of a CFT coupled to the dynamical AdS$_d$ region is only an effective one. Indeed, as emphasized in section \ref{face}, the Randall-Sundrum gravity is only valid down to  the short distance cutoff $\tilde\delta\sim L$, \ie see eqs.~\reef{ctoffplus} and \reef{ctoffminus}. Beyond this cutoff, gravity is no longer localized to the brane and the additional `Kaluza-Klein' modes of the graviton are strongly coupled to the brane and their contribution cannot be ignored. 

Further, this brane perspective also provides an effective description of the coupling to the defect CFT. That is, it only accounts for the couplings localized at the defect, which dominate at low energies, but ignores the subtle nonlocal couplings, which could be seen as coming through the bulk AdS geometry in the dual description. Of course, the quantum extremal islands in the effective description of the brane perspective are a clear example of this. These islands are a remnant of replica wormholes in the limit $n\to1$ \cite{Penington:2019kki,Hartman:2020swn}. However, in the replica trick construction of the corresponding Renyi entropies in the bath CFT, one can ask why the gravity on the different branes in the replica copies should connect with one another. However, these effective gravity theories are UV completed by a single theory of gravity in the bulk and so it is natural to consider geometries connecting the branes, \ie replica wormholes if the effective theory. Hence the connection of the brane and boundary through the bulk provides a simple explanation of these wormholes.  Given the simplicity of our construction, it may provide a useful framework in which to understand further subtleties in distinguishing the various expressions in the island rule.

As a final note here, we observe that the finite cutoff for the CFT on the brane has noticeable effects even for $d=2$, \eg see eq.~\reef{almost}. In contrast, the early discussions of \eg \cite{Almheiri:2019hni,Almheiri:2019psf, Almheiri:2019yqk, Chen:2019uhq, Penington:2019kki, Almheiri:2019qdq} assumed that one could use standard formulae for conformal transformations in the $d=2$ CFT in the gravitational region (\ie on the brane). It would be interesting to understand if the cutoff modifies any of this analysis in a significant way \cite{QEI}.\\

\hd{Brane geometry, Part I:} As described in section \ref{sec:branegravity}, we choose the brane tension to produce a negative cosmological constant in the gravity theory on the brane, in accord with eqs.~\reef{act3} and \reef{Newton2}. As a result, the $d$-dimensional geometry on the brane is AdS space. However, it is straightforward to consider the case where the brane tension takes its critical value, such that $1/\ell_\mt{eff}^2=0$, as is usually done in the Randall-Sundrum scenario \cite{Randall:1999ee,Randall:1999vf}. In this case, the analogous brane geometry is simply flat space, and the brane is easily embedded in the bulk AdS$_{d+1}$ geometry on a slice of constant radius (or constant $z$) in standard Poincar\'e coordinates. An interesting feature of this embedding is that the brane reaches the asymptotic AdS$_{d+1}$ boundary along the null boundaries of the flat space geometry (as well as a timelike and spacelike infinity) \eg see \cite{Karch:2001cw}. 
Hence we can naturally investigate quantum extremal surfaces and the island formula in flat space using the usual expressions for holographic entanglement entropy in this construction as long as we consider regions on null infinity. Notably this matches the approach pursued in \cite{Hartman:2020swn}, but contrasts with studies of \eg \cite{Gautason:2020tmk} which considered spacelike regions. It would, of course, be interesting to use this framework to study quantum extremal islands in the context of asymptotically flat braneworld black holes, \eg as described in \cite{Emparan:1999wa,Emparan:1999fd}. We should note however that there are undoubtedly subtleties with the proposed construction, \eg as the brane completely cuts out the asymptotic AdS$_{d+1}$ boundary (except for a single point) on constant time slices. 

Of course, one can also consider the case where the brane tension is chosen such that $1/\ell_\mt{eff}^2<0$. That is, the brane gravity theory would have a positive cosmological constant and the corresponding brane geometry becomes de Sitter space. In this case, one constructs a foliation of the bulk AdS$_{d+1}$ geometry in terms of $d$-dimensional de Sitter slices and the brane can be embedded along the slice with the appropriate curvature, \eg see \cite{Karch:2001cw}. In this case, the brane reaches the asymptotic AdS$_{d+1}$ boundary on the future and past timelike infinities of the de Sitter geometry. 
Hence, this construction provides a framework to use holographic entanglement entropy for investigating the island formula in de Sitter space as long as we consider regions on the timelike future of the latter geometry. Let us add that this would be similar to upcoming work of \cite{dSone,dStwo}, which studies related questions in the context of JT gravity with a positive cosmological constant 
\cite{Maldacena:2019cbz}. The de Sitter evolution of the Hartle-Hawking vacuum prepares a two-dimensional CFT state on circle and the entanglement entropy of various regions in the latter state are investigated, revealing new islands in the de Sitter geometry \cite{dSone,dStwo}.\\ 

\hd{Brane geometry, Part II:}

\begin{figure}[t]
\centering
\includegraphics[scale = 0.7]{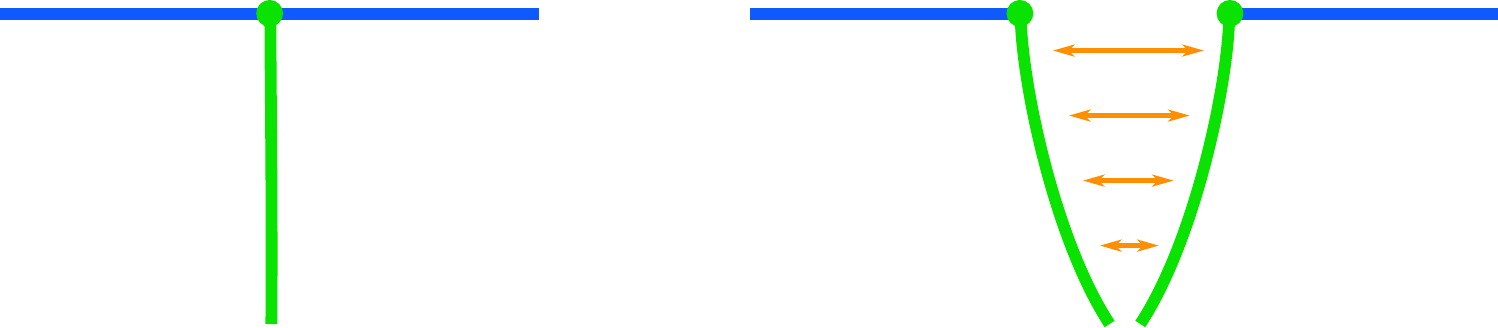}
\caption{Left: In the brane perspective, the bath CFT on the asymptotic boundary (blue) is connected to two copies of the effective CFT on the brane (green) but the resulting geometry is not a manifold. Right: For excitations below the effective CFT cutoff the system behaves as if it consists of two systems on a manifold which are weakly coupled in the gravitational region (green).}
\label{fig:no_mfld}
\end{figure}

The geometry of the setup presented in this paper might look unconventional. As seen from the brane perspective, we have the bath CFT on the asymptotic boundary with geometry $S^{d-1} \times \mathbb R$, and two copies of the same CFT on the brane with an AdS$_d$ geometry. These two geometries are joined by introducing a cutoff surface (with topology $S^{d-2} \times \mathbb R$) near the asymptotic boundary of the AdS$_d$ geometry and gluing it to the equator of the  $S^{d-1} \times \mathbb R$ geometry. In particular, the resulting geometry is not a manifold in the vicinity of the gluing region -- see the left panel of figure \ref{fig:no_mfld}. Of course, we can obtain a manifold by taking the $\mathbb Z_2$ quotient which identifies the two halves of the bath CFT, such that the theory is again defined on a manifold with topology $S^{d-1} \times \mathbb R$. However, we will ignore this simplification here. Rather, we want to comment on the theory before taking the $\mathbb Z_2$ quotient. 

First, we note that constructions where multiple CFTs are joined at a common defect are not rare. For example they appear in the study of boundary and interface CFTs (\eg see \cite{Chiodaroli:2012vc}), and sometimes seem to be required to remove anomalies \cite{Ooguri:2020sua}.

Second, we would like to argue that in the regime where the defect theory can be described by two copies of the boundary CFT coupled to Einstein gravity, we can approximately think of the full theory as two copies of the orbifolded theory (each living on a manifold), which are weakly coupled in the gravitational region -- see the right panel of figure \ref{fig:no_mfld}. This is particularly easy to see from the bulk perspective. For brevity we restrict ourselves to the discussion of graviton modes, but a similar story applies to all bulk fields. 

Let us begin by recalling that for $\veps \ll 1$, the spectrum of graviton fluctuations in the bulk is almost unchanged with respect to the modes in (two copies of) empty AdS space. Hence much of the corresponding physics should be very similar that of two copies of the the AdS$_{d+1}$, or to two copies of the dual CFT$_d$ on the boundaries of two independent AdS$_{d+1}$ geometries. Of course, one exception to the preceding is that upon gluing the two AdS$_{d+1}$ geometries together, a new set of very light graviton states localized in the vicinity of the brane \cite{Randall:1999vf,Randall:1999ee,Karch:2000ct,Karch:2001jb}, as discussed in section \ref{face}. For simplicity, we refer to the latter as the brane graviton modes, while we refer to the former as the standard normalizable modes.\footnote{These bulk modes are $\mathbb Z_2$ graded under reflection across the Planck brane, and the even modes survive the $\mathbb Z_2$ orbifold discussed above include the brane graviton states as well as half of the standard normalizable modes. However, this organization of the modes is not useful for the following discussion.}

On a fixed time slice, as shown in the right panel of figure \ref{fig:brane2}, the standard normalizable modes will describe stress energy excitations in the dual CFT on both the left and right halves of the asymptotic boundary. If we assume an approximate extrapolate dictionary \cite{Harlow:2011ke} for the brane theory as well, these normalizable modes will also describe analogous excitations for the effective CFT on the brane. However, there will be two sets of such excitations: those described by bulk excitations\footnote{We stress here that the localized excitations considered here do not correspond to individual energy eigenmodes, which were implicit in the previous paragraph. Rather they will consist of linear combinations of such eigenmodes evaluated on the fixed time slice being examined here. Of course, having superpositions of energy eigenmodes is what produces the complicated time evolution described below.} with support primarily in the right copy of the AdS$_{d+1}$ geometry, and those described by the analogous excitations primarily in the left AdS$_{d+1}$ geometry. Hence, the stress tensor on the brane can be decomposed into two pieces which correspond to subsectors of the brane theory, each of which is determined by bulk excitations which essentially live on one side of the brane. If these subsectors were truly superselection sectors (\eg as one might imagine arises in the limit $\veps\to0$), our brane theory would contain two independent copies of the boundary CFT  and each of these copies would only interact with the bath CFT on the corresponding half of the asymptotic boundary. That is, each of these systems would live on an independent manifold with topology $S^{d-1} \times \mathbb R$. 

However, this is not strictly correct and the two copies of the CFT on the brane are weakly coupled with $\veps\ll1$ but finite. In particular, localized stress energy excitations of the form considered above will not remain localized with time evolution. Rather they will eventually spread across the entire asymptotic boundary if time evolves for a sufficiently long time. For example, an excitation localized on the right asymptotic boundary will evolve to eventually produce excitations of the stress tensors on the left asymptotic boundary and on the brane as well. From the boundary perspective, excitations moving onto the brane correspond to excitations that are absorbed by the conformal defect (and remain there for a long time).

The spreading of the localized excitations can be seen to arise through two physical effects: First, the bulk excitations can tunnel between the two AdS$_{d+1}$ regions shown in figure \ref{fig:brane2}. Recall that (the radial part of) the linearized bulk equation of motion can be reduced to a Schroedinger equation with a double-well potential, where the height of the barrier is determined by the brane tension \cite{Karch:2000ct}. With $\veps\ll1$ but finite, the barrier height while large remains finite and there will be a finite probability for a bulk excitation on one side of the Planck brane to tunnel to the other. A second independent coupling comes because the stress tensors of the two copies of the CFT couple to the same gravity theory on the brane. From the bulk perspective, the nonlinear Einstein equation produces interactions between the brane graviton modes with excitations on either side of the brane. Hence bulk excitation excitations on one side can leak to the other side by scattering process involving the brane gravitons. However, we note that both effects become smaller as the brane tension approaches its critical value, \ie as $\veps$ approaches zero. Thus, to a good approximation, the brane theory can be treated at two copies of the boundary CFT which only interact weakly. \\

\hd{Entanglement wedge cross-sections:} Recent work \cite{Takayanagi:2017knl,Nguyen:2017yqw} has drawn attention to
the entanglement wedge cross-section, \ie for disconnected boundary regions, the codimension-two surfaces in the bulk which have minimal area and which split the entanglement wedge in two. In particular, there are a number of proposals relating these holographic surfaces to various entanglement measures: entanglement of purification \cite{Takayanagi:2017knl,Nguyen:2017yqw}, reflected entropy \cite{Dutta:2019gen},  odd entanglement entropy \cite{Tamaoka:2018ned,Kusuki:2019evw,Kusuki:2019rbk}, or entanglement negativity \cite{Kudler-Flam:2018qjo,Kusuki:2019zsp}. 

Turning to our model and examining figure \ref{fig:RTPhases}, we see that there are two such minimal surfaces in the connected phase, for which a quantum extremal island appears on the brane. These surfaces are simply disks of radius $P=P_0$ on either side of the brane, with area
\beq\label{reflw}
A=\frac{2\,L^{d-1}\, \Omega_{d-2}}{d-1} \ P_0^{d-1}\, {}_{2}F_1\left[ \frac{1}{2},\frac{d-1}{2},\frac{d+1}{2},-P_0^2 \right]\,,
\eeq
as can be seen from eq.~\reef{A_disc}. The fact that both disks have the same area results from the fact that the corresponding boundary regions are symmetric on either of the conformal defect -- see figure \ref{EEprob}. Of course, if one of the two caps comprising the boundary regions was smaller, the minimal area disk closer to this cap would provide the global minimum and hence become the entanglement wedge cross-section. It would be interesting to understand if the second minimal disk also plays an interesting role in characterizing the entanglement of the boundary state. In this vein, let us add that there are also two additional extremal disks which divide the entanglement wedge in two but their area is actually a local maximum. These disks again lie on either side of the brane but end on $\sigma_\xR$, the intersection of the RT surface with the brane. Again, it is natural to wonder if these surfaces have an interpretation in terms of the boundary entanglement. Let us note that similar surfaces appear in the following discussion.\\

\hd{RT Bubbles and Wormholes:} 

	In appendix \ref{bubble}, we consider a surprising class of RT surfaces with the topology of a sphere, \ie $S^{d-1}$ in the ($d+1$)-dimensional bulk. The appearance of these extremal `bubbles' is quite unusual as they are homologous to the entire boundary. Hence the standard RT prescription would assign an entropy to the ground state of the dual boundary system. Further, presence of a `zero mode' which allows the bubbles to be translated along the brane makes their interpretation even more puzzling. An essential feature for the appearance of the RT bubbles was that the gravitational coupling in the DGP term \reef{newbran} was negative, \ie $\lamb<0$. We also noted that the bubbles do not appear to be macroscopic objects in the brane theory. Rather, as shown in eq.~\reef{haiku2}, their size is always of order of the effective cutoff $\tilde\delta$.
	
Despite the unusual features of these RT bubbles, the discussion in appendix \ref{bubble} highlights a general feature of the quantum extremal islands in a simple way. In particular, as discussed below eq.~\reef{genbubble1}, there are two competing terms contributing to the generalized entropy of these surfaces: the bulk area which describes the entropy of the CFT fields on the brane enclosed by the bubble and the area of the boundary where they intersect the brane, which appears in the gravitational entropy of the DGP term. The bulk contribution naturally acts to contract the bubble but with $\lamb<0$, the brane contribution acts to expand the bubble. As described in the appendix, there is an equilibrium radius where these two effects balance one another. Of course, with $\lamb>0$, the brane contribution also acts to contract the boundary of the bubble and so no closed extremal surfaces appear, as expected.

As noted above, a similar competition is a general feature in the formation of quantum extremal islands. However, in this case as discussed in section \ref{sec:enzyme}, the bulk and brane contributions combine to produce a Bekenstein-Hawking term $\area(\sigma_\xR)/{4G_\mt{eff}}$ on the boundary of the island. This contribution, of course, imposes a large penalty to the formation of a large island and acts to contract the boundary towards a smaller (\ie vanishing) radius. For an island to appear, this contraction must be balanced by an expanding contribution. From the bulk perspective, this is simply coming from the remaining\footnote{We combined part of the bulk area into the Bekenstein-Hawking term above.} bulk area contribution of the RT surface, which we can ascribe to the quantum EE of the CFT state from the brane perspective. The point to be noted here is that for this to provide an expansion the RT surface must be anchored far from the island, \ie in the asymptotic (nongravitational) region associated with the boundary CFT. While perhaps self-evident, this discussion highlights the nonlocal nature of the physics producing the quantum extremal islands.

Let us add that the quantum extremal islands discussed here (as well as the RT bubbles) are remnants of replica wormholes in the limit $n\to1$. This follows from the fact that we are simply studying holographic EE with RT surfaces in a new bulk background, \ie with a back-reacted brane. Hence the analysis of \cite{Lewkowycz:2013nqa}\footnote{Following \cite{Dong:2016hjy,Faulkner:2017vdd}, the same applies for general time dependent situations.} introduces a smooth $n$-fold covering geometry for the corresponding Renyi entropies with positive integer indices. These covering geometries produce smooth wormhole geometries on brane analogous to those discussed in \cite{Almheiri:2019qdq,Penington:2019kki} for two dimensions. 

Now assuming replica symmetry, one can then take a $\mathbb Z_n$ orbifold quotient which leaves a single copy of the boundary geometry but the bulk solution now contains a codimension-two cosmic brane with tension $T_n=(n-1)/(4\Gbk\,n)$. In the presence of a DGP brane, we expect that there is an additional contribution where the two branes intersect, \ie the intersection surface carries an intrinsic tension $\widehat T_n=(n-1)/(4\Gbr\,n)$. In this setting, our discussion above for the formation of quantum extremal islands extends to the Renyi entropies in a relatively straightforward way. In particular, we expect that an area contribution associated with the boundary of the island now carries an effective tension $\tilde T_n=(n-1)/(4G_\mt{eff}\,n)$, which combines the intrinsic tension of this intersection surface and the contribution of the cosmic brane in the vicinity of the Planck brane. The contraction created by this term must be balance by the expansion provided by the remaining cosmic brane contributions. However, to provide an expansion the cosmic brane must be anchored by a twist operator in the asymptotic (nongravitational) boundary. Again, this highlights the nonlocal nature of the physics which implicitly supports the replica wormholes.

Of course, these dynamical considerations are emergent in the topological models considered in \cite{Marolf:2020xie,Penington:2019kki}. Hence it would be interesting to understand the implications of this dynamics to extend the new discussions of baby universes and ensembles to higher dimensions.\\

To conclude, let us comment that we will build on the holographic model constructed here to study the Page curve and the appearance of quantum extremal islands for higher dimensional black holes in \cite{QEI}. In particular, we study eternal black holes coming to equilibrium with an external heat bath (prepared at the same temperature) in a higher dimensional analog of the analysis appearing in \cite{Almheiri:2019yqk}. Let us reiterate that unconventional features (\ie Gauss-Bonnet and DGP couplings) introduced to favour quantum extremal islands here are unimportant in the discussion of higher dimensional black holes.\\


\section*{Acknowledgments}
We would like to thank Ahmed Almheiri, Raphael Bousso, Xi Dong, Netta Engelhardt,  Zach Fisher, Greg Gabadadze, Juan Hernandez, Don Marolf, Shan-Ming Ruan, Edgar Shaghoulian, Antony Speranza and Raman Sundrum for useful comments and discussions. Research at Perimeter Institute is supported in part by the Government of Canada through the Department of Innovation, Science and Economic Development Canada and by the Province of Ontario through the Ministry of Colleges and Universities. RCM is supported in part by a Discovery Grant from the Natural Sciences and Engineering Research Council of Canada, and by the BMO Financial Group. HZC is supported by the Province of Ontario and the University of Waterloo through an Ontario Graduate Scholarship. RCM and DN also received funding from the Simons Foundation through the ``It from Qubit'' collaboration. The work of IR is funded by the Gravity, Quantum Fields and Information group at AEI, which is generously supported by the Alexander von Humboldt Foundation and the Federal Ministry for Education and Research  through the Sofja Kovalevskaja Award. IR also acknowledges the support of the Perimeter Visiting Graduate Fellows program and the hospitality of Perimeter Institute, where part of this work was done. JS acknowledges the support of the Natural Sciences and Engineering Research Council of Canada (NSERC).

\appendix

\section{Generalized Entropy on the Brane}\label{generalE}
%

In sections \ref{sec:two-d} and \ref{sec:DGP}, we introduced intrinsic gravitational terms to the brane action. Following \cite{Almheiri:2019hni},\footnote{See also \cite{Almheiri:2019psf, Almheiri:2019yqk, Chen:2019uhq, Penington:2019kki, Almheiri:2019qdq}.} we assumed that these terms contribute to the generalized entropy, \eg see eq.~\reef{eq:sad0} or \reef{eq:sad}.
In this appendix, we present a extended version of an argument in \cite{Myers:2010tj}, which will support this assumption and our formula for generalized entropy. 

\begin{figure}[h]
	\def\svgwidth{0.6\linewidth}
	\centering{
\begingroup%
  \makeatletter%
  \providecommand\color[2][]{%
    \errmessage{(Inkscape) Color is used for the text in Inkscape, but the package 'color.sty' is not loaded}%
    \renewcommand\color[2][]{}%
  }%
  \providecommand\transparent[1]{%
    \errmessage{(Inkscape) Transparency is used (non-zero) for the text in Inkscape, but the package 'transparent.sty' is not loaded}%
    \renewcommand\transparent[1]{}%
  }%
  \providecommand\rotatebox[2]{#2}%
  \newcommand*\fsize{\dimexpr\f@size pt\relax}%
  \newcommand*\lineheight[1]{\fontsize{\fsize}{#1\fsize}\selectfont}%
  \ifx\svgwidth\undefined%
    \setlength{\unitlength}{800bp}%
    \ifx\svgscale\undefined%
      \relax%
    \else%
      \setlength{\unitlength}{\unitlength * \real{\svgscale}}%
    \fi%
  \else%
    \setlength{\unitlength}{\svgwidth}%
  \fi%
  \global\let\svgwidth\undefined%
  \global\let\svgscale\undefined%
  \makeatother%
  \begin{picture}(1,0.4125)%
    \lineheight{1}%
    \setlength\tabcolsep{0pt}%
    \put(0,0){\includegraphics[width=\unitlength,page=1]{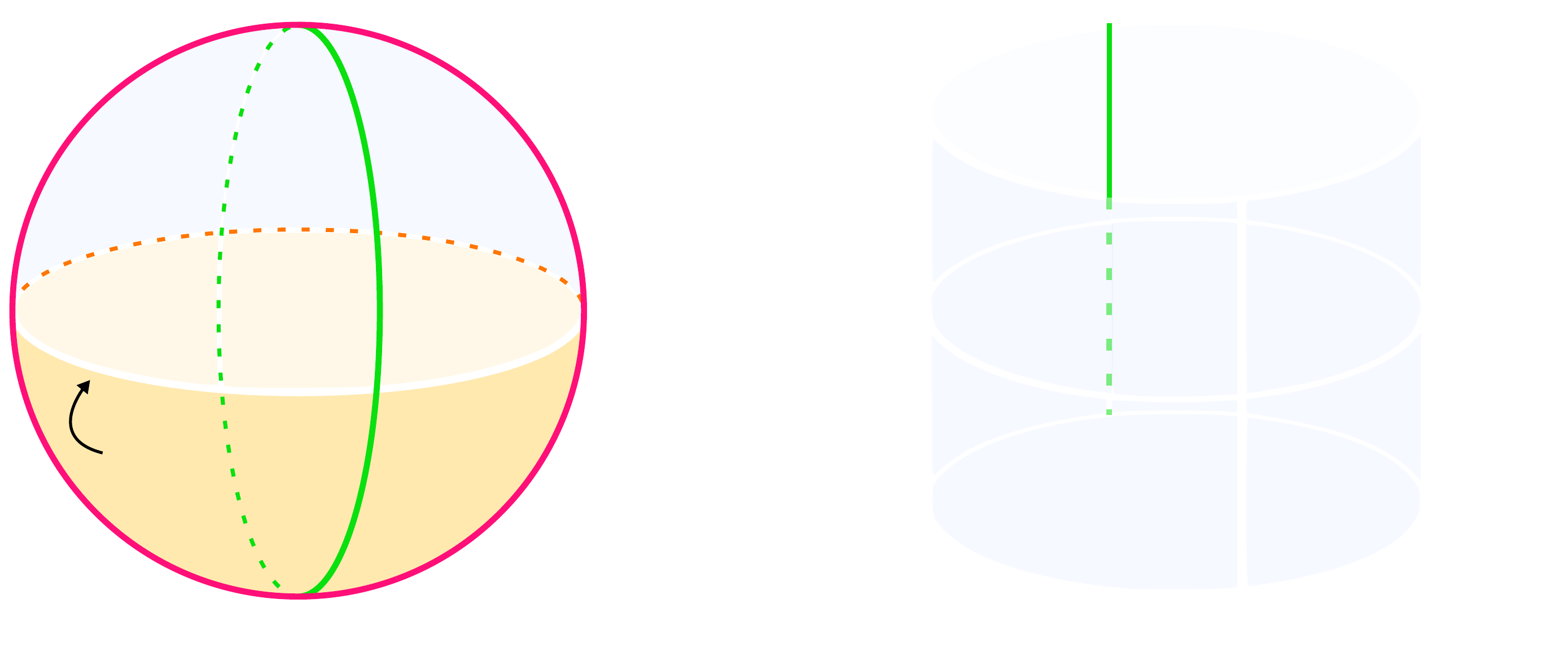}}%
    \put(0.0675595,0.10473836){\color[rgb]{0,0,0}\makebox(0,0)[lt]{\lineheight{1.25}\smash{\begin{tabular}[t]{l}$\Sigma_\mt{CFT}$\end{tabular}}}}%
    \put(0,0){\includegraphics[width=\unitlength,page=2]{defect31.pdf}}%
    \put(0.94500679,0.12140327){\color[rgb]{0,0,0}\makebox(0,0)[lt]{\lineheight{1.25}\smash{\begin{tabular}[t]{l}$t$\end{tabular}}}}%
    \put(0,0){\includegraphics[width=\unitlength,page=3]{defect31.pdf}}%
    \put(0.43559142,0.16104613){\color[rgb]{0,0,0}\makebox(0,0)[lt]{\lineheight{1.25}\smash{\begin{tabular}[t]{l}$t=0$\end{tabular}}}}%
  \end{picture}%
\endgroup%

		\caption{A timeslice of our $d$-dimensional CFT setup with entangling surface $\Sigma_\mt{CFT}$ and an equatorial conformal defect (the green line). In the right panel, one dimension is suppressed relative to the left panel.} \label{fig:defect}}
\end{figure}

As in the main text, we begin with a $d$-dimensional holographic CFT on $R\times S^{d-1}$ with a conformal defect on the equator of the sphere, sweeping out $R\times S^{d-2}$. On a fixed time-slice, we choose an entangling surface $\SCFT$ which divides the sphere into two equal halves along a maximal $S^{d-2}$ which lies orthogonal to the conformal defect. Now we wish to determine the entanglement entropy between the two halves
of the system, as sketched in figure \ref{fig:defect}. Recall that with the geometric approach \cite{Callan:1994py}, we must evaluate the partition function on a (Euclidean) background geometry with an infinitesimal conical defect. In order to construct a symmetric geometry where introducing such a defect is well-defined, we perform a Wick rotation on the boundary time (\ie $t_\mt{E}=it$) and then conformally transform the Euclidean background metric to a round $S^{d}$ with the conformal defect lying on a maximal $S^{d-1}$ on this background. Now $\SCFT$ remains a maximal $S^{d-2}$ which runs orthogonal to the defect and pierces the latter on a $S^{d-3}$. With this construction, there is a rotational symmetry in the two dimensions orthogonal to $\SCFT$. To evaluate the corresponding entanglement entropy, we construct $\mathcal{M}_{1-\eps}$, the `$n$-fold cover' with $n=1-\eps$, by introducing an infinitesimal conical defect at $\SCFT$. The entanglement entropy is then given by
\beq\label{entro9}
S = \lim_{\eps\to0}\left( \frac{\partial\ }{\partial\eps}+1 \right)\log
Z_{1-\eps}\,,
\eeq
where $Z_{1-\eps}$ is the partition function of the holographic CFT on the covering space $\mathcal{M}_{1-\eps}$. Of course, the latter has a dual description in terms of the bulk gravity, and using the usual saddle point approximation, eq.~\reef{entro9} becomes \cite{Myers:2010tj}
\beq\label{entropylimit}
S =-\lim_{\epsilon\to0}\Big(\frac{\partial}{\partial \epsilon} + 1\Big) I_{E, 1-\epsilon}\,,
\eeq 
where $I_{E,1-\epsilon}$ is the Euclidean bulk action evaluated on the appropriate dual solution.
%

\begin{figure}[h]
	\def\svgwidth{0.6\linewidth}
	\centering{
		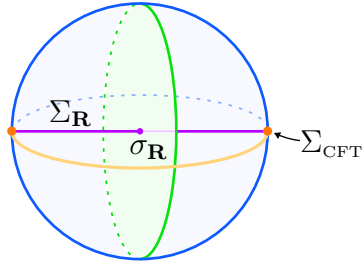
		\caption{A cross-section of the Euclidean geometry $\widetilde{\mathcal{M}}_{1}$. The orange semicircle and its complement along a time slice represent the orange shaded region of figure \ref{fig:defect} and its complement. The rotation that keeps $\Sigma_\mt{CFT}$ fixed represents euclidean time. An infinitesimal conical defect $\Sigma_\xR$ runs through the bulk and intersects the brane at $\sigma_\xR$.} \label{fig:defect2}
	}
\end{figure}

Setting $n=1$ for a moment, the bulk dual of $\mathcal{M}_{1}$ is simply the Euclidean version of the geometry constructed in section \ref{BranGeo}, which we denote $\widetilde{\mathcal{M}}_{1}$. Recall the boundary geometry is $S^{d}$ and the conformal defect runs around a maximal $S^{d-1}$. In the bulk, the geometry is locally EAdS$_{d+1}$ everywhere away from the brane, and the brane has a EAdS$_d$ geometry which extends out to the conformal defect at the asymptotic boundary and with the curvature scale given by eq.~\reef{curve1} -- see figure \ref{fig:defect2}. Now the  entangling surface $\SCFT$ on the asymptotic AdS boundary is the boundary of an extremal surface $\Sigma_\xR$ in the bulk, which runs  straight across the bulk solution and has a EAdS$_{d-1}$ geometry with curvature scale $L$.  This surface pierces the brane at a right angle and the intersection, another extremal surface $\sigma_\xR$, has the geometry of a EAdS$_{d-2}$ with curvature scale $\ell_\mt{B}$ -- see figure \ref{fig:defect2}. Now because of the symmetry of this configuration, the rotational symmetry about the entangling surface in the boundary extends to a rotational symmetry about $\Sigma_\xR$ in the bulk. Hence we can calculate the entanglement entropy with the same geometric approach as we applied in the boundary. That is, we construct $\widetilde{\mathcal{M}}_{1-\eps}$, the $n$-fold cover (with $n=1-\eps$) of the bulk solution  with a infinitesimal conical defect at $\Sigma_\xR$ and by extension, at $\sigma_\xR$ on the brane. 

That is, the angle around $\Sigma_\xR$ runs through a range $2\pi(1-\epsilon)$. Now  \cite{Fursaev:1994ea,Fursaev:1995ef} developed a description of such conical defects in which the singular geometry is replaced by a `regulator' geometry where the region
around the conical singularity is smoothed out. Applying their key result, we can write the bulk Riemann tensor  as a ``smooth" contribution away from $\Sigma_\xR$, the conical defect, and a singular order $\epsilon$ contribution at $\Sigma_\xR$,\footnote{This order $\eps$ contribution is universal, whereas the details of the regulator come into play at order $\eps^2$ and higher.}
\beq\label{separation}
^{(\epsilon)}R^{ab}{}_{cd} = R^{ab}{}_{cd}+2\pi \epsilon\, {\varepsilon}^{ab}{\varepsilon}_{cd}\,\delta_{\Sigma_\xR}\,,
\eeq 
where ${\varepsilon}_{ab}$ is the Euclidean volume form in the two-dimensional transverse space to $\Sigma_\xR$, and $R^{ab}{}_{cd}$ is the ``smooth" curvature piece. The $\delta_{\Sigma_\xR}$ is a two-dimensional delta function defined in \cite{Myers:2010tj}. The conical singularity intersects the brane at $\sigma_\xR$ and so we have a similar decomposition for the Riemann tensor on the brane,
\beq\label{separation2}
^{(\epsilon)}\tilde R^{ij}{}_{k\ell} = \tilde R^{ij}{}_{k\ell}+2\pi \epsilon \,\tilde{\varepsilon}^{ij}\tilde{\varepsilon}_{k\ell}\,\delta_{\sigma_\xR}\,.
\eeq 

Now recall that our aim is to evaluate the Euclidean action in eq.~\reef{entropylimit}. This action is the sum of the Euclidean versions\footnote{Note that the difference in signs in going between Minkowski and Euclidean signatures \cite{Myers:2010tj}.} of the bulk and brane actions in eqs.~\reef{act2} and \reef{newbran} (or perhaps eq.~\reef{JTee} for $d=2$), as well as the associated boundary terms. Equipped with eqs.~\reef{separation} and \reef{separation2}, it can be shown that in the limit of small $\epsilon$ that the Euclidean action can be expanded as
\beqa\label{epsilonaction}
I_{E,1-\epsilon}& =& (1-\epsilon)I_{E,1} +
 \int_\mt{bulk}\!\! d^{d+1}x \sqrt{g}\, 2\pi \epsilon {\varepsilon}^{ab}{\varepsilon}_{cd}\,\delta_{\Sigma_\xR}\, \frac{\partial \mathcal{L}_\mt{E,bulk}}{\partial R^{ab}{}_{cd}}\\
&&\qquad+\int_\mt{brane}\!\!\!\! d^{d}x \sqrt{\tilde g} \, 2\pi \epsilon \tilde{\varepsilon}^{ij}\tilde{\varepsilon}_{k\ell}\,\delta_{\sigma_\xR}\,  \frac{\partial \mathcal{L}_\mt{E,brane}}{\partial \tilde{R}^{ij}{}_{k\ell}} +\mathcal{O}(\epsilon^2)\,.
\eeqa
Noting the symmetry of our configuration, \ie the curvatures are constant everywhere along the surfaces $\Sigma_\xR$ and $\sigma_\xR$,
we then find the entropy in eq.~(\ref{entropylimit}) is given by
\beq\label{fish9}
S = -2\pi \frac{\partial \mathcal{L}_\mt{E,bulk}}{\partial R^{ab}{}_{cd}} {\varepsilon}^{ab}{\varepsilon}_{cd} \int_{\Sigma_\xR} d^{d-1}x \sqrt{h}-2\pi \frac{\partial \mathcal{L}_\mt{E,brane}}{\partial \tilde{R}^{ij}{}_{k\ell}} \tilde{\varepsilon}^{ij}\tilde{\varepsilon}_{k\ell} \int_{\sigma_\xR} d^{d-2}x \sqrt{h'}\,,
\eeq 
where $h$ and $h'$ are the induced metrics along the $\Sigma_\xR$ and $\sigma_\xR$, respectively. Hence we see that there is a contribution of the Wald entropy from both the bulk action and the brane action. Further, let us note that various signs appear upon analytically continuing back to Lorentzian spacetime, \ie in the Lagrangian and the transverse volume form \cite{Myers:2010tj}.

For the case where the Einstein-Hilbert action appears both in the bulk and on the brane, as in eqs.~\reef{act2} and \reef{newbran}, we find the formula for the generalized entropy \reef{fish9} becomes
\beq\label{fish88}
S = \frac{A(\Sigma_\xR)}{4 G_\mt{bulk}}+ \frac{A(\sigma_\xR)}{4 G_\mt{brane}}\,,
\eeq
as given in equation \reef{eq:sad}. The present derivation only applies to special symmetric configuration, as in \cite{Myers:2010tj}. The symmetry of this configuration preculdes finding any extrinsic curvature terms in eq.~\reef{fish9}, as would be expected for the Dong entropy \cite{Dong:2013qoa}. We note however that no such terms would correct eq.~\reef{fish88} for the generalized entropy coming from the Einstein-Hilbert term. It would, of course, be interesting to extend our derivation to more general configurations involving bulk DGP branes, along the lines of \cite{Lewkowycz:2013nqa} or \cite{Dong:2016hjy}.


\section{RT Bubbles}
\label{bubble}
%


In this appendix, we consider a simple but surprising class of RT surfaces. In particular, we show below that there are closed extremal surfaces with the topology of a sphere, \ie $S^{d-1}$ in the locally AdS$_{d+1}$ bulk geometry. In empty AdS space, one could consider such spherical surfaces, but their area would be extremized when they collapse to zero size. In the present case, we will show that in certain situations, the spherical RT surfaces can be supported at finite size by the brane.  To illustrate the situation, we continue with the special case of $d=3$ as in section \ref{sec:examples}, and afterwards comment on the situation with general $d$. 
%

\begin{figure}[h]
	\def\svgwidth{0.8\linewidth}
	\centering{
		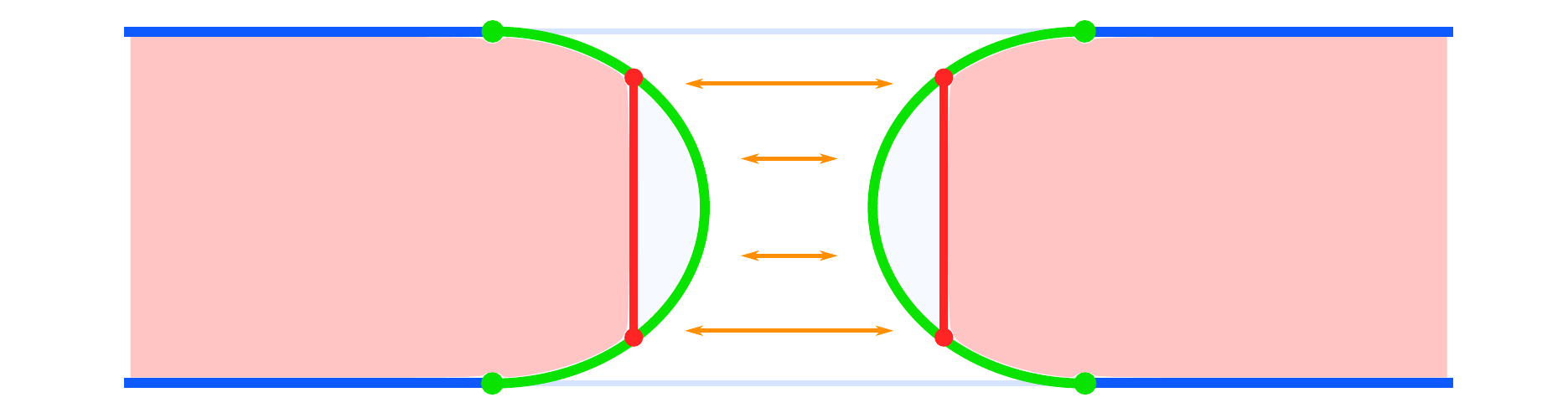
		\caption{An RT `bubble' on the brane: even for the vacuum, when $\Gbr<0$ the competing bulk and brane area terms can lead to a stable extremal surface, which is  homologous to the entire time slice for the boundary CFT. The entanglement wedge then corresponds to the shaded red region. Since the two sides of the brane are glued together, the RT surface has the topology of $S^{d-1}$. }
		\label{figbubble}
	}
\end{figure}

Consider the geometry illustrated in figure \ref{figbubble}. On either side of the brane, we have a disk satisfying $\zeta=$constant, \ie satisfying eq.~\reef{zetap} with $P_0=0$. Hence locally these surfaces extremize the entropy functional \reef{area} in the bulk. However, rather than extending out to the asymptotic boundary, as shown in the figure, the two disks intersect the brane and meet at some radius $\pb$.  Hence this RT surface has the topology of a sphere and we use the nomenclature `bubble' to describe these surfaces. For $d=3$, the generalised entropy \reef{eq:sad} of this bubble is
\begin{align}\label{Agenbubble}
\sgen&=\frac{\pi L^2}{\Gbk} \left(  \sqrt{1+\pb^2}-1  + \lamb\, \pb \right) +2\,\lgb 
\end{align}
with $\lamb$ defined in eq.~\eqref{newdefs}. We have also included the topological term introduced in eq.~\reef{Euler3}.
Of course, since these surfaces never reach the asymptotic boundary, this quantity is finite, \ie there are no UV divergences in eq.~\reef{Agenbubble}. 

Extremizing eq.~\reef{Agenbubble} with respect to the radius of the bubble, we find
\beq\label{gamdot}
\partial_{\pb}\sgen=0 \qquad
\implies\qquad \frac{\pb}{\sqrt{1+\pb^2}}=-\lamb=-\frac{\Gbk}{2L\,\Gbr}\,.
\eeq
Now recall that we will always have $\Gbk>0$, but considered the possibility of $\Gbr$ becoming negative in section \ref{sec:examples}. Let us first consider the case $\lamb\ge 0$, which implies $1/\Gbr\ge 0$. In this case, we can not satisfy eq.~\reef{gamdot}, since both the bulk and brane contributions to the generalised entropy \eqref{Agenbubble} are positive and monotonically increasing functions of $\pb$. Therefore the minimum lies at $\pb=0$, \ie where the bubble collapses to zero size -- see figure \reef{figAbubble}. 

Of course, the more interesting scenario is when $\lamb$, and hence
$1/\Gbr$, are negative. Then eq.~\reef{gamdot} has the solution
\begin{align}
\pbo=-\frac{\lamb}{\sqrt{1-\lamb^2}}\,,
\label{cookie}
\end{align}
for which the generalized entropy \reef{Agenbubble} becomes
\begin{align}\label{Sbubble}
\sgen= \frac{\pi L^2}{\Gbk} \left( \sqrt{1-\lamb^2} - 1 \right)+2\,\lgb .
\end{align}
We note that these expressions are only sensible for $-1<\lamb<0$. In fact, for $\lamb<-1$, there is no minimum for the generalized entropy \reef{Agenbubble}, \ie there is no solution for eq.~\reef{gamdot}, and rather $\pb$ runs off to infinity -- see figure \reef{figAbubble}.  This is, perhaps, not so surprising since we can see from eq.~\reef{Newton34} that this regime is pathological, with the graviton localized on the brane becoming a ghost.

Therefore we only consider the regime $-1<\lamb<0$ where eqs.~\reef{cookie} and \reef{Sbubble} apply. As illustrated in figure \reef{figAbubble}, eq.~\reef{cookie} is indeed the global minimum of the generalized entropy \eqref{Agenbubble}. We might note that the sum of the bulk and brane terms in eq.~\reef{Sbubble} is negative. That is, the combined contributions of the two area terms in eq.~\reef{eq:sad} is in fact {\it negative!} Hence we only get a sensible (\ie positive) result for the generalized entropy \reef{Agenbubble} with the inclusion of the topological term \reef{Euler3} and with $\lgb$ sufficiently positive, which was also favoured in section \ref{sec:examples}.
%

\begin{figure}[h]
	\def\svgwidth{0.8\linewidth}
	\centering{
		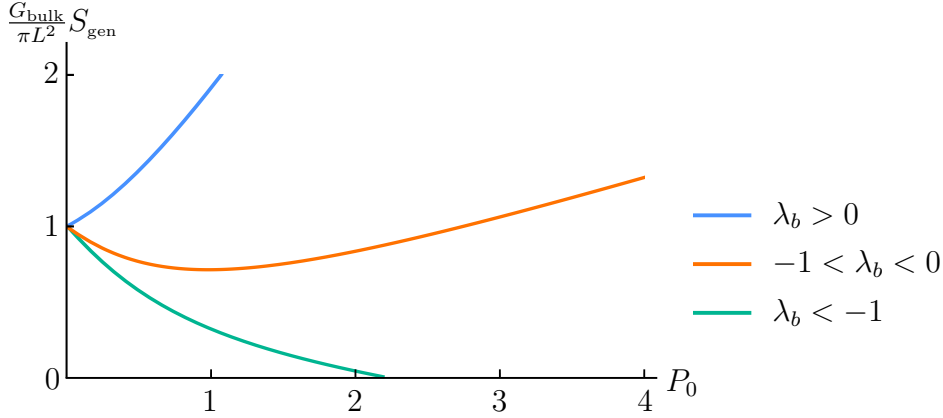
		\caption{The generalised area \eqref{Agenbubble} for a bubble as a function of its radius. For $\lamb>0$, the area is minimal for vanishing size, whereas for $-1<\lamb<0$ it has a finite size. For $\lamb<-1$, there is no global minimum, signalling an instability of the system. Further note that as $P_0$ approaches zero, $S_\mt{gen}\to \pi L^2/ G_\mt{bulk}$ since we have set $\lgb=\pi L^2/(2G_\mt{bulk})$.
		}
		\label{figAbubble}
	}
\end{figure}

These calculations are easily extended to higher dimensions, where eq.~\reef{Agenbubble} is replaced by
\beq\label{genbubble1}
S_\mt{gen}=\frac{L^{d-1}\,\Omega_{d-2}  }{2(d-1)\Gbk}\,\pb^{d-1}\ {}_{2}F_1\!\left[ \frac{1}{2},\frac{d-1}{2},\frac{d+1}{2},-\pb^2 \right]+\frac{L^{d-2}\, \Omega_{d-2}}{4\Gbr}\,\pb^{d-2}\,.
\eeq
We have not included contributions from any topological gravity terms in this expression for general $d$ -- see further comments below. To produce a qualitative understanding of this expression, 
we note that
\beq
  \hyperF\!\left[\frac{1}{2}, \frac{d-1}{2}, \frac{d+1}{2}, -\pb^2 \right]  \simeq\begin{cases}
  1 & \text{if $\pb\ll 1$}\,,
  \\
  \frac{d-1}{d-2}\frac{1}{\pb} &\text{if $\pb\gg 1$}\,.
\end{cases}
\label{eq:bubblegum}
\eeq
Now, we observe that for large $\pb$, the leading contribution in eq.~\reef{genbubble1} takes the expected form
\beq\label{expect4}
\sgen\simeq \frac{A(\sigma_\xR)}{4G_\mt{eff}} +\cdots \qquad{\rm where}\ \ \ \frac{A(\sigma_\xR)}{4G_\mt{eff}}=\frac{L^{d-1}\,\Omega_{d-2}  }{2(d-2)\Gbk}\,(1+\lamb)\,\pb^{d-2}\,,
\eeq
again using eq.~\eqref{newdefs}. Hence, there is a large penalty for having the RT surface meet the brane at a large radius $\pb$, which  will tend to push the intersection $\sigma_\xR$ to smaller radii. However, for small $\pb$, the bulk contribution to $\sgen$  grows like the volume, \ie it is proportional to $\pb^{d-1}$. Hence in this regime, the brane contribution dominates since it is proportional to $\lamb\pb^{d-2}$, and for $\lamb<0$, this term will favour  larger values of $\pb$. Hence for the interesting case of $\lamb<0$, we can expect that the generalized entropy for general $d$ is extremized at some finite value of $\pb$ of order $-\lamb$, just as we found for $d=3$. Of course, the denominator in eq.~\eqref{cookie} is also important for $\lambda_b$ close to $-1$, but this can not be seen with this simple qualitative analysis. Now, in fact, the extremality condition can in fact be solved exactly for any $d$. One finds
\beq\label{amazing}
  \partial_{\pb} S_\gen
  = \frac{L^{d-1}\,\Omega_{d-2}}{2 G_\bulk}\, \pb^{d-3}\left(
  \frac{\pb}{\sqrt{\pb^2+1}} + \lamb \right)=0\,.
\eeq
Of course, for $\lamb\ge0$, the only solution is $\pb=0$, \ie the bubble collapses to zero size, as expected. However, for $\lamb<0$, the minimum is given by $\pb=\pbo$, precisely the same critical  radius as in eq.~\eqref{cookie}. Substituting this critical radius into the generalized entropy \reef{genbubble1} does not yield any simplifications, however the result is easily evaluated numerically as a function of $\lamb$. Of course, the generalized entropy \reef{genbubble1} is negative at this minimum and so one would really need to add a topological term to the gravitational theory, either in the bulk or on the brane, to produce a sensible entropy, as we did for the $d=3$ example.

\subsection*{Wormholes and Cutoffs} 

The appearance of these extremal bubbles is quite unusual, of course. Since they are homologous to the entire boundary, this suggests that the ground state of the dual boundary system has an entropy by the standard RT prescription. The bulk construction makes clear that it is the conformal defect which introduces this large degeneracy of ground states.\footnote{As we see in figure \ref{figAbubble}, entropy associated with the zero-size bubble is nonvanishing and higher than that of the stable finite-size bubble due to the topological contribution. However, we note that it may be that the correct RT prescription is to choose `empty surface' in this case, giving zero entropy.} 

We should note, however, that the evaluation of this ground state entropy presented above is incomplete. In particular, there is a `zero mode' associated with these bubbles which allows them to be translated along the brane. Recall that while the empty AdS$_{d+1}$ geometry has an $SO(2,d)$ isometry (reflecting the conformal symmetry group of the boundary CFT), the backreacted brane geometry preserves an $SO(2,d-1)$ subgroup of these symmetries. Now our construction places the center of the bubbles at $P=0$, however, by acting with these symmetries, we can position the center anywhere on the brane. Further recall that one arrives at the RT prescription by evaluating (a particular limit of) a saddlepoint in the gravitational path integral \cite{Lewkowycz:2013nqa}. Hence we have discovered that there is a zero mode associated with the saddlepoints connected to the bubbles. Hence the integral over this zero mode would add a contribution to the entropy proportional to the logarithm of the (regulated) brane volume. It is interesting to speculate that this contribution may lift the negative value for $S_\mt{gen}(\pbo)$ to some positive entropy.

An essential feature required for the appearance of these bubbles was that the gravitational coupling associated with the DGP term \reef{newbran} was negative, \ie $1/G_\brane<0$. While this may seem unusual, let us note that integrating out quantum fields on the brane can produce either a positive or negative shift in Newton's constant. In particular, the shift is found to be negative for a $U(1)$ gauge field when $d<8$ \cite{Larsen:1995ax,Kabat:1995eq}. With the connection between the renormalization of Newton's constant and the area law contribution in entanglement entropy \cite{Callan:1994py,Susskind:1994sm}, this negative renormalization generates a puzzle which, however,  was finally resolved in terms of edge modes in \cite{Donnelly:2014fua,Donnelly:2015hxa}. There is a similar negative renormaliation for non-minimally coupled scalars \cite{Larsen:1995ax}, for which the resolution of the associated puzzle appears  in \cite{Faulkner:2013ana}. However, we should add that if we imagine $1/G_\brane<0$ is induced by additional quantum fields on the brane, then our entanglement entropy calculations are incomplete as they do not fully include the contributions of these extra fields. Hence our perspective here is to simply view the DGP term as a counterterm as would appear in the usual quantization of gravity on the brane, and in this context, the sign of $1/G_\brane$ is not proscribed but rather is chosen as needed to produce the `observed' value of $1/G_\mt{eff}$.

Another remark in this vein is that the bubble solutions appear as soon as $1/G_\brane$ is negative, \ie these solutions \reef{cookie} exist for very small values of $\lamb$ as long as $\lamb<0$. However, it is important to recall that the short distance cutoff is given by eq.~\reef{ctoffminus} in this regime. Hence combining  eqs.~\reef{cylindd} and \reef{cookie},  the areal size of the bubbles becomes
\beq
L\pbo=\frac{|\lamb|\,L}{\sqrt{1-\lamb^2}}\simeq \frac{|\lamb|}{\sqrt{1+|\lamb|}}\,\tilde\delta\,.
\label{haiku2}
\eeq 
where we have substituted eq.~\eqref{haiku} in the second expression.
This expression approaches the maximum size $\tilde\delta/\sqrt{2}$ as $\lamb\to-1$. That is, the radius of bubbles is always smaller than the cutoff scale $\tilde\delta$ on the brane! Therefore, these solutions are not reliable in the regime where Einstein gravity gives a good description of the brane.  On the other hand, our calculations in this appendix involved evaluating RT surfaces in the bulk, \ie they only depended on bulk perspective. Further, for $|\lamb|\gtrsim1/\sqrt{2}$, the corresponding RT surfaces grow much larger than the bulk AdS scale, and so would be seen as valid solutions. 
However, one may ask if there are physical constraints which will not allow us to realize theories with $\lamb$ which are that negative and so prevent us from considering scenarios where these bubbles have a macroscopic size. 

We close here with two final remarks: These bubbles are a remnant of replica wormholes in the limit $n\to1$ \cite{Penington:2019kki,Hartman:2020swn}. In the discussion section, we explore if there are any lessons that they may hold for the new discussions of baby universes and ensembles \cite{Marolf:2020xie}. Another comment is that the bubble surfaces produce an interesting
entanglement wedge, which extends to a band  covering a finite time interval on the boundary. Of course, this is reminiscent of the holographic construction of differential entropy \cite{Balasubramanian:2013rqa,Balasubramanian:2013lsa,Czech:2014wka,Myers:2014jia,Headrick:2014eia}, which can be used to evaluate the area of closed surfaces in the bulk. It would be interesting to examine these connections further.




\bibliography{bibliography}

\providecommand{\href}[2]{#2}\begingroup\raggedright\begin{thebibliography}{100}

\bibitem{Hawking:1974sw}
S.~Hawking, ``{Particle Creation by Black Holes},''
  \href{http://dx.doi.org/10.1007/BF02345020}{{\em Commun. Math. Phys.}
  {\bfseries 43} (1975) 199--220}. [Erratum: Commun.Math.Phys. 46, 206 (1976)].

\bibitem{Hawking:1974rv}
S.~Hawking, ``{Black hole explosions},''
  \href{http://dx.doi.org/10.1038/248030a0}{{\em Nature} {\bfseries 248} (1974)
  30--31}.

\bibitem{Hawking:1976de}
S.~Hawking, ``{Black Holes and Thermodynamics},''
  \href{http://dx.doi.org/10.1103/PhysRevD.13.191}{{\em Phys. Rev. D}
  {\bfseries 13} (1976) 191--197}.

\bibitem{Bekenstein:1972tm}
J.~Bekenstein, ``{Black holes and the second law},''
  \href{http://dx.doi.org/10.1007/BF02757029}{{\em Lett. Nuovo Cim.} {\bfseries
  4} (1972) 737--740}.

\bibitem{Bekenstein:1973ur}
J.~D. Bekenstein, ``{Black holes and entropy},''
  \href{http://dx.doi.org/10.1103/PhysRevD.7.2333}{{\em Phys. Rev. D}
  {\bfseries 7} (1973) 2333--2346}.

\bibitem{Ryu:2006ef}
S.~Ryu and T.~Takayanagi, ``{Aspects of Holographic Entanglement Entropy},''
  \href{http://dx.doi.org/10.1088/1126-6708/2006/08/045}{{\em JHEP} {\bfseries
  08} (2006) 045},
\href{http://arxiv.org/abs/hep-th/0605073}{{\ttfamily arXiv:hep-th/0605073
  [hep-th]}}.

\bibitem{Ryu:2006bv}
S.~Ryu and T.~Takayanagi, ``{Holographic derivation of entanglement entropy
  from AdS/CFT},'' \href{http://dx.doi.org/10.1103/PhysRevLett.96.181602}{{\em
  Phys. Rev. Lett.} {\bfseries 96} (2006) 181602},
\href{http://arxiv.org/abs/hep-th/0603001}{{\ttfamily arXiv:hep-th/0603001
  [hep-th]}}.

\bibitem{Hubeny:2007xt}
V.~E. Hubeny, M.~Rangamani, and T.~Takayanagi, ``{A Covariant holographic
  entanglement entropy proposal},''
  \href{http://dx.doi.org/10.1088/1126-6708/2007/07/062}{{\em JHEP} {\bfseries
  07} (2007) 062},
\href{http://arxiv.org/abs/0705.0016}{{\ttfamily arXiv:0705.0016 [hep-th]}}.

\bibitem{Rangamani:2016dms}
M.~Rangamani and T.~Takayanagi, ``{Holographic Entanglement Entropy},''
  \href{http://dx.doi.org/10.1007/978-3-319-52573-0}{{\em Lect. Notes Phys.}
  {\bfseries 931} (2017) pp.1--246},
\href{http://arxiv.org/abs/1609.01287}{{\ttfamily arXiv:1609.01287 [hep-th]}}.

\bibitem{Lewkowycz:2013nqa}
A.~Lewkowycz and J.~Maldacena, ``{Generalized gravitational entropy},''
  \href{http://dx.doi.org/10.1007/JHEP08(2013)090}{{\em JHEP} {\bfseries 08}
  (2013) 090},
\href{http://arxiv.org/abs/1304.4926}{{\ttfamily arXiv:1304.4926 [hep-th]}}.

\bibitem{Hawking:1976ra}
S.~Hawking, ``{Breakdown of Predictability in Gravitational Collapse},''
  \href{http://dx.doi.org/10.1103/PhysRevD.14.2460}{{\em Phys. Rev. D}
  {\bfseries 14} (1976) 2460--2473}.

\bibitem{Polchinski:2016hrw}
J.~Polchinski, \href{http://dx.doi.org/10.1142/9789813149441\_0006}{``{The
  Black Hole Information Problem},''} in {\em {Theoretical Advanced Study
  Institute in Elementary Particle Physics}: {New Frontiers in Fields and
  Strings}}, pp.~353--397.
\newblock 2017.
\newblock \href{http://arxiv.org/abs/1609.04036}{{\ttfamily arXiv:1609.04036
  [hep-th]}}.

\bibitem{Harlow:2014yka}
D.~Harlow, ``{Jerusalem Lectures on Black Holes and Quantum Information},''
  \href{http://dx.doi.org/10.1103/RevModPhys.88.015002}{{\em Rev. Mod. Phys.}
  {\bfseries 88} (2016) 015002},
  \href{http://arxiv.org/abs/1409.1231}{{\ttfamily arXiv:1409.1231 [hep-th]}}.

\bibitem{Page:1993wv}
D.~N. Page, ``{Information in black hole radiation},''
  \href{http://dx.doi.org/10.1103/PhysRevLett.71.3743}{{\em Phys. Rev. Lett.}
  {\bfseries 71} (1993) 3743--3746},
  \href{http://arxiv.org/abs/hep-th/9306083}{{\ttfamily arXiv:hep-th/9306083}}.

\bibitem{Bekenstein:1974ax}
J.~D. Bekenstein, ``{Generalized second law of thermodynamics in black hole
  physics},'' \href{http://dx.doi.org/10.1103/PhysRevD.9.3292}{{\em Phys. Rev.
  D} {\bfseries 9} (1974) 3292--3300}.

\bibitem{Wall:2009wm}
A.~C. Wall, ``{Ten Proofs of the Generalized Second Law},''
  \href{http://dx.doi.org/10.1088/1126-6708/2009/06/021}{{\em JHEP} {\bfseries
  06} (2009) 021}, \href{http://arxiv.org/abs/0901.3865}{{\ttfamily
  arXiv:0901.3865 [gr-qc]}}.

\bibitem{Wall:2011hj}
A.~C. Wall, ``{A proof of the generalized second law for rapidly changing
  fields and arbitrary horizon slices},''
  \href{http://dx.doi.org/10.1103/PhysRevD.85.104049}{{\em Phys. Rev. D}
  {\bfseries 85} (2012) 104049},
  \href{http://arxiv.org/abs/1105.3445}{{\ttfamily arXiv:1105.3445 [gr-qc]}}.
  [Erratum: Phys.Rev.D 87, 069904 (2013)].

\bibitem{Faulkner:2013ana}
T.~Faulkner, A.~Lewkowycz, and J.~Maldacena, ``{Quantum corrections to
  holographic entanglement entropy},''
  \href{http://dx.doi.org/10.1007/JHEP11(2013)074}{{\em JHEP} {\bfseries 11}
  (2013) 074},
\href{http://arxiv.org/abs/1307.2892}{{\ttfamily arXiv:1307.2892 [hep-th]}}.

\bibitem{Engelhardt:2014gca}
N.~Engelhardt and A.~C. Wall, ``{Quantum Extremal Surfaces: Holographic
  Entanglement Entropy beyond the Classical Regime},''
  \href{http://dx.doi.org/10.1007/JHEP01(2015)073}{{\em JHEP} {\bfseries 01}
  (2015) 073},
\href{http://arxiv.org/abs/1408.3203}{{\ttfamily arXiv:1408.3203 [hep-th]}}.

\bibitem{Almheiri:2019psf}
A.~Almheiri, N.~Engelhardt, D.~Marolf, and H.~Maxfield, ``{The entropy of bulk
  quantum fields and the entanglement wedge of an evaporating black hole},''
  \href{http://dx.doi.org/10.1007/JHEP12(2019)063}{{\em JHEP} {\bfseries 12}
  (2019) 063},
\href{http://arxiv.org/abs/1905.08762}{{\ttfamily arXiv:1905.08762 [hep-th]}}.

\bibitem{Penington:2019npb}
G.~Penington, ``{Entanglement Wedge Reconstruction and the Information
  Paradox},''
\href{http://arxiv.org/abs/1905.08255}{{\ttfamily arXiv:1905.08255 [hep-th]}}.

\bibitem{Almheiri:2019hni}
A.~Almheiri, R.~Mahajan, J.~Maldacena, and Y.~Zhao, ``{The Page curve of
  Hawking radiation from semiclassical geometry},''
  \href{http://dx.doi.org/10.1007/JHEP03(2020)149}{{\em JHEP} {\bfseries 03}
  (2020) 149}, \href{http://arxiv.org/abs/1908.10996}{{\ttfamily
  arXiv:1908.10996 [hep-th]}}.

\bibitem{Almheiri:2019yqk}
A.~Almheiri, R.~Mahajan, and J.~Maldacena, ``{Islands outside the horizon},''
\href{http://arxiv.org/abs/1910.11077}{{\ttfamily arXiv:1910.11077 [hep-th]}}.

\bibitem{Almheiri:2019psy}
A.~Almheiri, R.~Mahajan, and J.~E. Santos, ``{Entanglement islands in higher
  dimensions},''
\href{http://arxiv.org/abs/1911.09666}{{\ttfamily arXiv:1911.09666 [hep-th]}}.

\bibitem{Almheiri:2019qdq}
A.~Almheiri, T.~Hartman, J.~Maldacena, E.~Shaghoulian, and A.~Tajdini,
  ``{Replica Wormholes and the Entropy of Hawking Radiation},''
\href{http://arxiv.org/abs/1911.12333}{{\ttfamily arXiv:1911.12333 [hep-th]}}.

\bibitem{Penington:2019kki}
G.~Penington, S.~H. Shenker, D.~Stanford, and Z.~Yang, ``{Replica wormholes and
  the black hole interior},''
\href{http://arxiv.org/abs/1911.11977}{{\ttfamily arXiv:1911.11977 [hep-th]}}.

\bibitem{Akers:2019nfi}
C.~Akers, N.~Engelhardt, and D.~Harlow, ``{Simple holographic models of black
  hole evaporation},'' \href{http://arxiv.org/abs/1910.00972}{{\ttfamily
  arXiv:1910.00972 [hep-th]}}.

\bibitem{Rozali:2019day}
M.~Rozali, J.~Sully, M.~Van~Raamsdonk, C.~Waddell, and D.~Wakeham,
  ``{Information radiation in BCFT models of black holes},''
  \href{http://dx.doi.org/10.1007/JHEP05(2020)004}{{\em JHEP} {\bfseries 05}
  (2020) 004}, \href{http://arxiv.org/abs/1910.12836}{{\ttfamily
  arXiv:1910.12836 [hep-th]}}.

\bibitem{Chen:2019uhq}
H.~Z. Chen, Z.~Fisher, J.~Hernandez, R.~C. Myers, and S.-M. Ruan,
  ``{Information Flow in Black Hole Evaporation},''
\href{http://arxiv.org/abs/1911.03402}{{\ttfamily arXiv:1911.03402 [hep-th]}}.

\bibitem{Bousso:2019ykv}
R.~Bousso and M.~Toma\v~sevi\'c, ``{Unitarity From a Smooth Horizon?},''
  \href{http://arxiv.org/abs/1911.06305}{{\ttfamily arXiv:1911.06305
  [hep-th]}}.

\bibitem{Gautason:2020tmk}
F.~F. Gautason, L.~Schneiderbauer, W.~Sybesma, and L.~Thorlacius, ``{Page Curve
  for an Evaporating Black Hole},''
  \href{http://arxiv.org/abs/2004.00598}{{\ttfamily arXiv:2004.00598
  [hep-th]}}.

\bibitem{Hartman:2020swn}
T.~Hartman, E.~Shaghoulian, and A.~Strominger, ``{Islands in Asymptotically
  Flat 2D Gravity},'' \href{http://arxiv.org/abs/2004.13857}{{\ttfamily
  arXiv:2004.13857 [hep-th]}}.

\bibitem{Marolf:2020xie}
D.~Marolf and H.~Maxfield, ``{Transcending the ensemble: baby universes,
  spacetime wormholes, and the order and disorder of black hole information},''
\href{http://arxiv.org/abs/2002.08950}{{\ttfamily arXiv:2002.08950 [hep-th]}}.

\bibitem{Hollowood:2020cou}
T.~J. Hollowood and S.~P. Kumar, ``{Islands and Page Curves for Evaporating
  Black Holes in JT Gravity},''
  \href{http://arxiv.org/abs/2004.14944}{{\ttfamily arXiv:2004.14944
  [hep-th]}}.

\bibitem{Anegawa:2020ezn}
T.~Anegawa and N.~Iizuka, ``{Notes on islands in asymptotically flat 2d dilaton
  black holes},'' \href{http://arxiv.org/abs/2004.01601}{{\ttfamily
  arXiv:2004.01601 [hep-th]}}.

\bibitem{Hashimoto:2020cas}
K.~Hashimoto, N.~Iizuka, and Y.~Matsuo, ``{Islands in Schwarzschild black
  holes},'' \href{http://arxiv.org/abs/2004.05863}{{\ttfamily arXiv:2004.05863
  [hep-th]}}.

\bibitem{Sully:2020pza}
J.~Sully, M.~Van~Raamsdonk, and D.~Wakeham, ``{BCFT entanglement entropy at
  large central charge and the black hole interior},''
  \href{http://arxiv.org/abs/2004.13088}{{\ttfamily arXiv:2004.13088
  [hep-th]}}.

\bibitem{Balasubramanian:2020hfs}
V.~Balasubramanian, A.~Kar, O.~Parrikar, G.~Sárosi, and T.~Ugajin,
  ``{Geometric secret sharing in a model of Hawking radiation},''
  \href{http://arxiv.org/abs/2003.05448}{{\ttfamily arXiv:2003.05448
  [hep-th]}}.

\bibitem{Alishahiha:2020qza}
M.~Alishahiha, A.~Faraji~Astaneh, and A.~Naseh, ``{Island in the Presence of
  Higher Derivative Terms},'' \href{http://arxiv.org/abs/2005.08715}{{\ttfamily
  arXiv:2005.08715 [hep-th]}}.

\bibitem{Geng:2020qvw}
H.~Geng and A.~Karch, ``{Massive Islands},''
  \href{http://arxiv.org/abs/2006.02438}{{\ttfamily arXiv:2006.02438
  [hep-th]}}.

\bibitem{Krishnan:2020oun}
C.~Krishnan, V.~Patil, and J.~Pereira, ``{Page Curve and the Information
  Paradox in Flat Space},'' \href{http://arxiv.org/abs/2005.02993}{{\ttfamily
  arXiv:2005.02993 [hep-th]}}.

\bibitem{Randall:1999ee}
L.~Randall and R.~Sundrum, ``{A Large mass hierarchy from a small extra
  dimension},'' \href{http://dx.doi.org/10.1103/PhysRevLett.83.3370}{{\em Phys.
  Rev. Lett.} {\bfseries 83} (1999) 3370--3373},
\href{http://arxiv.org/abs/hep-ph/9905221}{{\ttfamily arXiv:hep-ph/9905221
  [hep-ph]}}.

\bibitem{Randall:1999vf}
L.~Randall and R.~Sundrum, ``{An Alternative to compactification},''
  \href{http://dx.doi.org/10.1103/PhysRevLett.83.4690}{{\em Phys. Rev. Lett.}
  {\bfseries 83} (1999) 4690--4693},
\href{http://arxiv.org/abs/hep-th/9906064}{{\ttfamily arXiv:hep-th/9906064
  [hep-th]}}.

\bibitem{Karch:2000ct}
A.~Karch and L.~Randall, ``{Locally localized gravity},''
  \href{http://dx.doi.org/10.1088/1126-6708/2001/05/008}{{\em JHEP} {\bfseries
  05} (2001) 008}, \href{http://arxiv.org/abs/hep-th/0011156}{{\ttfamily
  arXiv:hep-th/0011156}}.

\bibitem{Dvali:2000hr}
G.~R. Dvali, G.~Gabadadze, and M.~Porrati, ``{4-D gravity on a brane in 5-D
  Minkowski space},''
  \href{http://dx.doi.org/10.1016/S0370-2693(00)00669-9}{{\em Phys. Lett.}
  {\bfseries B485} (2000) 208--214},
\href{http://arxiv.org/abs/hep-th/0005016}{{\ttfamily arXiv:hep-th/0005016
  [hep-th]}}.

\bibitem{Myers:2010tj}
R.~C. Myers and A.~Sinha, ``{Holographic c-theorems in arbitrary dimensions},''
  \href{http://dx.doi.org/10.1007/JHEP01(2011)125}{{\em JHEP} {\bfseries 01}
  (2011) 125},
\href{http://arxiv.org/abs/1011.5819}{{\ttfamily arXiv:1011.5819 [hep-th]}}.

\bibitem{QEI}
H.~Z. Chen, R.~C. Myers, D.~Neuenfeld, I.~A. Reyes, and J.~Sandor, ``{Quantum
  Extremal Islands Made Easy, Part II: Black Holes on the Brane},'' {\em in
  preparation} .

\bibitem{PhysRevLett.28.1082}
J.~W. York, ``Role of conformal three-geometry in the dynamics of
  gravitation,'' \href{http://dx.doi.org/10.1103/PhysRevLett.28.1082}{{\em
  Phys. Rev. Lett.} {\bfseries 28} (Apr, 1972) 1082--1085}.
  \url{http://link.aps.org/doi/10.1103/PhysRevLett.28.1082}.

\bibitem{Gibbons:1976ue}
G.~W. Gibbons and S.~W. Hawking, ``{Action Integrals and Partition Functions in
  Quantum Gravity},''
\href{http://dx.doi.org/10.1103/PhysRevD.15.2752}{{\em Phys. Rev.} {\bfseries
  D15} (1977) 2752--2756}.

\bibitem{Emparan:1999pm}
R.~Emparan, C.~V. Johnson, and R.~C. Myers, ``{Surface terms as counterterms in
  the AdS / CFT correspondence},''
  \href{http://dx.doi.org/10.1103/PhysRevD.60.104001}{{\em Phys. Rev.}
  {\bfseries D60} (1999) 104001},
\href{http://arxiv.org/abs/hep-th/9903238}{{\ttfamily arXiv:hep-th/9903238
  [hep-th]}}.

\bibitem{FG}
C.~Fefferman and C.~R. Graham, ``{Conformal invariants},'' {\em in {\it The
  Mathematical Heritage of \'Elie Cartan (Lyon, 1984)}, Ast\'erisque,}
  {\bfseries {\rm Numero Hors Serie}} (1985) 95--116.

\bibitem{Fefferman:2007rka}
C.~Fefferman and C.~R. Graham, ``{The ambient metric},'' {\em Ann. Math. Stud.}
  {\bfseries 178} (2011) 1--128,
\href{http://arxiv.org/abs/0710.0919}{{\ttfamily arXiv:0710.0919 [math.DG]}}.

\bibitem{israel1966singular}
W.~Israel, ``Singular hypersurfaces and thin shells in general relativity,''
  {\em Il Nuovo Cimento B} {\bfseries 44} no.~1, (1966) 1--14.

\bibitem{Misner:1974qy}
C.~W. Misner, K.~S. Thorne, and J.~A. Wheeler, {\em {Gravitation}}.
\newblock W. H. Freeman, San Francisco,
1973.
\newblock

\bibitem{Skenderis:2002wp}
K.~Skenderis, ``{Lecture notes on holographic renormalization},''
  \href{http://dx.doi.org/10.1088/0264-9381/19/22/306}{{\em Class. Quant.
  Grav.} {\bfseries 19} (2002) 5849--5876},
\href{http://arxiv.org/abs/hep-th/0209067}{{\ttfamily arXiv:hep-th/0209067
  [hep-th]}}.

\bibitem{deHaro:2000vlm}
S.~de~Haro, S.~N. Solodukhin, and K.~Skenderis, ``{Holographic reconstruction
  of space-time and renormalization in the AdS/CFT correspondence},''
  \href{http://dx.doi.org/10.1007/s002200100381}{{\em Commun. Math. Phys.}
  {\bfseries 217} (2001) 595--622},
\href{http://arxiv.org/abs/hep-th/0002230}{{\ttfamily arXiv:hep-th/0002230
  [hep-th]}}.

\bibitem{deHaro:2000wj}
S.~de~Haro, K.~Skenderis, and S.~N. Solodukhin, ``{Gravity in warped
  compactifications and the holographic stress tensor},''
  \href{http://dx.doi.org/10.1088/0264-9381/18/16/307}{{\em Class. Quant.
  Grav.} {\bfseries 18} (2001) 3171--3180},
\href{http://arxiv.org/abs/hep-th/0011230}{{\ttfamily arXiv:hep-th/0011230
  [hep-th]}}.

\bibitem{Shiromizu:1999wj}
T.~Shiromizu, K.-i. Maeda, and M.~Sasaki, ``{The Einstein equation on the
  3-brane world},'' \href{http://dx.doi.org/10.1103/PhysRevD.62.024012}{{\em
  Phys. Rev.} {\bfseries D62} (2000) 024012},
\href{http://arxiv.org/abs/gr-qc/9910076}{{\ttfamily arXiv:gr-qc/9910076
  [gr-qc]}}.

\bibitem{Verlinde:1999fy}
H.~L. Verlinde, ``{Holography and compactification},''
  \href{http://dx.doi.org/10.1016/S0550-3213(00)00224-8}{{\em Nucl. Phys.}
  {\bfseries B580} (2000) 264--274},
\href{http://arxiv.org/abs/hep-th/9906182}{{\ttfamily arXiv:hep-th/9906182
  [hep-th]}}.

\bibitem{Gubser:1999vj}
S.~S. Gubser, ``{AdS / CFT and gravity},''
  \href{http://dx.doi.org/10.1103/PhysRevD.63.084017}{{\em Phys. Rev.}
  {\bfseries D63} (2001) 084017},
\href{http://arxiv.org/abs/hep-th/9912001}{{\ttfamily arXiv:hep-th/9912001
  [hep-th]}}.

\bibitem{Karch:2001jb}
A.~Karch, E.~Katz, and L.~Randall, ``{Absence of a VVDZ discontinuity in
  AdS(AdS)},'' \href{http://dx.doi.org/10.1088/1126-6708/2001/12/016}{{\em
  JHEP} {\bfseries 12} (2001) 016},
  \href{http://arxiv.org/abs/hep-th/0106261}{{\ttfamily arXiv:hep-th/0106261}}.

\bibitem{Porrati:2001gx}
M.~Porrati, ``{Mass and gauge invariance 4. Holography for the Karch-Randall
  model},'' \href{http://dx.doi.org/10.1103/PhysRevD.65.044015}{{\em Phys. Rev.
  D} {\bfseries 65} (2002) 044015},
  \href{http://arxiv.org/abs/hep-th/0109017}{{\ttfamily arXiv:hep-th/0109017}}.

\bibitem{Emparan:2006ni}
R.~Emparan, ``{Black hole entropy as entanglement entropy: A Holographic
  derivation},'' \href{http://dx.doi.org/10.1088/1126-6708/2006/06/012}{{\em
  JHEP} {\bfseries 06} (2006) 012},
\href{http://arxiv.org/abs/hep-th/0603081}{{\ttfamily arXiv:hep-th/0603081
  [hep-th]}}.

\bibitem{Myers:2013lva}
R.~C. Myers, R.~Pourhasan, and M.~Smolkin, ``{On Spacetime Entanglement},''
  \href{http://dx.doi.org/10.1007/JHEP06(2013)013}{{\em JHEP} {\bfseries 06}
  (2013) 013},
\href{http://arxiv.org/abs/1304.2030}{{\ttfamily arXiv:1304.2030 [hep-th]}}.

\bibitem{Skenderis:1999nb}
K.~Skenderis and S.~N. Solodukhin, ``{Quantum effective action from the AdS /
  CFT correspondence},''
  \href{http://dx.doi.org/10.1016/S0370-2693(99)01467-7}{{\em Phys. Lett.}
  {\bfseries B472} (2000) 316--322},
\href{http://arxiv.org/abs/hep-th/9910023}{{\ttfamily arXiv:hep-th/9910023
  [hep-th]}}.

\bibitem{Henningson:1998gx}
M.~Henningson and K.~Skenderis, ``{The Holographic Weyl anomaly},''
  \href{http://dx.doi.org/10.1088/1126-6708/1998/07/023}{{\em JHEP} {\bfseries
  07} (1998) 023},
\href{http://arxiv.org/abs/hep-th/9806087}{{\ttfamily arXiv:hep-th/9806087
  [hep-th]}}.

\bibitem{Henningson:1998ey}
M.~Henningson and K.~Skenderis, ``{Holography and the Weyl anomaly},''
  \href{http://dx.doi.org/10.1002/(SICI)1521-3978(20001)48:1/3<125::AID-PROP125>3.0.CO;2-B,
  10.1002/(SICI)1521-3978(20001)48:1/3<125::AID-PROP125>3.3.CO;2-2}{{\em
  Fortsch. Phys.} {\bfseries 48} (2000) 125--128},
\href{http://arxiv.org/abs/hep-th/9812032}{{\ttfamily arXiv:hep-th/9812032
  [hep-th]}}.

\bibitem{Burgess:1999vb}
C.~P. Burgess, N.~R. Constable, and R.~C. Myers, ``{The Free energy of N=4
  superYang-Mills and the AdS / CFT correspondence},''
  \href{http://dx.doi.org/10.1088/1126-6708/1999/08/017}{{\em JHEP} {\bfseries
  08} (1999) 017},
\href{http://arxiv.org/abs/hep-th/9907188}{{\ttfamily arXiv:hep-th/9907188
  [hep-th]}}.

\bibitem{Alvarez:1982zi}
O.~Alvarez, ``{Theory of Strings with Boundaries: Fluctuations, Topology, and
  Quantum Geometry},''
\href{http://dx.doi.org/10.1016/0550-3213(83)90490-X}{{\em Nucl. Phys.}
  {\bfseries B216} (1983) 125--184}.

\bibitem{Frolov:1996hd}
V.~P. Frolov, W.~Israel, and S.~N. Solodukhin, ``{On one loop quantum
  corrections to the thermodynamics of charged black holes},''
  \href{http://dx.doi.org/10.1103/PhysRevD.54.2732}{{\em Phys. Rev.} {\bfseries
  D54} (1996) 2732--2745},
\href{http://arxiv.org/abs/hep-th/9602105}{{\ttfamily arXiv:hep-th/9602105
  [hep-th]}}.

\bibitem{Duff:1977ay}
M.~J. Duff, ``{Observations on Conformal Anomalies},''
\href{http://dx.doi.org/10.1016/0550-3213(77)90410-2}{{\em Nucl. Phys.}
  {\bfseries B125} (1977) 334--348}.

\bibitem{Duff:1993wm}
M.~J. Duff, ``{Twenty years of the Weyl anomaly},''
  \href{http://dx.doi.org/10.1088/0264-9381/11/6/004}{{\em Class. Quant. Grav.}
  {\bfseries 11} (1994) 1387--1404},
\href{http://arxiv.org/abs/hep-th/9308075}{{\ttfamily arXiv:hep-th/9308075
  [hep-th]}}.

\bibitem{Chen:2019iro}
Y.~Chen, ``{Pulling Out the Island with Modular Flow},''
\href{http://arxiv.org/abs/1912.02210}{{\ttfamily arXiv:1912.02210 [hep-th]}}.

\bibitem{Jackiw:1984je}
R.~Jackiw, ``{Lower Dimensional Gravity},''
\href{http://dx.doi.org/10.1016/0550-3213(85)90448-1}{{\em Nucl. Phys.}
  {\bfseries B252} (1985) 343--356}.

\bibitem{Teitelboim:1983ux}
C.~Teitelboim, ``{Gravitation and Hamiltonian Structure in Two Space-Time
  Dimensions},''
\href{http://dx.doi.org/10.1016/0370-2693(83)90012-6}{{\em Phys. Lett.}
  {\bfseries 126B} (1983) 41--45}.

\bibitem{Maldacena:2016upp}
J.~Maldacena, D.~Stanford, and Z.~Yang, ``{Conformal symmetry and its breaking
  in two dimensional Nearly Anti-de-Sitter space},''
  \href{http://dx.doi.org/10.1093/ptep/ptw124}{{\em PTEP} {\bfseries 2016}
  no.~12, (2016) 12C104},
\href{http://arxiv.org/abs/1606.01857}{{\ttfamily arXiv:1606.01857 [hep-th]}}.

\bibitem{domino}
D.~Neuenfeld and Others {\em in preparation} .

\bibitem{Takayanagi:2011zk}
T.~Takayanagi, ``{Holographic Dual of BCFT},''
  \href{http://dx.doi.org/10.1103/PhysRevLett.107.101602}{{\em Phys. Rev.
  Lett.} {\bfseries 107} (2011) 101602},
\href{http://arxiv.org/abs/1105.5165}{{\ttfamily arXiv:1105.5165 [hep-th]}}.

\bibitem{Fujita:2011fp}
M.~Fujita, T.~Takayanagi, and E.~Tonni, ``{Aspects of AdS/BCFT},''
  \href{http://dx.doi.org/10.1007/JHEP11(2011)043}{{\em JHEP} {\bfseries 11}
  (2011) 043}, \href{http://arxiv.org/abs/1108.5152}{{\ttfamily arXiv:1108.5152
  [hep-th]}}.

\bibitem{Porrati:2001db}
M.~Porrati, ``{Higgs phenomenon for 4-D gravity in anti-de Sitter space},''
  \href{http://dx.doi.org/10.1088/1126-6708/2002/04/058}{{\em JHEP} {\bfseries
  04} (2002) 058}, \href{http://arxiv.org/abs/hep-th/0112166}{{\ttfamily
  arXiv:hep-th/0112166}}.

\bibitem{Miemiec:2000eq}
A.~Miemiec, ``{A Power law for the lowest eigenvalue in localized massive
  gravity},''
  \href{http://dx.doi.org/10.1002/1521-3978(200107)49:7<747::AID-PROP747>3.0.CO;2-T}{{\em
  Fortsch. Phys.} {\bfseries 49} (2001) 747--755},
  \href{http://arxiv.org/abs/hep-th/0011160}{{\ttfamily arXiv:hep-th/0011160}}.

\bibitem{Schwartz:2000ip}
M.~D. Schwartz, ``{The Emergence of localized gravity},''
  \href{http://dx.doi.org/10.1016/S0370-2693(01)00152-6}{{\em Phys. Lett. B}
  {\bfseries 502} (2001) 223--228},
  \href{http://arxiv.org/abs/hep-th/0011177}{{\ttfamily arXiv:hep-th/0011177}}.

\bibitem{Buchel:2009sk}
A.~Buchel, J.~Escobedo, R.~C. Myers, M.~F. Paulos, A.~Sinha, and M.~Smolkin,
  ``{Holographic GB gravity in arbitrary dimensions},''
  \href{http://dx.doi.org/10.1007/JHEP03(2010)111}{{\em JHEP} {\bfseries 03}
  (2010) 111},
\href{http://arxiv.org/abs/0911.4257}{{\ttfamily arXiv:0911.4257 [hep-th]}}.

\bibitem{Dvali:2007hz}
G.~Dvali, ``{Black Holes and Large N Species Solution to the Hierarchy
  Problem},'' \href{http://dx.doi.org/10.1002/prop.201000009}{{\em Fortsch.
  Phys.} {\bfseries 58} (2010) 528--536},
\href{http://arxiv.org/abs/0706.2050}{{\ttfamily arXiv:0706.2050 [hep-th]}}.

\bibitem{Dvali:2007wp}
G.~Dvali and M.~Redi, ``{Black Hole Bound on the Number of Species and Quantum
  Gravity at LHC},'' \href{http://dx.doi.org/10.1103/PhysRevD.77.045027}{{\em
  Phys. Rev.} {\bfseries D77} (2008) 045027},
\href{http://arxiv.org/abs/0710.4344}{{\ttfamily arXiv:0710.4344 [hep-th]}}.

\bibitem{Reeb:2009rm}
D.~Reeb, ``{Running of Newton's Constant and Quantum Gravitational Effects},''
  \href{http://dx.doi.org/10.1142/9789814340212_0029}{{\em Subnucl. Ser.}
  {\bfseries 46} (2011) 651--660},
\href{http://arxiv.org/abs/0901.2963}{{\ttfamily arXiv:0901.2963 [hep-th]}}.

\bibitem{Aharony:2003qf}
O.~Aharony, O.~DeWolfe, D.~Z. Freedman, and A.~Karch, ``{Defect conformal field
  theory and locally localized gravity},''
  \href{http://dx.doi.org/10.1088/1126-6708/2003/07/030}{{\em JHEP} {\bfseries
  07} (2003) 030}, \href{http://arxiv.org/abs/hep-th/0303249}{{\ttfamily
  arXiv:hep-th/0303249}}.

\bibitem{EW1}
B.~Czech, J.~L. Karczmarek, F.~Nogueira, and M.~Van~Raamsdonk, ``{The Gravity
  Dual of a Density Matrix},''
  \href{http://dx.doi.org/10.1088/0264-9381/29/15/155009}{{\em Class. Quant.
  Grav.} {\bfseries 29} (2012) 155009},
\href{http://arxiv.org/abs/1204.1330}{{\ttfamily arXiv:1204.1330 [hep-th]}}.

\bibitem{EW2}
M.~Headrick, V.~E. Hubeny, A.~Lawrence, and M.~Rangamani, ``{Causality \&
  holographic entanglement entropy},''
  \href{http://dx.doi.org/10.1007/JHEP12(2014)162}{{\em JHEP} {\bfseries 12}
  (2014) 162},
\href{http://arxiv.org/abs/1408.6300}{{\ttfamily arXiv:1408.6300 [hep-th]}}.

\bibitem{EW3}
A.~C. Wall, ``{Maximin Surfaces, and the Strong Subadditivity of the Covariant
  Holographic Entanglement Entropy},''
  \href{http://dx.doi.org/10.1088/0264-9381/31/22/225007}{{\em Class. Quant.
  Grav.} {\bfseries 31} no.~22, (2014) 225007},
\href{http://arxiv.org/abs/1211.3494}{{\ttfamily arXiv:1211.3494 [hep-th]}}.

\bibitem{Jafferis:2015del}
D.~L. Jafferis, A.~Lewkowycz, J.~Maldacena, and S.~J. Suh, ``{Relative entropy
  equals bulk relative entropy},''
  \href{http://dx.doi.org/10.1007/JHEP06(2016)004}{{\em JHEP} {\bfseries 06}
  (2016) 004},
\href{http://arxiv.org/abs/1512.06431}{{\ttfamily arXiv:1512.06431 [hep-th]}}.

\bibitem{Dong:2016eik}
X.~Dong, D.~Harlow, and A.~C. Wall, ``{Reconstruction of Bulk Operators within
  the Entanglement Wedge in Gauge-Gravity Duality},''
  \href{http://dx.doi.org/10.1103/PhysRevLett.117.021601}{{\em Phys. Rev.
  Lett.} {\bfseries 117} no.~2, (2016) 021601},
\href{http://arxiv.org/abs/1601.05416}{{\ttfamily arXiv:1601.05416 [hep-th]}}.

\bibitem{Faulkner:2017vdd}
T.~Faulkner and A.~Lewkowycz, ``{Bulk locality from modular flow},''
  \href{http://dx.doi.org/10.1007/JHEP07(2017)151}{{\em JHEP} {\bfseries 07}
  (2017) 151},
\href{http://arxiv.org/abs/1704.05464}{{\ttfamily arXiv:1704.05464 [hep-th]}}.

\bibitem{Cotler:2017erl}
J.~Cotler, P.~Hayden, G.~Penington, G.~Salton, B.~Swingle, and M.~Walter,
  ``{Entanglement Wedge Reconstruction via Universal Recovery Channels},''
  \href{http://dx.doi.org/10.1103/PhysRevX.9.031011}{{\em Phys. Rev.}
  {\bfseries X9} no.~3, (2019) 031011},
\href{http://arxiv.org/abs/1704.05839}{{\ttfamily arXiv:1704.05839 [hep-th]}}.

\bibitem{Sorkin_1983}
R.~D. Sorkin, ``{On the Entropy of the vacuum outside a horizon},'' in {\em
  Tenth International Conference on General Relativity and Gravitation},
  B.~Bertotti, F.~de~Felice, and A.~Pascolini, eds., pp.~734--736.
\newblock Consiglio Nazionale Delle Ricerche, Roma (held in Padova, 4-9 July,
  1983).
\newblock
\href{http://arxiv.org/abs/arXiv:1402.3589}{{\ttfamily arXiv:1402.3589}}.
\newblock

\bibitem{Bombelli:1986rw}
L.~Bombelli, R.~K. Koul, J.~Lee, and R.~D. Sorkin, ``{A Quantum Source of
  Entropy for Black Holes},''
\href{http://dx.doi.org/10.1103/PhysRevD.34.373}{{\em Phys. Rev.} {\bfseries
  D34} (1986) 373--383}.

\bibitem{Srednicki:1993im}
M.~Srednicki, ``{Entropy and area},''
  \href{http://dx.doi.org/10.1103/PhysRevLett.71.666}{{\em Phys. Rev. Lett.}
  {\bfseries 71} (1993) 666--669},
\href{http://arxiv.org/abs/hep-th/9303048}{{\ttfamily arXiv:hep-th/9303048
  [hep-th]}}.

\bibitem{Wald:1993nt}
R.~M. Wald, ``{Black hole entropy is the Noether charge},''
  \href{http://dx.doi.org/10.1103/PhysRevD.48.R3427}{{\em Phys. Rev.}
  {\bfseries D48} no.~8, (1993) R3427--R3431},
\href{http://arxiv.org/abs/gr-qc/9307038}{{\ttfamily arXiv:gr-qc/9307038
  [gr-qc]}}.

\bibitem{Iyer:1994ys}
V.~Iyer and R.~M. Wald, ``{Some properties of Noether charge and a proposal for
  dynamical black hole entropy},''
  \href{http://dx.doi.org/10.1103/PhysRevD.50.846}{{\em Phys. Rev.} {\bfseries
  D50} (1994) 846--864},
\href{http://arxiv.org/abs/gr-qc/9403028}{{\ttfamily arXiv:gr-qc/9403028
  [gr-qc]}}.

\bibitem{Jacobson:1993vj}
T.~Jacobson, G.~Kang, and R.~C. Myers, ``{On black hole entropy},''
  \href{http://dx.doi.org/10.1103/PhysRevD.49.6587}{{\em Phys. Rev.} {\bfseries
  D49} (1994) 6587--6598},
\href{http://arxiv.org/abs/gr-qc/9312023}{{\ttfamily arXiv:gr-qc/9312023
  [gr-qc]}}.

\bibitem{Dong:2013qoa}
X.~Dong, ``{Holographic Entanglement Entropy for General Higher Derivative
  Gravity},'' \href{http://dx.doi.org/10.1007/JHEP01(2014)044}{{\em JHEP}
  {\bfseries 01} (2014) 044},
\href{http://arxiv.org/abs/1310.5713}{{\ttfamily arXiv:1310.5713 [hep-th]}}.

\bibitem{Hung:2011ta}
L.-Y. Hung, R.~C. Myers, and M.~Smolkin, ``{Some Calculable Contributions to
  Holographic Entanglement Entropy},''
  \href{http://dx.doi.org/10.1007/JHEP08(2011)039}{{\em JHEP} {\bfseries 08}
  (2011) 039}, \href{http://arxiv.org/abs/1105.6055}{{\ttfamily arXiv:1105.6055
  [hep-th]}}.

\bibitem{Myers:1994sg}
R.~C. Myers, ``{Black hole entropy in two-dimensions},''
  \href{http://dx.doi.org/10.1103/PhysRevD.50.6412}{{\em Phys.\ Rev.\ D}
  {\bfseries 50} (1994) 6412--6421},
  \href{http://arxiv.org/abs/hep-th/9405162}{{\ttfamily arXiv:hep-th/9405162}}.

\bibitem{Krtous:2014pva}
P.~Krtous and A.~Zelnikov, ``{Minimal surfaces and entanglement entropy in
  anti-de Sitter space},''
  \href{http://dx.doi.org/10.1007/JHEP10(2014)077}{{\em JHEP} {\bfseries 10}
  (2014) 077},
\href{http://arxiv.org/abs/1406.7659}{{\ttfamily arXiv:1406.7659 [hep-th]}}.

\bibitem{Myers:1987yn}
R.~C. Myers, ``{Higher Derivative Gravity, Surface Terms and String Theory},''
\href{http://dx.doi.org/10.1103/PhysRevD.36.392}{{\em Phys. Rev.} {\bfseries
  D36} (1987) 392}.

\bibitem{Hung:2011xb}
L.-Y. Hung, R.~C. Myers, and M.~Smolkin, ``{On Holographic Entanglement Entropy
  and Higher Curvature Gravity},''
  \href{http://dx.doi.org/10.1007/JHEP04(2011)025}{{\em JHEP} {\bfseries 04}
  (2011) 025}, \href{http://arxiv.org/abs/1101.5813}{{\ttfamily arXiv:1101.5813
  [hep-th]}}.

\bibitem{Gradshteyn:1702455}
I.~S. Gradshteyn, I.~M. Ryzhik, D.~Zwillinger, and V.~Moll,
  \href{http://dx.doi.org/0123849330}{{\em {Table of integrals, series, and
  products; 8th ed.}}}
\newblock Academic Press, Amsterdam, Sep, 2014.
\newblock \url{https://cds.cern.ch/record/1702455}.

\bibitem{Larsen:1995ax}
F.~Larsen and F.~Wilczek, ``{Renormalization of black hole entropy and of the
  gravitational coupling constant},''
  \href{http://dx.doi.org/10.1016/0550-3213(95)00548-X}{{\em Nucl. Phys.}
  {\bfseries B458} (1996) 249--266},
\href{http://arxiv.org/abs/hep-th/9506066}{{\ttfamily arXiv:hep-th/9506066
  [hep-th]}}.

\bibitem{Kabat:1995eq}
D.~N. Kabat, ``{Black hole entropy and entropy of entanglement},''
  \href{http://dx.doi.org/10.1016/0550-3213(95)00443-V}{{\em Nucl. Phys.}
  {\bfseries B453} (1995) 281--299},
\href{http://arxiv.org/abs/hep-th/9503016}{{\ttfamily arXiv:hep-th/9503016
  [hep-th]}}.

\bibitem{Headrick:2010zt}
M.~Headrick, ``{Entanglement Renyi entropies in holographic theories},''
  \href{http://dx.doi.org/10.1103/PhysRevD.82.126010}{{\em Phys. Rev.}
  {\bfseries D82} (2010) 126010},
\href{http://arxiv.org/abs/1006.0047}{{\ttfamily arXiv:1006.0047 [hep-th]}}.

\bibitem{Dong:2016hjy}
X.~Dong, A.~Lewkowycz, and M.~Rangamani, ``{Deriving covariant holographic
  entanglement},'' \href{http://dx.doi.org/10.1007/JHEP11(2016)028}{{\em JHEP}
  {\bfseries 11} (2016) 028},
\href{http://arxiv.org/abs/1607.07506}{{\ttfamily arXiv:1607.07506 [hep-th]}}.

\bibitem{Maldacena:2016hyu}
J.~Maldacena and D.~Stanford, ``{Remarks on the Sachdev-Ye-Kitaev model},''
  \href{http://dx.doi.org/10.1103/PhysRevD.94.106002}{{\em Phys. Rev. D}
  {\bfseries 94} no.~10, (2016) 106002},
  \href{http://arxiv.org/abs/1604.07818}{{\ttfamily arXiv:1604.07818
  [hep-th]}}.

\bibitem{Sachdev:1992fk}
S.~Sachdev and J.~Ye, ``{Gapless spin fluid ground state in a random, quantum
  Heisenberg magnet},''
  \href{http://dx.doi.org/10.1103/PhysRevLett.70.3339}{{\em Phys. Rev. Lett.}
  {\bfseries 70} (1993) 3339},
  \href{http://arxiv.org/abs/cond-mat/9212030}{{\ttfamily
  arXiv:cond-mat/9212030}}.

\bibitem{Sachdev:2010um}
S.~Sachdev, ``{Holographic metals and the fractionalized Fermi liquid},''
  \href{http://dx.doi.org/10.1103/PhysRevLett.105.151602}{{\em Phys. Rev.
  Lett.} {\bfseries 105} (2010) 151602},
  \href{http://arxiv.org/abs/1006.3794}{{\ttfamily arXiv:1006.3794 [hep-th]}}.

\bibitem{Ktalks}
A.~Kitaev, ``{A simple model of quantum holography},'' {\em Talks at KITP on
  April 7, 2015 and May 27, 2015} ,
  \href{http://arxiv.org/abs/http://online.kitp.ucsb.edu/online/entangled15/kitaev/,
  http://online.kitp.ucsb.edu/online/entangled15/kitaev2/}{{\ttfamily
  http://online.kitp.ucsb.edu/online/entangled15/kitaev/,
  http://online.kitp.ucsb.edu/online/entangled15/kitaev2/}}.

\bibitem{Karch:2001cw}
A.~Karch and L.~Randall, ``{Localized gravity in string theory},''
  \href{http://dx.doi.org/10.1103/PhysRevLett.87.061601}{{\em Phys. Rev. Lett.}
  {\bfseries 87} (2001) 061601},
  \href{http://arxiv.org/abs/hep-th/0105108}{{\ttfamily arXiv:hep-th/0105108}}.

\bibitem{DeWolfe:2001pq}
O.~DeWolfe, D.~Z. Freedman, and H.~Ooguri, ``{Holography and defect conformal
  field theories},'' \href{http://dx.doi.org/10.1103/PhysRevD.66.025009}{{\em
  Phys. Rev. D} {\bfseries 66} (2002) 025009},
  \href{http://arxiv.org/abs/hep-th/0111135}{{\ttfamily arXiv:hep-th/0111135}}.

\bibitem{DHoker:2007hhe}
E.~D'Hoker, J.~Estes, and M.~Gutperle, ``{Exact half-BPS Type IIB interface
  solutions. II. Flux solutions and multi-Janus},''
  \href{http://dx.doi.org/10.1088/1126-6708/2007/06/022}{{\em JHEP} {\bfseries
  06} (2007) 022}, \href{http://arxiv.org/abs/0705.0024}{{\ttfamily
  arXiv:0705.0024 [hep-th]}}.

\bibitem{DHoker:2008rje}
E.~D'Hoker, J.~Estes, M.~Gutperle, and D.~Krym, ``{Exact Half-BPS Flux
  Solutions in M-theory II: Global solutions asymptotic to AdS(7) x S**4},''
  \href{http://dx.doi.org/10.1088/1126-6708/2008/12/044}{{\em JHEP} {\bfseries
  12} (2008) 044}, \href{http://arxiv.org/abs/0810.4647}{{\ttfamily
  arXiv:0810.4647 [hep-th]}}.

\bibitem{Chiodaroli:2009yw}
M.~Chiodaroli, M.~Gutperle, and D.~Krym, ``{Half-BPS Solutions locally
  asymptotic to AdS(3) x S**3 and interface conformal field theories},''
  \href{http://dx.doi.org/10.1007/JHEP02(2010)066}{{\em JHEP} {\bfseries 02}
  (2010) 066}, \href{http://arxiv.org/abs/0910.0466}{{\ttfamily arXiv:0910.0466
  [hep-th]}}.

\bibitem{Chiodaroli:2011nr}
M.~Chiodaroli, E.~D'Hoker, Y.~Guo, and M.~Gutperle, ``{Exact half-BPS
  string-junction solutions in six-dimensional supergravity},''
  \href{http://dx.doi.org/10.1007/JHEP12(2011)086}{{\em JHEP} {\bfseries 12}
  (2011) 086}, \href{http://arxiv.org/abs/1107.1722}{{\ttfamily arXiv:1107.1722
  [hep-th]}}.

\bibitem{Chiodaroli:2012vc}
M.~Chiodaroli, E.~D'Hoker, and M.~Gutperle, ``{Holographic duals of Boundary
  CFTs},'' \href{http://dx.doi.org/10.1007/JHEP07(2012)177}{{\em JHEP}
  {\bfseries 07} (2012) 177}, \href{http://arxiv.org/abs/1205.5303}{{\ttfamily
  arXiv:1205.5303 [hep-th]}}.

\bibitem{Emparan:1999wa}
R.~Emparan, G.~T. Horowitz, and R.~C. Myers, ``{Exact description of black
  holes on branes},''
  \href{http://dx.doi.org/10.1088/1126-6708/2000/01/007}{{\em JHEP} {\bfseries
  01} (2000) 007}, \href{http://arxiv.org/abs/hep-th/9911043}{{\ttfamily
  arXiv:hep-th/9911043}}.

\bibitem{Emparan:1999fd}
R.~Emparan, G.~T. Horowitz, and R.~C. Myers, ``{Exact description of black
  holes on branes. 2. Comparison with BTZ black holes and black strings},''
  \href{http://dx.doi.org/10.1088/1126-6708/2000/01/021}{{\em JHEP} {\bfseries
  01} (2000) 021}, \href{http://arxiv.org/abs/hep-th/9912135}{{\ttfamily
  arXiv:hep-th/9912135}}.

\bibitem{dSone}
J.~Maldacena, ``{Entropies of subregions in two-dimensional cosmological
  models}.'' {Black Hole Microstructure Conference, June 8-12, 2020}.
\newblock \url{https://youtu.be/wzVMDrfaRXA?t=5406}.

\bibitem{dStwo}
Y.~Cheng, V.~Gorbenko, and J.~Maldacena {\em in preparation} .

\bibitem{Maldacena:2019cbz}
J.~Maldacena, G.~J. Turiaci, and Z.~Yang, ``{Two dimensional Nearly de Sitter
  gravity},'' \href{http://arxiv.org/abs/1904.01911}{{\ttfamily
  arXiv:1904.01911 [hep-th]}}.

\bibitem{Ooguri:2020sua}
H.~Ooguri and T.~Takayanagi, ``{Cobordism Conjecture in AdS},''
  \href{http://arxiv.org/abs/2006.13953}{{\ttfamily arXiv:2006.13953
  [hep-th]}}.

\bibitem{Harlow:2011ke}
D.~Harlow and D.~Stanford, ``{Operator Dictionaries and Wave Functions in
  AdS/CFT and dS/CFT},'' \href{http://arxiv.org/abs/1104.2621}{{\ttfamily
  arXiv:1104.2621 [hep-th]}}.

\bibitem{Takayanagi:2017knl}
T.~Takayanagi and K.~Umemoto, ``{Entanglement of purification through
  holographic duality},''
  \href{http://dx.doi.org/10.1038/s41567-018-0075-2}{{\em Nature Phys.}
  {\bfseries 14} no.~6, (2018) 573--577},
  \href{http://arxiv.org/abs/1708.09393}{{\ttfamily arXiv:1708.09393
  [hep-th]}}.

\bibitem{Nguyen:2017yqw}
P.~Nguyen, T.~Devakul, M.~G. Halbasch, M.~P. Zaletel, and B.~Swingle,
  ``{Entanglement of purification: from spin chains to holography},''
  \href{http://dx.doi.org/10.1007/JHEP01(2018)098}{{\em JHEP} {\bfseries 01}
  (2018) 098}, \href{http://arxiv.org/abs/1709.07424}{{\ttfamily
  arXiv:1709.07424 [hep-th]}}.

\bibitem{Dutta:2019gen}
S.~Dutta and T.~Faulkner, ``{A canonical purification for the entanglement
  wedge cross-section},'' \href{http://arxiv.org/abs/1905.00577}{{\ttfamily
  arXiv:1905.00577 [hep-th]}}.

\bibitem{Tamaoka:2018ned}
K.~Tamaoka, ``{Entanglement Wedge Cross Section from the Dual Density
  Matrix},'' \href{http://dx.doi.org/10.1103/PhysRevLett.122.141601}{{\em Phys.
  Rev. Lett.} {\bfseries 122} no.~14, (2019) 141601},
  \href{http://arxiv.org/abs/1809.09109}{{\ttfamily arXiv:1809.09109
  [hep-th]}}.

\bibitem{Kusuki:2019evw}
Y.~Kusuki and K.~Tamaoka, ``{Entanglement Wedge Cross Section from CFT:
  Dynamics of Local Operator Quench},''
  \href{http://dx.doi.org/10.1007/JHEP02(2020)017}{{\em JHEP} {\bfseries 02}
  (2020) 017}, \href{http://arxiv.org/abs/1909.06790}{{\ttfamily
  arXiv:1909.06790 [hep-th]}}.

\bibitem{Kusuki:2019rbk}
Y.~Kusuki and K.~Tamaoka, ``{Dynamics of Entanglement Wedge Cross Section from
  Conformal Field Theories},''
  \href{http://arxiv.org/abs/1907.06646}{{\ttfamily arXiv:1907.06646
  [hep-th]}}.

\bibitem{Kudler-Flam:2018qjo}
J.~Kudler-Flam and S.~Ryu, ``{Entanglement negativity and minimal entanglement
  wedge cross sections in holographic theories},''
  \href{http://dx.doi.org/10.1103/PhysRevD.99.106014}{{\em Phys. Rev. D}
  {\bfseries 99} no.~10, (2019) 106014},
  \href{http://arxiv.org/abs/1808.00446}{{\ttfamily arXiv:1808.00446
  [hep-th]}}.

\bibitem{Kusuki:2019zsp}
Y.~Kusuki, J.~Kudler-Flam, and S.~Ryu, ``{Derivation of Holographic Negativity
  in AdS$_3$/CFT$_2$},''
  \href{http://dx.doi.org/10.1103/PhysRevLett.123.131603}{{\em Phys. Rev.
  Lett.} {\bfseries 123} no.~13, (2019) 131603},
  \href{http://arxiv.org/abs/1907.07824}{{\ttfamily arXiv:1907.07824
  [hep-th]}}.

\bibitem{Callan:1994py}
C.~G. Callan, Jr. and F.~Wilczek, ``{On geometric entropy},''
  \href{http://dx.doi.org/10.1016/0370-2693(94)91007-3}{{\em Phys. Lett.}
  {\bfseries B333} (1994) 55--61},
\href{http://arxiv.org/abs/hep-th/9401072}{{\ttfamily arXiv:hep-th/9401072
  [hep-th]}}.

\bibitem{Fursaev:1994ea}
D.~V. Fursaev and S.~N. Solodukhin, ``{On one loop renormalization of black
  hole entropy},'' \href{http://dx.doi.org/10.1016/0370-2693(95)01290-7}{{\em
  Phys. Lett.} {\bfseries B365} (1996) 51--55},
\href{http://arxiv.org/abs/hep-th/9412020}{{\ttfamily arXiv:hep-th/9412020
  [hep-th]}}.

\bibitem{Fursaev:1995ef}
D.~V. Fursaev and S.~N. Solodukhin, ``{On the description of the Riemannian
  geometry in the presence of conical defects},''
  \href{http://dx.doi.org/10.1103/PhysRevD.52.2133}{{\em Phys. Rev.} {\bfseries
  D52} (1995) 2133--2143},
\href{http://arxiv.org/abs/hep-th/9501127}{{\ttfamily arXiv:hep-th/9501127
  [hep-th]}}.

\bibitem{Susskind:1994sm}
L.~Susskind and J.~Uglum, ``{Black hole entropy in canonical quantum gravity
  and superstring theory},''
  \href{http://dx.doi.org/10.1103/PhysRevD.50.2700}{{\em Phys.\ Rev.\ D}
  {\bfseries 50} (1994) 2700--2711},
  \href{http://arxiv.org/abs/hep-th/9401070}{{\ttfamily arXiv:hep-th/9401070}}.

\bibitem{Donnelly:2014fua}
W.~Donnelly and A.~C. Wall, ``{Entanglement entropy of electromagnetic edge
  modes},'' \href{http://dx.doi.org/10.1103/PhysRevLett.114.111603}{{\em Phys.
  Rev. Lett.} {\bfseries 114} no.~11, (2015) 111603},
\href{http://arxiv.org/abs/1412.1895}{{\ttfamily arXiv:1412.1895 [hep-th]}}.

\bibitem{Donnelly:2015hxa}
W.~Donnelly and A.~C. Wall, ``{Geometric entropy and edge modes of the
  electromagnetic field},''
  \href{http://dx.doi.org/10.1103/PhysRevD.94.104053}{{\em Phys. Rev.}
  {\bfseries D94} no.~10, (2016) 104053},
\href{http://arxiv.org/abs/1506.05792}{{\ttfamily arXiv:1506.05792 [hep-th]}}.

\bibitem{Balasubramanian:2013rqa}
V.~Balasubramanian, B.~Czech, B.~D. Chowdhury, and J.~de~Boer, ``{The entropy
  of a hole in spacetime},''
  \href{http://dx.doi.org/10.1007/JHEP10(2013)220}{{\em JHEP} {\bfseries 10}
  (2013) 220}, \href{http://arxiv.org/abs/1305.0856}{{\ttfamily arXiv:1305.0856
  [hep-th]}}.

\bibitem{Balasubramanian:2013lsa}
V.~Balasubramanian, B.~D. Chowdhury, B.~Czech, J.~de~Boer, and M.~P. Heller,
  ``{Bulk curves from boundary data in holography},''
  \href{http://dx.doi.org/10.1103/PhysRevD.89.086004}{{\em Phys.\ Rev.\ D}
  {\bfseries 89} no.~8, (2014) 086004},
  \href{http://arxiv.org/abs/1310.4204}{{\ttfamily arXiv:1310.4204 [hep-th]}}.

\bibitem{Czech:2014wka}
B.~Czech, X.~Dong, and J.~Sully, ``{Holographic Reconstruction of General Bulk
  Surfaces},'' \href{http://dx.doi.org/10.1007/JHEP11(2014)015}{{\em JHEP}
  {\bfseries 11} (2014) 015}, \href{http://arxiv.org/abs/1406.4889}{{\ttfamily
  arXiv:1406.4889 [hep-th]}}.

\bibitem{Myers:2014jia}
R.~C. Myers, J.~Rao, and S.~Sugishita, ``{Holographic Holes in Higher
  Dimensions},'' \href{http://dx.doi.org/10.1007/JHEP06(2014)044}{{\em JHEP}
  {\bfseries 06} (2014) 044}, \href{http://arxiv.org/abs/1403.3416}{{\ttfamily
  arXiv:1403.3416 [hep-th]}}.

\bibitem{Headrick:2014eia}
M.~Headrick, R.~C. Myers, and J.~Wien, ``{Holographic Holes and Differential
  Entropy},'' \href{http://dx.doi.org/10.1007/JHEP10(2014)149}{{\em JHEP}
  {\bfseries 10} (2014) 149}, \href{http://arxiv.org/abs/1408.4770}{{\ttfamily
  arXiv:1408.4770 [hep-th]}}.

\end{thebibliography}\endgroup
\bibliographystyle{utphys}

\end{document}